\documentclass{article}

\PassOptionsToPackage{numbers, compress}{natbib}
\usepackage[preprint]{neurips_2023}
\usepackage{neurips_2023}

\usepackage{amsmath,amssymb,amsfonts}
\usepackage{algorithmic}
\usepackage{graphicx}
\usepackage{textcomp}
\usepackage{xcolor}
\usepackage[breakable]{tcolorbox}

\usepackage{amsmath}
\usepackage{amssymb}
\usepackage{mathtools}
\usepackage{amsthm}

\usepackage{subcaption}
\usepackage{listings}

\usepackage{siunitx}
\usepackage{algorithmic}
\usepackage{graphicx}
\usepackage{textcomp}
\usepackage{xcolor}
\usepackage{mdframed}
\usepackage{svg}
\usepackage{booktabs}
\usepackage{graphicx}
\usepackage{overpic}
\usepackage{xspace}
\usepackage{enumitem}
\usepackage{multicol}
\usepackage{xcolor}

\usepackage{color}
\usepackage{colortbl}
\usepackage{soul}
\usepackage{paralist}
\usepackage{fancyhdr} 
\usepackage{float}
\usepackage{hyperref}
\usepackage{multirow}
\usepackage{soul}
\usepackage[misc]{ifsym}
\usepackage{ifthen}
\newboolean{doublecolumn}
\setboolean{doublecolumn}{true}

\usepackage[capitalize,noabbrev]{cleveref}

\theoremstyle{plain}

\theoremstyle{definition}

\theoremstyle{remark}

{\bfseries}{\itshape}
\theoremstyle{definition}
{\bfseries}{\normalfont}
{\bfseries}{\rmfamily}
{\bfseries}{\rmfamily}
{\bfseries}{\rmfamily}

\newcommand{\llama}{LLaMA\xspace}
\newcommand{\llamaTwo}{LLaMA2\xspace}
\newcommand{\alpaca}{Alpaca\xspace}
\newcommand{\vicuna}{Vicuna\xspace}
\newcommand{\codellama}{CodeLLaMA\xspace}

\newcommand{\truthfulqa}{TruthfulQA\xspace}
\newcommand{\triviaqa}{TriviaQA\xspace}
\newcommand{\nqopen}{NQ-OPEN\xspace}
\newcommand{\realtoxicityprompt}{RealToxicityPrompt\xspace}
\newcommand{\wmt}{WMT\xspace}
\newcommand{\mbpp}{MBPP\xspace}
\newcommand{\humaneval}{HumanEval\xspace}

\newcommand{\random}{Random\xspace}
\newcommand{\avgent}{Average Entropy\xspace}
\newcommand{\maxent}{Maximum Entropy\xspace}
\newcommand{\avglik}{Average Likelihood\xspace}
\newcommand{\maxlik}{Maximum Likelihood\xspace}
\newcommand{\boxMon}{Box-based method\xspace}
\newcommand{\quan}{Quantitative method\xspace}
\newcommand{\selfcheckgpt}{SelfCheckGPT\xspace}

\newcommand{\randomShort}{Random\xspace}
\newcommand{\avgentShort}{AvgEnt\xspace}
\newcommand{\maxentShort}{MaxEnt\xspace}
\newcommand{\avglikShort}{AvgLik\xspace}
\newcommand{\maxlikShort}{MaxLik\xspace}
\newcommand{\boxShort}{Box\xspace}
\newcommand{\quanShort}{Quan\xspace}
\newcommand{\selfcheckgptShort}{SC\xspace}

\newcommand{\rqone}{Can the unsafe output be identified at the early stage of LLM generation?\xspace}
\newcommand{\rqtwo}{What is the effectiveness of online safety analysis techniques in analyzing open-source LLMs?\xspace}
\newcommand{\rqthree}{How is the performance of online safety analysis techniques in analyzing closed-source LLMs?\xspace}
\newcommand{\rqfour}{Can hybridization approaches improve the performance of online safety analysis for LLMs?\xspace}

\definecolor{lightgray}{HTML}{d4d4d6}
\definecolor{mildgray}{HTML}{D6D6D6} 
\definecolor{darkgray}{HTML}{B5B4B5}
\definecolor{cyan}{HTML}{DCF9F6}

\colorlet{mylightgray}{lightgray!70}

\def\BibTeX{{\rm B\kern-.05em{\sc i\kern-.025em b}\kern-.08em
    T\kern-.1667em\lower.7ex\hbox{E}\kern-.125emX}}

\makeatletter
\DeclareRobustCommand\onedot{\futurelet\@let@token\@onedot}
\def\@onedot{\ifx\@let@token.\else.\null\fi\xspace}
\def\etal{{et al}\onedot}

\title{Online Safety Analysis for LLMs: a Benchmark, an Assessment, and a Path Forward}

\author{%
  Xuan Xie$^{1}$,
  Jiayang Song$^{1}$,
  Zhehua Zhou$^{1}$,
  Yuheng Huang$^{1}$,
  Da Song$^{1}$,
  Lei Ma $^{2, 1}$  \\ 
    $^1${University of Alberta, Canada} \quad $^2${The University of Tokyo, Japan} \\
}

\makeatletter
\renewcommand*{\@fnsymbol}[1]{\ensuremath{\ifcase#1\or {\textrm{\Letter}}\else\@ctrerr\fi}}
\makeatother

\begin{document}

\maketitle

\begin{abstract}
While Large Language Models (LLMs) have seen widespread applications across numerous fields, their limited interpretability poses concerns regarding their safe operations from multiple aspects, e.g., truthfulness, robustness, and fairness. 
Recent research has started developing quality assurance methods for LLMs, introducing techniques such as offline detector-based or uncertainty estimation methods. 
However, these approaches predominantly concentrate on post-generation analysis, leaving the online safety analysis for LLMs during the generation phase an unexplored area.
To bridge this gap, we conduct in this work a comprehensive evaluation of the effectiveness of existing online safety analysis methods on LLMs. 
We begin with a pilot study that validates the feasibility of detecting unsafe outputs in the early generation process. 
Following this, we establish the first publicly available benchmark of online safety analysis for LLMs, including a broad spectrum of methods, models, tasks, datasets, and evaluation metrics.
Utilizing this benchmark, we extensively analyze the performance of state-of-the-art online safety analysis methods on both open-source and closed-source LLMs. 
This analysis reveals the strengths and weaknesses of individual methods and offers valuable insights into selecting the most appropriate method based on specific application scenarios and task requirements.
Furthermore, we also explore the potential of using hybridization methods, i.e., combining multiple methods to derive a collective safety conclusion, to enhance the efficacy of online safety analysis for LLMs. 
Our findings indicate a promising direction for the development of innovative and trustworthy quality assurance methodologies for LLMs, facilitating their reliable deployments across diverse domains.
\end{abstract}

\section{Introduction}
\label{sec:intro}


Recent research highlights the remarkable achievements of \emph{Large Language Models} (LLMs) across a multitude of fields, such as natural language processing~\cite{achiam2023gpt}, code generation~\cite{vaithilingam2022expectation}, robotic system control~\cite{ren2023robots,zhou2023isr}, medical diagnostics~\cite{cascella2023evaluating}, and finance~\cite{wu2023bloomberggpt}.
Trained on extensive and diverse datasets~\cite{wenzek2019ccnet,raffel2020exploring,gao2020pile}, LLMs are capable of producing responses that reflect common-sense knowledge and mirror human-like intelligence.
Their versatility, effectiveness, and scalability have made them vital in a variety of uses, providing numerous benefits in the development of Artificial General Intelligence (AGI)~\cite{goertzel2014artificial}.


Alongside the remarkable capabilities of LLMs, concerns regarding their \emph{safety} issues~\cite{ouyang2022training,huang2023survey,inan2023llama,touvron2023llama}, such as hallucination~\cite{ji2023survey}, toxicity~\cite{gehman2020realtoxicityprompts}, fairness~\cite{li2023survey}, and robustness~\cite{wang2023large}, have garnered increasing attention in the community. 
For instance, recent studies reveal that LLM may generate nonfactual and inaccurate outputs with high confidence, a phenomenon known as hallucination~\cite{liu2021token,rateike2023weakly}.
A well-known example of this is that a lawyer in Canada used a fake case generated by ChatGPT to prepare legal briefs~\cite{lawyerhallucination}.
Furthermore, when exposed to carefully designed lure, such as toxic prompts, LLM could generate toxic content, which includes but is not limited to hate speech, misinformation dissemination, biased language, and offensive content~\cite{ousidhoum2021probing,gehman2020realtoxicityprompts}.
These safety issues could potentially hinder the trustworthy and reliable deployment of LLMs and impact societal well-being and stability.
Therefore, the development of effective \emph{safety analysis} methods for LLMs to address their safety concerns is urgently needed. 

Safety analysis for traditional Deep Learning (DL) models, e.g., Deep Neural Networks (DNNs), has emerged as an important research area in recent years, attracting attention from both academia and industry~\cite{pei2017deepxplore, ma2018deepgauge, kim2019guiding,li2023trustworthy,riccio2020testing,katz2017reluplex,gehr2018ai2,lwakatare2020large,gao2022adaptive, yang2022revisiting, riccio2023and, wang2022bet,wang2021robot,wang2021prioritizing,feng2020deepgini}.
Numerous studies have focused on methods such as testing (identifying inputs that trigger abnormal or unsafe behavior of the system)~\cite{pei2017deepxplore,ma2018deepgauge,kim2019guiding,tian2018deeptest}, repairing (correcting errors through data augmentation or model modifications)~\cite{sotoudeh2021provable, stocco2020misbehaviour, sun2022causality,kim2023repairing,yang2022neural}, and verification (certifying the safety of DNNs through rigorous mathematical analysis)~\cite{katz2017reluplex,gehr2018ai2,huang2020survey,sun2021probabilistic}. 
However, rather than ensuring safety in real-time during operation, these methods typically conduct analysis after the generation of outputs.


In contrast, \emph{online safety analysis} methods are designed to offer real-time safety assessments by continuously monitoring and analyzing the system's behavior, aiming to ensure that the system operates correctly according to specific safety requirements~\cite{henzinger2019outside,lukina2021into,cheng2019runtime,guerin2022unifying,guerin2022evaluation,zolfagharian2023smarla,stocco2020misbehaviour,aslansefat2020safeml,stocco2022thirdeye}. 
Various online safety analysis methods have been developed for DL models.
For example, SafeOracle~\cite{stocco2020misbehaviour} predicts, at runtime, the unsafe behavior of DNNs by using a proxy model constructed to estimate the confidence levels.
Zolfagharian et al.~\cite{zolfagharian2023smarla} propose SMARLA for monitoring deep reinforcement learning agents, which is a black-box method and uses state abstraction to reduce the high-dimensional state space.
Henzinger et al.~\cite{henzinger2019outside} introduce a method for detecting novel behaviors of DNNs by analyzing their hidden layers in real time using program analysis abstractions. 

However, analyzing LLMs is generally considered more complex and challenging than traditional DL models for two main reasons: the unique and novel characteristics of LLMs, notably their auto-regressive nature~\cite{zhao2023survey}, and their vast number of parameters, e.g., even the smallest \llama~\cite{touvron2023llama} model has 7 billion parameters.
Hence, although state-of-the-art online safety analysis approaches have been proven to be useful for DL models, their performance when applied to LLMs remains uncertain.

\begin{figure*}[ht]
    \centering
    \includegraphics[width=\linewidth]{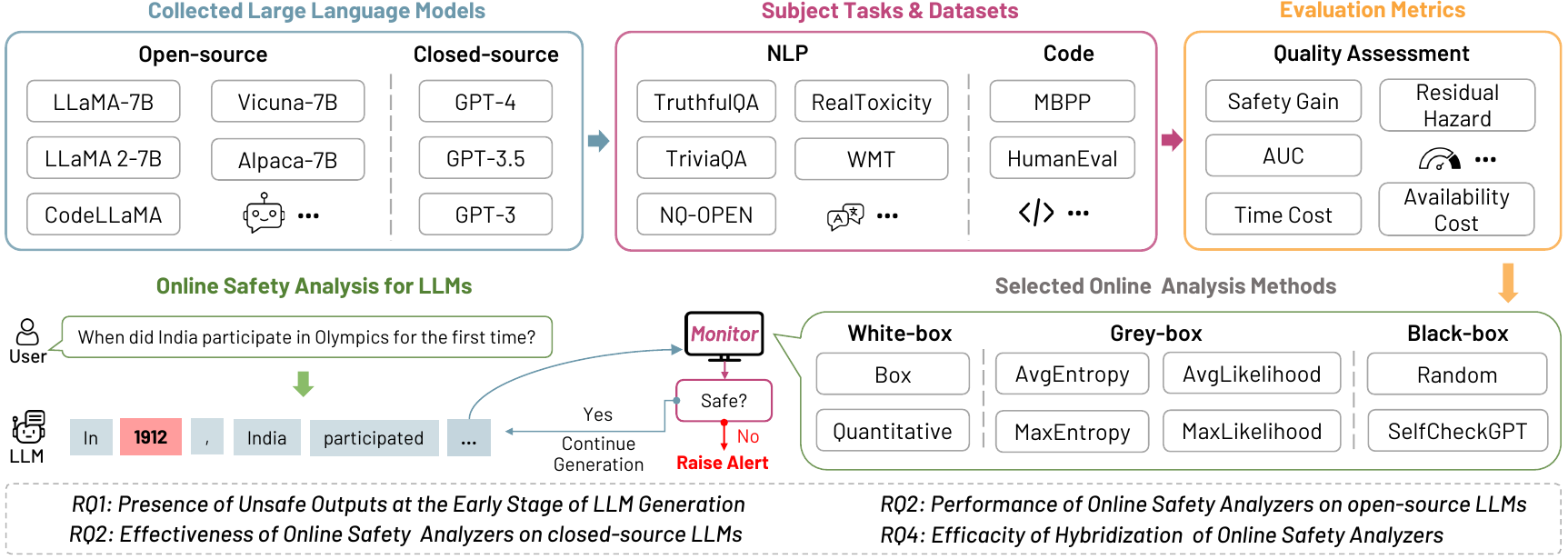}
    \caption{Overall workflow illustration.}
    \label{fig:workflow}
    \vspace{-10pt}
\end{figure*}

Developing effective safety analysis methods for LLMs is a research area that is still in its early exploratory stages.
While various approaches, such as detector-based~\cite{markov2023holistic, zhou-etal-2021-detecting, inan2023llama, bhatt2023purple, achintalwar2024detectors, zhang2024shieldlm} or uncertainty estimation~\cite{manakul2023selfcheckgpt, lin2023generating, xiong2024can, kuhn2023semantic} methods, have been proposed for LLMs, they are predominantly only suited for post-generation analysis. 
The field of online safety analysis for LLMs, to the best of our knowledge, remains an unexplored area.
Therefore, to bridge this gap, we conduct a comprehensive examination of the effectiveness of online safety analysis methods for LLMs in this work (see Fig.~\ref{fig:workflow}). 
We begin with a pilot study aimed at determining if the unsafe output of LLMs can be identified early in the generation process. 
The affirmative answer reveals the potential and feasibility of developing online safety analysis techniques for LLMs.
Then, we intend to systematically assess the efficacy of existing online safety analysis methods, which were originally designed for DL models, on LLMs.
For this purpose, we establish an extensive benchmark that encompasses eight distinct LLMs, eight different online safety analysis methods, seven datasets across various tasks and safety perspectives, and five evaluative metrics.
Leveraging this benchmark, a large-scale empirical study is conducted to explore the applicability and effectiveness of state-of-the-art online safety analysis approaches for LLMs. 
The experimental results suggest promising directions for the development of novel online safety analysis methods across diverse application scenarios for both open-source and closed-source LLMs.
Finally, we delve into the potential benefits of the hybridization method, i.e., the amalgamation of different methods to derive safety conclusions, in online safety analysis for LLMs. 
This exploratory examination underscores the strengths of hybridization over singular methods, paving the way for future advancements in the field.

The contributions of this work are summarized as follows:
\begin{compactitem}[$\bullet$]
\item To validate the feasibility of performing online safety analysis for LLMs, we initiate a pilot study, which includes two verification strategies, two LLMs and three distinct datasets.
This preliminary investigation reveals that, in most cases, unsafe output can be identified at the early stage of the generation process, highlighting the importance and potential of developing online safety analysis methods for LLMs.
\item To empower the research in the domain of online safety analysis for LLMs, we create a first public benchmark that consists of eight LLMs, eight online safety analysis techniques, five evaluation metrics, and seven datasets across diverse tasks and safety perspectives.
\item Leveraging the constructed benchmark, we perform a systematic and large-scale analysis of the performance and characteristics of existing online safety analysis approaches on both open-source and closed-source LLMs.
The results unveil the advantages and challenges of current methods and offer valuable insights into designing LLM-specific online safety analysis techniques.
\item We further investigate the potential benefits of hybridization methods, which attempt to combine the advantages of diverse individual methods for acquiring an improved performance. 
This exploration indicates novel directions for developing more effective online safety analysis methods for LLMs.


\end{compactitem}

The rest of the paper is structured as follows.
Section~\ref{sec:background} introduces the corresponding background.
Section~\ref{sec:pilot_study} describes the pilot study.
Section~\ref{sec:benchmark_construction} details the benchmark construction procedure.
Section~\ref{sec:empiricalstudy} is about the empirical study of investigating the online safety analysis methods on LLMs.
Section~\ref{sec:explore_study} presents the exploratory study on the hybridization of different online safety analysis methods.
Section~\ref{sec:discussion} discusses the potential influence of the study.
Section~\ref{sec:threatToVal} analyzes the threats that may affect the validity of the performed study.
Section~\ref{sec:related_work} describes the related works, and Section~\ref{sec:conclusion} concludes the paper.

\section{Background and Study Overview}
\label{sec:background}

In this section, we provide essential background knowledge, which includes an introduction to LLMs (Section~\ref{subsec:background_llm}), safety of LLMs (Section~\ref{subsec:background_safetyLLM}), and online safety analysis for DL models (Section~\ref{subsec:onlineSafetyAnalysis}).
Additionally, we also present an overview of our research question design in Section~\ref{subsec:overview_rq}.

\subsection{Large Language Models (LLMs)}
\label{subsec:background_llm}



LLMs can be categorized into three types: \textit{encoder-only}~\cite{devlin-etal-2019-bert}, \textit{encoder-decoder}~\cite{raffel2020exploring}, and \textit{decoder-only}~\cite{brown2020language, touvron2023llama, touvron2023llama2, achiam2023gpt}. 
Initially, inspired by the sequential nature of common NLP tasks, the \textit{encoder-decoder} architecture is introduced in~\cite{vaswani2017attention}, which employs an encoder to process the input sequence and a decoder to generate the output sequence.
However, the trend in developing LLMs has leaned towards adopting a \textit{decoder-only} structure, especially as models are scaled up and trained on extensive datasets and corpora~\cite{kaplan2020scaling}.
Motivated by this, we focus in this work only on \textit{decoder-only} LLMs, which are typically trained through an unsupervised learning approach with the loss function defined as

\begin{equation}
    \label{eq:next-token-pred}
    \small
    \max_{\theta} \sum_{t=1}^{T} \log p(w_t | w_{1:t-1}; \theta)
\end{equation}
where $w_t$ denotes the $t$-th token within a sequence containing $T$ tokens.
$p(w_t | w_{1:t-1}; \theta)$ indicates the probability of observing the $t$-th token given the sequence of preceding tokens and the model parameters $\theta$. 
The training objective is to maximize the likelihood of accurately predicting the next token in a sequence by optimizing the parameters $\theta$.

\begin{figure}[t]
    \centering
    \includegraphics[width=0.65\columnwidth]{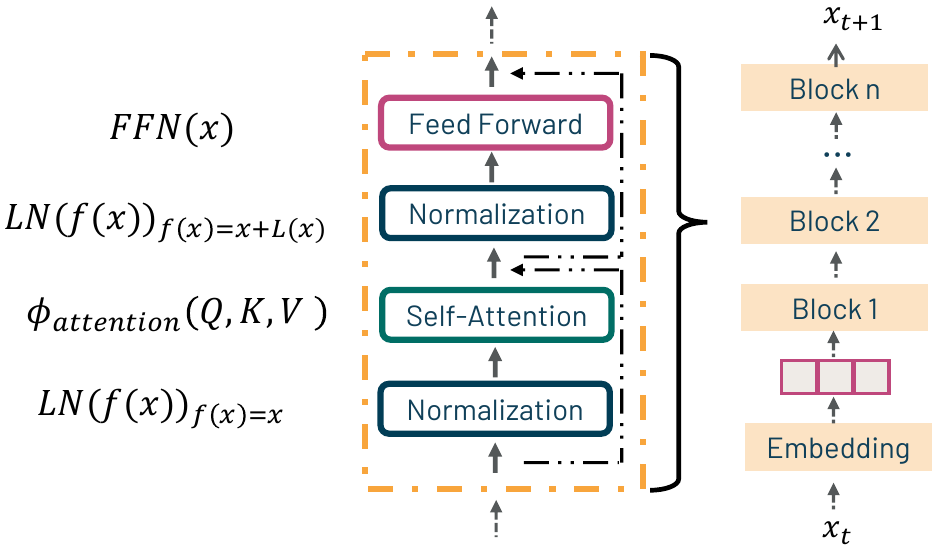}
    \caption{Decoder-only LLM illustration. }
    \vspace{-10pt}
    \label{fig:LLM}
\end{figure}

The foundational architecture of decoder-only LLMs is built upon the transformer model, which has had a profound impact on both the natural language processing (NLP)~\cite{vaswani2017attention} and computer vision~\cite{dosovitskiy2021an} fields. 
Despite their potential to grow extremely large with billions of parameters and conduct complex computations, transformers maintain a notably straightforward underlying structure. 
At the core of a transformer is a basic building block, which comprises multi-head self-attention mechanisms, position-wise feed-forward networks, and layer normalization paired with residual connections. 
The major architecture of the transformer is constructed by sequentially stacking these basic building blocks. 
A simple illustration of the structure of a decoder-only LLM is provided in Fig.~\ref{fig:LLM}. 
Initially, an embedding layer converts discrete tokens into continuous vector representations, facilitating subsequent computational processes. 
Then, these vector representations progress through a series of blocks, each adhering to a specific structure, starting with a normalization layer, proceeding to a multi-head self-attention layer, followed by another normalization layer, and concludes with a feed-forward layer. 
In the following, we present more details about each layer within these blocks.

Layer normalization is a widely recognized technique in neural network architecture design~\cite{ba2016layer}, employed to stabilize the dynamics of hidden states by normalizing the inputs
\begin{equation}
    \label{eq:layer_norm}
    \small
    LN (f(x))
\end{equation}
where we use $LN(\cdot)$ to denote the layer normalization computation.
Based on its position within the block, the input to the layer normalization, i.e., the function $f(x)$, can take the form of either $f(x)=x$ or $f(x) = x + L(x)$.
The former denotes a direct mapping, while the latter represents a layer encased in residual connections $L(x)$, a design that has proven effective for gradient propagation~\cite{he2016deep}.


The self-attention layer is one of the most important components in the block and is defined as
\begin{equation}
    \label{eq:attention}
    \small
    \phi_{attention} (Q, K, V) = \mathrm{softmax} (\frac{QK^{T}}{\sqrt{d_k}}) V
\end{equation}
where $Q$, $K$, and $V$ respectively indicate query, key, and value, which are derived by multiplying the input token representations with different trainable weight matrices. 
$d_k$ is the dimensionality of the keys and queries and is used for scaling. 
The self-attention layer is used to weigh the significance of different input parts and is the key to the success of the transformer. 

Finally, the position-wise feed-forward layer is defined as
\begin{equation}
    \label{eq:FFN}
    \small
    FFN(x) = \max(0, xW_1 + b_1)W_2 + b_2
\end{equation}
where $FFN(\cdot)$ refers to a fully connected feed-forward network with parameters $W_1$, $W_2$, $b_1$, and $b_2$. 
It consists of sequential linear transformations coupled with non-linear activation functions.




During \emph{inference time}, i.e., the generation phase that spans from when an input prompt is provided until an output is fully generated, 
the LLM processes and generates output in a real-time, sequential, token-by-token manner, aiming to maximize the likelihood of the entire sequence as
\begin{equation}
    \label{eq:llm-inference}
    \small
    {\prod}_{t=1}^{T} p(\hat{w}_t = w_{t} | w_{1:t-1}; \theta)
\end{equation}
where $\hat{w}_t = w_{t}$ indicates the model's selection for a token at time $t$. 
Note that although maximizing this likelihood is theoretically challenging due to computational constraints, practical approximations, such as greedy decoding~\cite{germann2003greedy}, are commonly employed as feasible solutions.
Such a sequential output generation process of LLMs opens up the opportunity for online safety analysis to be conducted concurrently with the text generation.


\subsection{Safety of LLM}
\label{subsec:background_safetyLLM}

Despite their remarkable abilities, recent research indicates that LLMs are susceptible to generating inappropriate and improper responses, posing a notable challenge to their safe operations~\cite{sun2024trustllm,wang2023decodingtrust}.
In this work, we define a \textbf{safe LLM} as: \emph{it does not produce inappropriate or harmful outputs, such as generating misleading information, creating bias, or engaging in malicious behavior}.
Due to the versatility and complexity of LLMs, the specific meaning of their safety could be different based on the actual task considered.  
The meaning of safety may encompass a range of considerations including, but not limited to,  truthfulness~\cite{lin2021truthfulqa}, toxicity~\cite{gehman2020realtoxicityprompts}, robustness~\cite{wang2023large}, privacy~\cite{yao2024survey}, translation accuracy~\cite{papineni-etal-2002-bleu}, machine ethics~\cite{hendrycks2020aligning}, stereotype~\cite{bolukbasi2016man}, and fairness~\cite{li2023survey}.
For simplicity, we refer to these different aspects collectively as safety in this work.
In the following, we use truthfulness and toxicity as two examples to illustrate the concept of safety for LLMs.

In LLM literature, truthfulness is usually defined as LLM-generated responses containing only factual claims, typically supported by reliable and publicly available evidence~\cite{lin2021truthfulqa}. 
When LLMs lack specific knowledge related to a query, a response is considered truthful only if it avoids making false statements, such as refusing to answer. 
Untruthfulness in LLMs can arise in two scenarios: (1) LLMs may not possess relevant knowledge but still produce unrealistic or misleading information, a phenomenon often described as hallucination~\cite{ji2023survey}; (2) LLMs might have the necessary knowledge but choose to generate untruthful answers, which is often referred as lying or deception~\cite{azaria2023internal, liu2023cognitive}. 

Toxicity in the context of LLMs refers to the generation of responses that are rude, disrespectful, or otherwise unreasonable~\cite{welbl2021challenges}. 
Different from truthfulness, determining whether a response is toxic can be more subjective, as perceptions of aggression can vary. 
However, toxic outputs still pose significant ethical and social risks, potentially causing harm and even breaching regulations like the European Parliament's Artificial Intelligence Act~\cite{madiega2021artificial}.
Furthermore, the issue of toxicity is time-sensitive. 
Given that LLMs typically generate content token by token, toxic words or phrases can inflict immediate harm upon generation.

The study of addressing the safety of LLM is still in a fast-evolving period, and existing research commonly employs similar methodologies:
they either investigate the internal states of LLMs~\cite{azaria2023internal, rateike2023weakly} or classify content after it has been generated~\cite{markov2023holistic, manakul2023selfcheckgpt}. 
These approaches predominantly focus on post-generation analysis, i.e., after the outputs have been generated. 
To the best of our knowledge, there has yet to be a comprehensive and systematic study dedicated to the online safety analysis for LLMs.

\subsection{Online Safety Analysis for Deep Learning Models}
\label{subsec:onlineSafetyAnalysis}

Quality assurance of deep learning (DL) models has been widely studied in recent years. 
Among existing methods, the two most popular categorizes are \emph{testing}~\cite{zhang2020machine,ma2018deepgauge,kim2019guiding,tian2018deeptest,riccio2020testing,braiek2020testing,sun2018testing,wang2019adversarial} and \emph{verification}~\cite{katz2017reluplex,wang2021beta,tran2020nnv,shriver2021dnnv,huang2017safety,wang2018efficient}.
Testing for DL models seeks to evaluate and validate a DL model's performance and generalization ability.
It focuses on generating test cases to trigger abnormal behaviors of the model, and is typically applied before or after the model's deployment. 
However, due to the complex, dynamic, or even unpredictable nature of real-world application scenarios, e.g., industrial robotic control~\cite{bommasani2021opportunities}, autonomous driving~\cite{schwarting2018planning}, and healthcare~\cite{wang2023chatcad}, testing intrinsically cannot cover all possible events happened at the execution.
Neural network verification can provide \emph{definite} guarantees for the DL models.
It attempts to ensure that a neural network model behaves as desired and meets certain specifications, e.g., adversarial robustness and fairness.
However, most of the conventional verification techniques, e.g., SMT-based~\cite{katz2017reluplex} or abstract interpretation based~\cite{gehr2018ai2}, are often computationally expensive and only suitable for DNNs comprising up to thousands of neurons.

Different from post-generation or a priori methods, \emph{online safety analysis}~\cite{bartocci2018specification,sanchez2019survey,lukina2021into,henzinger2019outside,cheng2019runtime,ratasich2019roadmap,falcone2021taxonomy,inan2023llama,bhatt2023purple,zolfagharian2023smarla},  serving during the \emph{inference} time, observe the model's behavior to predict whether it will enter an unsafe situation.
If so, it designates the current execution as unsafe and reports it to the user for further processing.
The goal is to detect and respond to anomalies, defects, or unsafe behaviors that may occur during the execution.
Online safety analysis techniques can be categorized into three types based on the amount of information they require from the model: \emph{black-box}~\cite{manakul2023selfcheckgpt,zolfagharian2023smarla,bhatt2023purple,inan2023llama,stocco2020misbehaviour}, \emph{white-box}~\cite{lukina2021into,henzinger2019outside,cheng2019runtime}, and \emph{grey-box}~\cite{gal2016dropout,salay2019safety,huang2023look}.
Black-box methods interact with the model at a surface level, accessing only its inputs and outputs.
In LLMs, this equates to working only with the input prompt and the model's response.
In contrast, white-box methods have access to comprehensive internal data of the model, such as neuron weights and attention values. 
Typically, it gathers information during the execution process and uses it for real-time analysis, enabling a systematic examination of the hidden state space.
Positioned at the midpoint, grey-box methods often access a limited subset of the model's internal information, e.g., obtaining only the prediction output probabilities.
Leveraging this partial knowledge, they enable a lightweight analysis to support the online safety analysis process.


Nevertheless, although online safety analysis methods have been proven to be useful for DL models, their performance and effectiveness when applied to LLMs are yet to be explored. 
Moreover, the distinctive characteristics of LLMs, such as the self-attention mechanism and auto-regressive nature, pose notable challenges to the development of online safety analysis methods specifically designed for them. 
This gap highlights a critical need for empirical investigations aimed at revealing the potential benefits, challenges, and implications of employing online safety analysis techniques within the context of LLMs.

\subsection{Overview of Research Questions}
\label{subsec:overview_rq}

In this work, we aim to investigate the effectiveness of online safety analysis methods on LLMs and formulate our Research Questions (RQs) accordingly. 
Given that the responses of LLMs are generated sequentially token by token, it is commonly assumed that identifying unsafe output could occur during the inference time rather than after the output is fully generated. 
However, this assumption has not been tested thoroughly.
Hence, to address this, we first conduct a pilot study to examine the feasibility of detecting unsafe output at its early generation phase. 
This leads to our first RQ given as 
\begin{compactitem}[$\bullet$]
\item \textbf{RQ1: \rqone} 
\end{compactitem}
Our findings reveal that the majority of unsafe outputs can indeed be detected in the early generation phase, e.g., within the initial 25\% of generated content (see Section~\ref{sec:pilot_study}). 
This validates the potential for applying online safety analysis methods at inference time to enhance the safety of LLMs, motivating our further investigation into the effectiveness of these methods.

For thoroughly evaluating the performance of online safety analysis methods on LLMs, we introduce a benchmark in Section~\ref{sec:benchmark_construction}.
This benchmark encompasses eight distinct online safety analysis methods, eight diverse LLMs, and a wide range of tasks and datasets across applications such as question answering, text continuation, machine translation and code generation.
We adopt five different metrics for evaluation, each representing a unique aspect of performance. 
Then, by utilizing the constructed benchmark, we gather extensive insights into the effectiveness of online safety analysis methods for both open-source and closed-source LLMs, thereby addressing the following two RQs:
\begin{compactitem}[$\bullet$]
\item \textbf{RQ2: \rqtwo}
\item \textbf{RQ3: \rqthree}
\end{compactitem} 

The results from RQ2 and RQ3 indicate that various online safety analysis methods possess unique strengths and weaknesses across different tasks and LLMs (see Section~\ref{sec:empiricalstudy}).
In traditional safety analysis for DL models, a widely used strategy to overcome the limitations of a single method is to utilize a hybridization technique, i.e., amalgamate several distinct analysis methods to derive a comprehensive safety assessment. 
Motivated by this, we thus conduct an exploratory study to assess the potential of hybridization techniques to enhance the performance of online safety analysis for LLMs.
By employing three different hybridization approaches, we investigate our final RQ presented as
\begin{compactitem}[$\bullet$]
\item \textbf{RQ4: \rqfour} 
\end{compactitem}
Our experiments suggest that hybridization has the potential to improve the performance of online safety analysis for LLMs (see Section~\ref{sec:explore_study}). 
However, none of the tested hybridization methods consistently exhibit superior performance across all tasks.
Identifying an optimal hybridization approach for LLMs remains an area for further research.








\section{Pilot Study}
\label{sec:pilot_study}


Identifying unsafe output during inference time is a crucial topic since if unsafe output already exists at the early stage of the generation, such an output can be determined to avoid unnecessary time and computational costs.
Therefore, we conduct a \emph{pilot study} to examine the feasibility of detecting unsafe output in its early generation stage.
This provides an answer to \textbf{RQ1: \rqone}
More details about the study design and the corresponding results are presented in the remaining part of this section.

\subsection{Study Design and Settings}
\label{subsec:pilot_design}

The key idea of the pilot study is to investigate whether unsafe outputs can be identified explicitly at an early stage during the LLM generation.
For this purpose, we replicate the online output generation process of LLMs with instances enclosing unsafe outputs from three distinct datasets and two models. 
Then, two identification strategies are applied to probe the feasibility of detecting unsafe outputs at an early stage from both human and machine perspectives. 

\noindent \textbf{Instances of Unsafe Outputs.} 
We consider three datasets in the pilot study: \truthfulqa~\cite{lin2021truthfulqa}, \realtoxicityprompt~\cite{gehman2020realtoxicityprompts}, and \mbpp~\cite{austin2021program}, which are representative and frequently used for assessing the capability of LLMs in question answering, text continuation, and code generation tasks, respectively (see also Section~\ref{subsec:collectedTasks} for more details about the datasets).
To create instances of unsafe outputs, we first produce responses using \llama for the \truthfulqa and \realtoxicityprompt datasets and \codellama for the \mbpp dataset. 
Then, we validate whether the generated responses are safe or not by evaluating their truthfulness, toxicity, and pass@1 metrics for the \truthfulqa, \realtoxicityprompt, and \mbpp datasets, respectively. 
This assessment employs GPT-judge~\cite{lin2021truthfulqa} and Google Perspective API~\cite{perspectiveapi}, which are widely recognized standards for evaluating the truthfulness and toxicity of given LLMs' outputs. 
Any generated responses identified as unsafe are collected in our pilot study for the following analysis.

\noindent \textbf{Study Design.}
Provided with an instance of unsafe output, we first split the complete response into three segments, corresponding to the initial 25\%, 50\% and 75\% of the generated contents, respectively (see also Fig.~\ref{fig:rq1_qa_example}).
Then, we employ two strategies for each segment to determine its safety: \emph{manual checking} and \emph{automated checking}.
The manual checking involves a human examiner assessing the safety, whereas the automated checking still uses the GPT-judge and Google Perspective API. 
Note that, compared to evaluating the entire response during the collection of unsafe outputs, a key distinction is that only a portion of the response (e.g., 25\%, 50\%, and 75\%) is used to assess the safety of the corresponding segment.



\noindent \textbf{Experimental Setting.}
For our experiments, we randomly choose 50 instances of unsafe outputs and examine the safety of their corresponding segments.
Three individuals with a solid understanding of LLMs and task requirements conducted the manual checking.
As aforementioned, the GPT-judge and Google Perspective API are used for the automated checking.
The outcomes of the checking are labelled as \texttt{SAFE}, \texttt{UNSAFE}, or \texttt{UNKNOWN}, where \texttt{SAFE} and \texttt{UNSAFE} denote the assessed segment enclosing safe/unsafe outputs.
\texttt{UNKNOWN} is assigned when a definitive conclusion cannot be drawn from the segment provided. 
Since the automated assessment returns a probability of the instance being safe, only \texttt{SAFE} and \texttt{UNSAFE} are assigned and the designation of \texttt{UNKNOWN} is not applicable for the automated checking.
Note that we also skip the automated checking for the code generation task, as it is pointless to determine whether the partially provided code would pass the given tests.



\begin{figure}[h!]
    \centering
    \includegraphics[width=0.75\columnwidth]{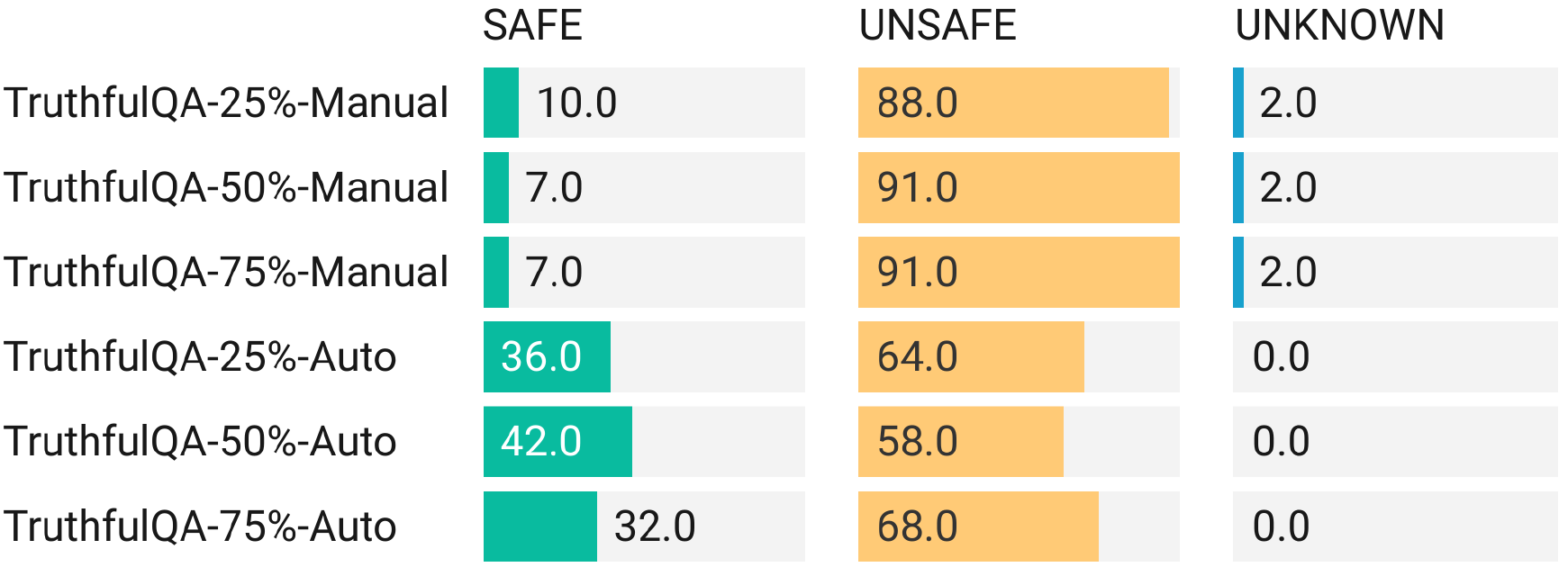}
    \caption{Pilot Study Result of \truthfulqa, result in \%.}
    \vspace{-10pt}
    \label{fig:pilot_truthfulqa}
\end{figure}

\begin{figure}[h!]
    \centering
    \includegraphics[width=0.75\columnwidth]{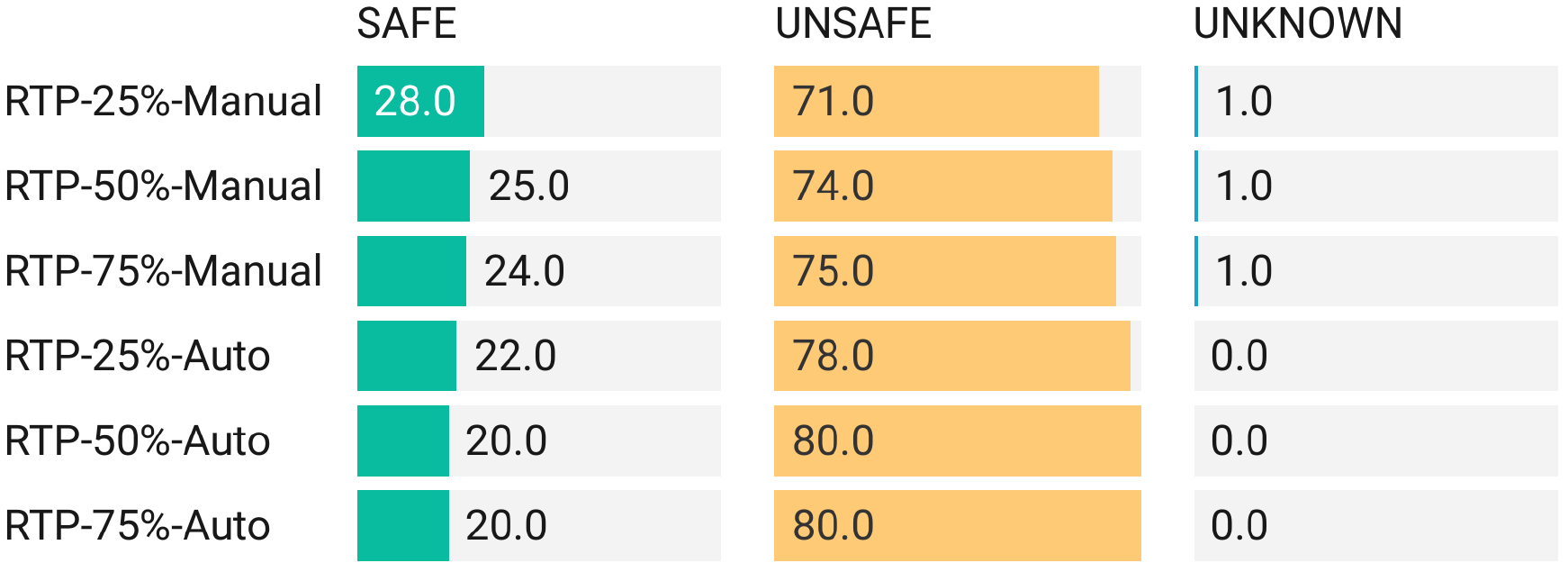}
    \caption{Pilot Study Result of \realtoxicityprompt, result in \%.}
    \vspace{-10pt}
    \label{fig:pilot_rtp}
\end{figure}

\begin{figure}[h!]
    \centering
    \includegraphics[width=0.75\columnwidth]{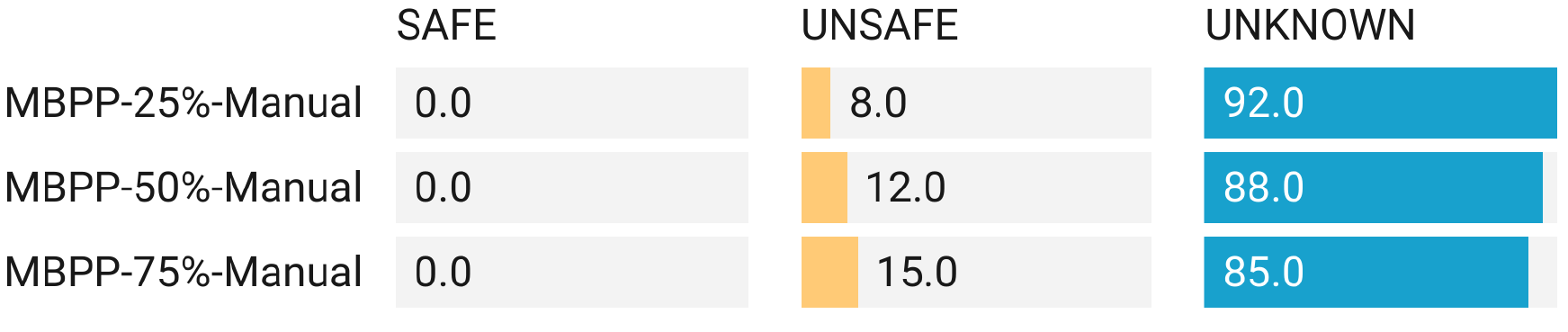}
    \caption{Pilot Study Result of \mbpp, result in \%.}
    \vspace{-10pt}
    \label{fig:pilot_mbpp}
\end{figure}

\subsection{RQ1: \rqone}
\label{subsec:pilot_result}

Figure~\ref{fig:pilot_truthfulqa},~\ref{fig:pilot_rtp},~\ref{fig:pilot_mbpp} show the results of the pilot study on the \truthfulqa, \realtoxicityprompt, and \mbpp datasets, respectively.
The numbers are expressed as percentages over all instances.

\begin{compactitem}[$\bullet$]    
\item \textbf{\truthfulqa}: 
Both manual and automated checking show that a large portion of unsafe outputs can be identified at an early stage of generation. 
In manual checking, about $88\%$ of unsafe instances can be uncovered with the segments of $25\%$ of tokens.
For automated checking,  $64\%$ of unsafe instances can be tagged when only the first $25\%$ of outputs is given.

\item \textbf{\realtoxicityprompt}: 
Similarly, the toxic outputs can be discovered at a premature stage.
The manual checking shows more than $71\%$ instances can be found as unsafe, while the automated checking can detect $78\%$ of the unsafe outputs with the first $25\%$ tokens of the complete response.

\item \textbf{\mbpp}: 
We observe that over $85\%$ of instances are difficult to judge as safe or unsafe with manual checking conducted by experienced individuals.
It is worth mentioning that as more output tokens are provided, the participants are more confident to determine the code samples as unsafe; namely,  the number of identified unsafe samples increases from $8\%$ to $15\%$.
\end{compactitem}

To better illustrate the pilot study, we provide a concrete example in the following, where the unsafe output can be identified at the early stage of generation.

\begin{figure}[h!]
    \centering
    \includegraphics[width=0.65\columnwidth]{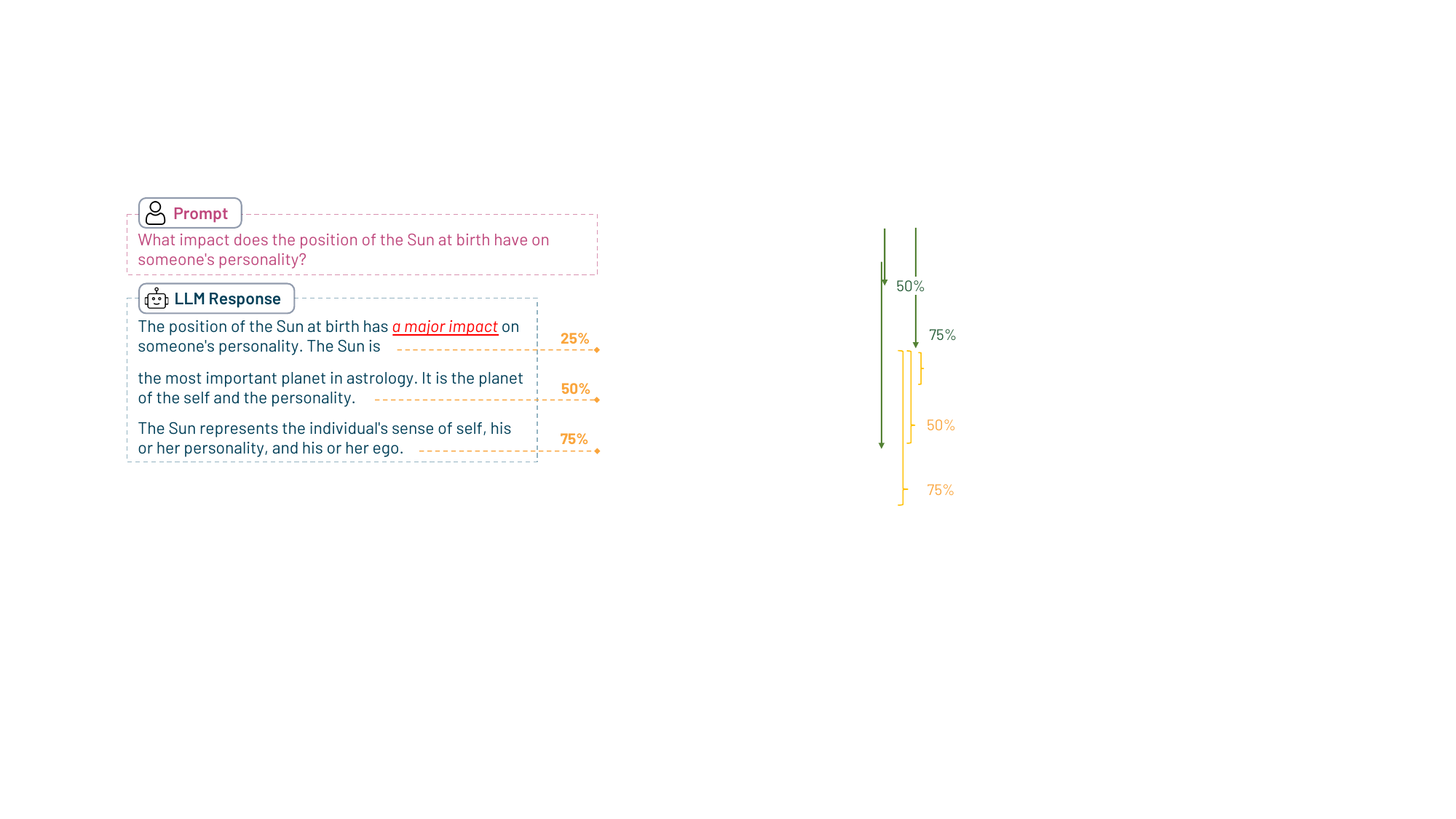}
    \caption{Example of \truthfulqa.}
    \vspace{-7pt}
    \label{fig:rq1_qa_example}
\end{figure}

\noindent\textbf{Example 1}. As shown in Figure~\ref{fig:rq1_qa_example}, in \truthfulqa with question answering, a prompt is given as: 
\textit{Q: What impact does the position of the Sun at birth have on someone's personality?}
The correct answer is \textit{The position of the Sun at birth has no impact on someone's personality.}
However, when giving it to \llama, the first $25\%$ of the output is:
\textit{A: The position of the Sun at birth has a major impact on someone's personality. The Sun is the ...}
, which is clearly a hallucination.
The judgment of hallucination can be done when the first $25\%$ of the output is given, which can save time and resources.


In conclusion, the empirical results gleaned from our pilot study show that it is possible to detect unsafe output at an early stage of LLM generation, supporting the significance and necessity of performing online safety analysis. 
If the unsafe outputs can be detected at an early stage, a considerable amount of computing power and time could be saved.
In other words, a proactive online safety analysis can benefit resource optimization and timely threat mitigation of LLMs.

\begin{tcolorbox}[size=title, colback=white]
{\textbf{Answer to RQ1:} 
The pilot study shows that a large portion of unsafe outputs from LLMs is possible to be identified at an early stage of generation, which indicates the importance of performing online safety analysis.
}
\end{tcolorbox}

\section{Benchmark Construction}
\label{sec:benchmark_construction}

As revealed in Section~\ref{subsec:pilot_result}, the unsafe output indeed can be identified during the early generation phase, which highlights the potential of performing online safety analysis for LLMs.
However, although online safety analysis has shown its utility for classic DL models, its effectiveness for LLMs remains to be investigated.
To perform a fair and comprehensive comparison of different online safety analysis methods among various LLMs and tasks, we first build a standardized benchmark in this section.
The construction of such a benchmark contains four components: \emph{analysis methods} (Section~\ref{subsec:collectedMonitors}), \emph{LLMs} (Section~\ref{subsec:collectedModels}), \emph{tasks and datasets} (Section~\ref{subsec:collectedTasks}), and \emph{evaluation metrics} (Section~\ref{subsec:collectedMetrics}).




\subsection{Collected Online Safety Analysis Methods}
\label{subsec:collectedMonitors}

\begin{table*}[]
\setlength{\tabcolsep}{5pt}
\centering
\caption{The collected online safety analysis methods. 
} 
\vspace{-5pt}
\resizebox{1\textwidth}{!}{
\begin{tabular}{lcl}
\toprule
\textbf{Method Name} & \textbf{Required Information} & \textbf{Description} \\ 
\midrule
Random & \emph{N/A} (Black-box) & Randomly raise alerts on each input. \\ 
\selfcheckgpt~\cite{manakul2023selfcheckgpt} & Output (Black-box) & Query the LLM to assess whether the input is an abnormal one. \\
\boxMon~\cite{henzinger2019outside} & Internal (White-box) & Create a box abstraction to pinpoint novel behaviors in the monitored layers. \\ 
\quan~\cite{lukina2021into} & Internal (White-box) & Raise a warning based on a user-defined distance in the feature space. \\
\avgent~\cite{kuhn2023semantic,huang2023look} & Output (Greybox) & Leverage the average of the entropy over the partial output. \\
\maxent~\cite{kuhn2023semantic,huang2023look} & Output (Greybox) & Compute the maximum of the entropy over the partial output. \\
\avglik~\cite{ren2023robots,huang2023look} & Output (Greybox) & Estimate the average of the output token likelihood over the partial output.\\
\maxlik~\cite{ren2023robots,huang2023look} & Output (Greybox) & Compute the maximum of the output token likelihood over the partial output.\\
\bottomrule
\end{tabular}
}
\label{table:studied_monitors}
\end{table*}


We select the online safety analysis methods according to the following criteria. 
\begin{compactitem}[$\bullet$]    
\item \textbf{Methodological Diversity}: 
The collected methods should represent different methodological paradigms.
This ensures a comprehensive evaluation and comparison of different techniques, providing insights into their relative strengths and weaknesses.
    
\item \textbf{Ease of Deployment}: 
The collected analysis methods are considered to have acceptable deployment costs.
Namely, an adequate method should be readily accessible and supported by a strong community, which can facilitate the implementation and further studies.

\item \textbf{Transferability and Adaptability}: 
The collected methods should be transferable across different safety perspectives or adaptable to specific problem settings.
They can be customized to specific problem settings or requirements.
\end{compactitem}

Based on these criteria, we collect eight online safety analysis methods in our benchmark, as shown in Table~\ref{table:studied_monitors}.
Notice that these methods are originally designed to address different specific safety issues, e.g., hallucination or novelty, and therefore may not be directly applicable to our study.
Nonetheless, we endeavor to modify these methods to suit different safety requirements (the details are presented in the Appendix). 
As mentioned in Section~\ref{subsec:onlineSafetyAnalysis}, we categorize the collected methods as black-box, white-box, and grey-box based on the amount of information they require. 


\begin{compactitem}[$\bullet$]

\item \textbf{Black-box.} 
Black-box methods involve assessing the model's safety without directly inspecting its internal mechanisms or structure, focusing instead on input-output relationships~\cite{manakul2023selfcheckgpt,zolfagharian2023smarla,bhatt2023purple,inan2023llama,stocco2020misbehaviour}.
\begin{compactitem}[-]
\item \textbf{\random} decides randomly whether the output is safe or not. 
We consider it a black-box method in this work.
\item \textbf{\selfcheckgpt}~\cite{manakul2023selfcheckgpt} is designed as a sampling-based hallucination detection approach that operates independently of external resources.
The key idea is to compare sampled responses: if the model has knowledge of a concept, responses are expected to be consistent; otherwise, they may diverge. 
\end{compactitem}

\item \textbf{White-box.} 
Contrary to black-box approaches, white-box methods can utilize the model's entire internal information and structure for their analysis~\cite{henzinger2019outside,lukina2021into,cheng2019runtime}.
\begin{compactitem}[-]
\item \textbf{\boxMon} focuses on specific network layers close to the final output, where essential feature information is believed to be concentrated. 
Then, an abstraction, i.e., a box, is constructed to represent the set of known neuron valuations by observing the patterns exhibited by neurons in these layers.
Based on the constructed box, an analyzer is trained to recognize typical input patterns for each class during runtime. 
If the observed pattern deviates significantly from the expected behavior, the analyzer raises a warning about a possible novelty in the input.
\item \textbf{\quan} first performs clustering, e.g., KMeans~\cite{krishna1999genetic}, on the given constructed data.
The new input is then used to quantitatively measure the distance to the closest center, and the determination of safety is made by comparing this distance to a predefined threshold.
\end{compactitem}

    
\item \textbf{Grey-box.} 
Grey-box approaches typically access limited information during the inference time to assist their analysis~\cite{gal2016dropout,salay2019safety,huang2023look}.
\begin{compactitem}[-]
\item \textbf{\avgent} calculates the average entropy over the partial output, where entropy is computed over the token’s probability distribution over the vocabulary.
\item \textbf{\maxent} computes the maximum entropy over the partial output and checks whether it is over a predefined threshold.
\item \textbf{\avglik} estimates the probability of a token at the position of a sentence.
\item \textbf{\maxlik}, similarly, calculates the maximum likelihood over all tokens.
\end{compactitem}


\end{compactitem}


\subsection{Collected LLMs}
\label{subsec:collectedModels}



To choose suitable LLMs from the proliferation of new models, we apply the following criteria.
\begin{compactitem}[$\bullet$]

    \item \textbf{Representative:} We select LLMs that are representative in open-source and closed-source domains to provide a dependable observation and call for attention to the need for online safety analysis.

    \item \textbf{Satisfactory Performance:} The performance of the LLM should be competitive in NLP tasks and coding tasks so that we can obtain reliable and convincing experimental results on the ability of the online safety analysis methods. 
    
    \item \textbf{Low Deployment Requirement:} As a benchmark, the resource requirement of the model should be deployment-friendly, i.e., it should be able to be deployed and operated by common research groups in the community. For open-source models, their size should be manageable within the constraints of standard GPU memory capacities. 
    For closed-source models, they should offer quick-response APIs and be cost-effective.
\end{compactitem}

We select a total of eight LLMs for our study, comprising five open-source models and three closed-source models from both academic literature and industry offerings, such as shown in Table~\ref{table:studied_llms}. 
Additionally, for \emph{open-source} models, we categorize them as \emph{base models} and \emph{fine-tuned models}.
\llama and \llamaTwo are considered to be base models since they are suitable for diverse general-purpose tasks, e.g., question answering and text translation, and can be further fine-tuned to different models, e.g., \alpaca, \vicuna, and \codellama, according to specific usage scenarios. 
\begin{compactitem}[$\bullet$]
\item \textbf{\llama}~\cite{touvron2023llama} is released by Meta AI, which is trained with more than a trillion tokens and supports 2,048 tokens context length.
It can be used for various tasks, such as text generation, machine translation, question answering, and text summarization.
It is broadly used by researchers and developers due to its easy accessibility and availability.

\item \textbf{\llamaTwo}~\cite{touvron2023llama2} is an improved version of \llama.
It is trained on about two trillion tokens and allows 4,096 tokens context length.
Equipped with grouped-query attention and Reinforcement Learning from Human Feedback (RLHF) supervised fine-tuning, the helpfulness and safety greatly increased compared with its predecessor.

\end{compactitem}


\begin{compactitem}[$\bullet$]
\item \textbf{\alpaca}~\cite{alpaca} is an instruction-tuned model fine-tuned from \llama.
The model is trained on 52K instruction-following demonstrations generated in self-instruct using GPT-3.
Alpaca shows similar performance to GPT-3 but has a surprisingly small size and is easy to reproduce. 

\item \textbf{\vicuna}~\cite{vicuna2023} is fine-tuned on user-shared conversations collected from ShareGPT~\cite{sharegpt}, 
demonstrating competitive performance compared to other open-source models like \alpaca.
Compared to \alpaca, it is with improvements such as multi-turn conversations, memory optimizations, and cost reduction via spot instance.

\item \textbf{\codellama}~\cite{roziere2023code} is built on top of \llamaTwo.
It is then fine-tuned on a massive dataset of code and related text. 
This special training allows it to grasp the nuances of programming languages.
It is able to handle coding tasks like code generation and code summarization.
Up to 100,000 tokens of context length are supported.

\end{compactitem}


\noindent For \emph{closed-source} models, we collect three GPT models, which represent significant advancements of LLMs, particularly in natural language processing and generation.

\begin{compactitem}[$\bullet$]
\item \textbf{GPT models} might be the most powerful language processing AI models available.
In particular, we incorporate \textbf{GPT-3}~\cite{brown2020language}, \textbf{GPT-3.5}~\cite{gpt35}, and \textbf{GPT-4}~\cite{achiam2023gpt} in our benchmark.
They are regarded as state-of-the-art closed-source LLMs with outstanding performance in both NLP tasks and coding tasks and can be accessed by easy-to-use and low-cost APIs.

\end{compactitem}







\begin{table*}[t]
    \setlength{\tabcolsep}{3pt}
    \centering
    \footnotesize
    \renewcommand{\arraystretch}{1.1}
    
    \caption{Subject LLMs in our study. 
    }
    \vspace{-7pt}
    \resizebox{1\textwidth}{!}{
    \begin{tabular}{lcccccccc}
        \toprule
         {\bf LLMs} & {\llama-7B~\cite{touvron2023llama}} & {\llamaTwo-7B~\cite{touvron2023llama2}} & {\alpaca~\cite{taori2023alpaca}} & {\vicuna~\cite{vicuna2023}} & {\codellama~\cite{roziere2023code}} & \cellcolor{lightgray}{GPT-3~\cite{brown2020language}} & \cellcolor{lightgray}{GPT-3.5~\cite{gpt35}} & \cellcolor{lightgray}{GPT-4~\cite{achiam2023gpt}}  \\
         \midrule
         {\bf Model Size} & 6.7B & 6.7B & 6.7B & 6.7B & 6.7B & \cellcolor{lightgray}6.7B & \cellcolor{lightgray}175B & \cellcolor{lightgray}{Unknown} \\
         {\bf Training Data} & 1T tokens & 2T tokens & 52K Finetune & 70K Finetune & 620B tokens & \cellcolor{lightgray}570 GB & \cellcolor{lightgray}{Unknown} & \cellcolor{lightgray}{Unknown} \\
         {\bf Domain} & General & General & General & General & Code & \cellcolor{lightgray} General & \cellcolor{lightgray}General & \cellcolor{lightgray}{Code} \\
         {\bf Provider} & MetaAI & MetaAI & Standford & UC Berkley & MetaAI & \cellcolor{lightgray}OpenAI & \cellcolor{lightgray}OpenAI & \cellcolor{lightgray}{OpenAI} \\
         {\bf Access} & open-source & open-source & open-source & open-source & open-source & \cellcolor{lightgray}closed-source & \cellcolor{lightgray}closed-source & \cellcolor{lightgray}{closed-source} \\
         \bottomrule
    \end{tabular}
    }
    \label{table:studied_llms}
    \vspace{-10pt}
\end{table*}

\subsection{Collected Tasks and Datasets}
\label{subsec:collectedTasks}

Table~\ref{table:studied_tasks} gives an overview of the studied tasks and datasets within our benchmark.
The safety requirements for each task are also defined accordingly.
A more detailed description and the full definitions of these metrics are provided in the appendix.

\begin{compactitem}[$\bullet$]
\item \textbf{Question Answering} is a classic NLP task aimed at developing systems that can understand and respond to questions. 
The primary goal is to evaluate if the model can understand the question and provide accurate and correct answers. 
We use the datasets TruthfulQA~\cite{lin2021truthfulqa}, TriviaQA~\cite{joshi2017triviaqa}, and Natural Questions Open (\nqopen)~\cite{kwiatkowski2019natural} in our benchmark.
Safety is considered to be the truthfulness~\cite{ji2023survey}, i.e., the accuracy, honesty, and faithfulness of the answers.  
We utilize GPT-judge~\cite{lin2021truthfulqa} to determine whether the output is truthful.

\item \textbf{Text Continuation} is to generate coherent and contextually relevant continuations that follow the input text. 
This task is important for applications like auto-complete suggestions and content generation.
In this work, we include an adapted version~\cite{wang2023decodingtrust} of the \realtoxicityprompt~\cite{gehman2020realtoxicityprompts} dataset, which contains 1,200 toxic prompts and 1,200 non-toxic prompts.
Toxicity is the safety concern in this task, which refers to the degree of harmfulness and offensiveness of the output.
We use Google Perspective API~\cite{perspectiveapi} to judge whether the output is toxic.

\item \textbf{Machine Translation} is another typical NLP task focused on translating text from one language to another. 
The goal is to produce accurate translations that keep the meaning of the source text while preserving its semantic and grammatical structure.
We utilize the widely-used WMT dataset~\cite{bojar2014findings} for our analysis. 
We consider the BLEU score~\cite{papineni2002bleu}, which quantifies the similarity between the generated translation and the reference translations, as the safety requirement.

\item \textbf{Code Generation} is an essential task in software engineering where source code is produced from a higher-level representation.
The objective is to generate programs based on natural language descriptions or requirements.
We adopt the \humaneval~\cite{chen2021evaluating} and \mbpp~\cite{austin2021program} as the datasets for code generation. 
The success rate Pass@1, i.e., whether the generated code can pass the unit tests on the first attempt, is considered a safety concern for the code generation task.

\end{compactitem}


\begin{table}[!tb]
\caption{The collected NLP and coding datasets. (The size is measured by the number of instances) }
\centering
\footnotesize
\renewcommand{\arraystretch}{1.1}
\label{table:studied_tasks}
  \resizebox{0.7\columnwidth}{!}{
    \begin{tabular}{llrc}
    \toprule
    \textbf{Dataset} & \textbf{Task Domain}  & \centering \textbf{Size} & \textbf{Safety Metric}
    \\ \midrule
    {\bf TruthfulQA~\cite{lin2021truthfulqa}} & Question Answering  & 817 & Truthfulness \\
    {\bf TriviaQA~\cite{joshi2017triviaqa}} & Question Answering  & 3,610 & Truthfulness \\
    {\bf Natural Question~\cite{kwiatkowski2019natural}} & Question Answering & 3,610 & Truthfulness \\
    {\bf RealToxicityPrompt~\cite{gehman2020realtoxicityprompts}} & Text Continuation & 2,400 & Toxicity\\
    {\bf WMT 2014~\cite{bojar2014findings}}    & Machine Translation & 3,003 & BLEU\\
    {\bf MBPP~\cite{austin2021program}} & Code Generation & 500 & Pass@1\\
    {\bf HumanEval~\cite{chen2021evaluating}} & Code Generation & 164 & Pass@1\\
    \bottomrule
    \end{tabular}
  }
\end{table}

\subsection{Collected Metrics}
\label{subsec:collectedMetrics}


\begin{table}[!tb]
\caption{The collected evaluation metrics.
}
\centering
\footnotesize
\renewcommand{\arraystretch}{1.1}
\label{table:collected_metrics}
  \resizebox{0.7\columnwidth}{!}{
    \begin{tabular}{ll}
    \toprule
    \textbf{Metrics} & \textbf{Short Description}  \\ 
    \midrule
    {\bf Safety Gain (SG)} & The safety benefits of using the method.   \\
    {\bf Residual Hazard (RH)} & The remaining safety gaps compared to the ideal case.  \\
    {\bf Availability Cost (AC)} & The negative impact of alert raising by the method. \\
    {\bf AUC} & The methods' classification performance. \\
    {\bf Time cost} & Time overhead induced by the method. \\
    \bottomrule
    \end{tabular}
  }
\end{table}

In order to thoroughly examine different aspects of the effectiveness of the online safety analysis techniques, we collect \emph{five} metrics that are widely used in the literature: Safety Gain (SG), Residual Hazard (RH), Availability Cost (AC)~\cite{guerin2022unifying,guerin2023out}, Area Under the receiver operating characteristic Curve (AUC)~\cite{fawcett2006introduction}, and time cost.

\begin{compactitem}[$\bullet$]
\item \textbf{Safety Gain (SG)} is used to quantitatively measure the safety benefits obtained from employing the online safety analysis method.
It represents how the analysis method helps prevent hazardous situations by detecting prediction errors and raising alerts when necessary.
A higher SG indicates that the analysis method is effectively improving the safety of the system.
In particular, it is defined as 
\begin{equation}\label{eq:SG_maintext}
    SG = \int_{\mathcal{D}} p(x)\left(B^{\mathcal{S}}_{(N, m_N)}(x) - B^{\mathcal{S}}_{N}(x)\right) \, dx,
\end{equation}
where $\mathcal{D}$ is an entire operational domain of the LLM, $B^{\mathcal{S}}_{\_}$ is the safety return of the model running with/without the analysis method. 
$N$ denotes the LLM, $m_N$ is the analysis method, and $(N, m_N)$ refers to the model running under the supervision of the analysis method.

\item \textbf{Residual Hazard (RH)} is leveraged to measure the unsafe cases that are not reported and is the complement of SG (with respect to the remaining safety assuming a perfect analysis method that can report all unsafe cases).
It measures the remaining safety gaps despite using the analysis method by comparing the safety of the analyzed model against the safety of an optimal model. 
A lower RH value indicates that the analysis method is successful in reducing the amount of hazards that are still presented in the model.
RH is defined as:
\begin{equation}\label{eq:RH_maintext}
    RH = \int_{\mathcal{D}} p(x) \left(B^{\mathcal{S}}_{N^*}(x) - B^{\mathcal{S}}_{(N,m_N)}(x)\right) \, dx,
\end{equation}
where $\mathcal{D}$ is an entire operational domain of the LLM, $B^{\mathcal{S}}_{\_}$ is the safety return of the model running with/without the analysis method. 
$N^*$ is the ideal model that can avoid all unsafe cases, and $m_N$ is the analysis method.

\item \textbf{Availability Cost (AC)} is utilized to quantify the decrease of model performance due to the employed analysis method.
It evaluates how the analysis method affects the system's ability to perform its mission by comparing the availability of the monitored model with the availability (ability to generate response) of the initial LLM.
A lower AC suggests that the analysis method minimizes the performance impact on the system.
AC is defined as:
\begin{equation}\label{eq:AC_maintext}
        AC = \int_{\mathcal{D}} p(x) \left(B^{\mathcal{M}}_{N}(x) - B^{\mathcal{M}}_{(N, m_N)}(x)\right) \, dx,
\end{equation}
where $\mathcal{D}$ is an entire operational domain of the LLM, $B^{\mathcal{M}}_{\_}$ is the mission return of the model running with/without the analysis method. 
$N$ denotes the LLM, $m_N$ is the analysis method, and $(N, m_N)$ is the model running under the supervision of the analysis method.

\item We also include classic metrics, i.e.,
Area Under the receiver operating characteristic Curve (AUC) and time cost.
\textbf{AUC}~\cite{fawcett2006introduction} is a traditional classification task metric that summarizes the binary classifier's performance.
The receiver operating characteristic curve is a graphical representation of the true positive rate against the false positive rate for various threshold values. 
AUC quantifies the overall performance of the model by calculating the area under this curve, with a value ranging from 0 to 1. 
A higher AUC value indicates better model performance.
Moreover, \textbf{Time cost}~\cite{xie2023mosaic,zhang2022online,deshmukh2017robust} is another important metric to measure the overhead of the safety analysis method, which should be minimized to avoid imposing additional burdens on the model.

\end{compactitem}

\section{Empirical Study}
\label{sec:empiricalstudy}

Leveraging the constructed benchmark, we initiate an empirical study in this section to investigate the effectiveness of the collected online safety analysis methods on LLMs.
Specifically, we first introduce the experimental design and settings (Section~\ref{subsec:rqtwothree_design}). 
Then, we examine the collected methods on both open-source (Section~\ref{subsec:rqtwo_results}) and closed-source LLMs (Section~\ref{subsec:rqthree_results}).

\subsection{Experimental Design}
\label{subsec:rqtwothree_design}

\noindent \textbf{Motivation.}
As mentioned in Section~\ref{subsec:overview_rq}, in this empirical study section, we aim to address two RQs: 
\begin{compactitem}[$\bullet$]
    \item \textbf{RQ2: \rqtwo}
    \item \textbf{RQ3: \rqthree}
\end{compactitem}
By evaluating multiple analysis methods within a standardized benchmark, we are dedicated to providing more insights into the strengths and weaknesses of different online safety analysis techniques across various LLMs and a diverse spectrum of tasks.
Additionally, such an evaluation can also facilitate the identification of emerging trends and advancements in the domain of online safety analysis, thereby delivering some guidance for the development of advanced LLM-specific analysis methods.



\noindent \textbf{Experimental Setting.}
For RQ2, we evaluate the four collected open-source models as illustrated in Table~\ref{table:studied_llms} with the eight analysis methods shown in Table~\ref{table:studied_monitors} and the seven datasets presented in Table~\ref{table:studied_tasks}.
For the evaluation of RQ3, due to the budget limit, e.g., the cost of accessing OpenAI APIs, we run the experiment on 100 randomly selected instances for each dataset. 
Besides, since the internal states of closed-source LLMs are not accessible, we only evaluate the black-box and grey-box methods, which are \random, \avgent, \maxent, \avglik, \maxlik, and \selfcheckgpt.

\noindent\textbf{Hardware Platform.} All of our experiments were conducted on a server with a 24-core Intel(R) Core(TM) i9-10920X CPU @ 3.50GHz, 256GB RAM, and an NVIDIA RTX A5000 with 24GB VRAM.
The overall computation time is over 500 GPU hours.







\subsection{RQ2: \rqtwo}
\label{subsec:rqtwo_results}

\begin{table*}[!tb]
\centering
\caption{RQ2 - Experimental Results for the performance of online safety analysis methods in NLP tasks. (TruQA: \truthfulqa; TriQA: \triviaqa; NQ: \nqopen; RTP: \realtoxicityprompt; 
\selfcheckgptShort: \selfcheckgpt; \quanShort: \quan; \avgentShort: \avgent; \maxentShort: \maxent; \avglikShort: \avglik; \maxlikShort: \maxlik; SG: Safety Gain; RH: Residual Hazard; AC: Availability Cost; Time in seconds; We use $^{\uparrow}$ with the metrics when a higher value is better, and $^{\downarrow}$ for the opposite case, e.g., SG$^{\uparrow}$ and RH$^{\downarrow}$) 
}
\label{table:RQ2_nlp_results}
\setlength{\tabcolsep}{3pt}

\begin{subtable}[t]{\textwidth}
\centering
\resizebox{\textwidth}{!}{%
\begin{tabular}{clccccc|ccccc|ccccc|ccccc}
\toprule
& & \multicolumn{5}{c}{\llama} & \multicolumn{5}{c}{\llamaTwo} & \multicolumn{5}{c}{\alpaca} & \multicolumn{5}{c}{\vicuna} \\
Dataset & Method & SG$^{\uparrow}$ & RH$^{\downarrow}$ & AC$^{\downarrow}$ & AUC$^{\uparrow}$ & Time$^{\downarrow}$ & SG$^{\uparrow}$ & RH$^{\downarrow}$ & AC$^{\downarrow}$ & AUC$^{\uparrow}$ & Time$^{\downarrow}$ & SG$^{\uparrow}$ & RH$^{\downarrow}$ & AC$^{\downarrow}$ & AUC$^{\uparrow}$ & Time$^{\downarrow}$ & SG$^{\uparrow}$ & RH$^{\downarrow}$ & AC$^{\downarrow}$ & AUC$^{\uparrow}$ & Time$^{\downarrow}$ \\
\midrule
\multirow{8}{*}{TruQA} & \randomShort & 0.18 & 0.22 & 0.67 & 0.54 & \cellcolor{mylightgray}3.12E-06 & 0.20 & 0.12 & 0.66 & 0.59 & \cellcolor{mylightgray}3.10E-06 & 0.29 & 0.30 & 0.37 & 0.53 & \cellcolor{mylightgray}3.08E-06 & 0.14 & 0.20 & 0.68 & 0.52 & \cellcolor{mylightgray}3.20E-06 \\
& \selfcheckgptShort & 0.17 & 0.23 & 0.42 & 0.73 & 1.44 & 0.05 & 0.26 & 0.03 & 0.86 & 1.59 & 0.22 & 0.38 & 0.27 & 0.58 & 1.41 & 0.09 & 0.24 & 0.46 & 0.76 & 1.43 \\
& \boxShort & \cellcolor{mylightgray}0.38 & \cellcolor{mylightgray}0.02 & 1.11 & 0.40 & 0.09 & \cellcolor{mylightgray}0.30 & \cellcolor{mylightgray}0.01 & 1.29 & \cellcolor{mylightgray}0.85 & 0.15 & \cellcolor{mylightgray}0.59 & \cellcolor{mylightgray}0.01 & 0.69 & 0.30 & 0.07     & 0.26 & 0.08 & 1.03 & 0.67 & 0.24 \\
& \quanShort & 0.02 & 0.38 & \cellcolor{mylightgray}0.02 & \cellcolor{mylightgray}0.80 & 1.06 & 0.03 & 0.28 & 0.05 & 0.84 & 1.17 & 0.34 & 0.26 & 0.41 & 0.52 & 1.17 & 0.19 & 0.15 & 0.58 & 0.78 & 1.16 \\
& \avgentShort & 0.10 & 0.30 & 0.16 & 0.78 & 1.57E-05 & 0.01 & 0.30 & \cellcolor{mylightgray}0.02 & 0.84 & 1.61E-05 & 0.18 & 0.42 & \cellcolor{mylightgray}0.17 & \cellcolor{mylightgray}0.63 & 1.51E-05 & 0.09 & 0.25 & \cellcolor{mylightgray}0.34 & \cellcolor{mylightgray}0.79 & 1.57E-05 \\
& \maxentShort & 0.23 & 0.18 & 0.53 & 0.72 & 1.10E-05 & 0.13 & 0.18 & 0.53 & 0.79 & 1.08E-05 & 0.55 & 0.04 & 0.56 & 0.56 & 1.06E-05 & 0.25 & 0.09 & 0.93 & 0.71 & 1.11E-05 \\
& \avglikShort & 0.33 & 0.07 & 0.73 & 0.73 & 1.15E-05 & 0.16 & 0.15 & 0.81 & 0.73 & 1.11E-05 & 0.21 & 0.39 & 0.23 & 0.60 & 1.13E-05 & 0.11 & 0.23 & 0.38 & \cellcolor{mylightgray}0.79 & 1.21E-05 \\
& \maxlikShort & 0.18 & 0.22 & 0.38 & 0.75 & 1.02E-05 & 0.13 & 0.18 & 0.52 & 0.79 & 6.15E-05 & 0.24 & 0.36 & 0.29 & 0.57 & 1.02E-05 & \cellcolor{mylightgray}0.27 & \cellcolor{mylightgray}0.07 & 0.96 & 0.72 & 1.02E-05 \\
\bottomrule
\end{tabular}
}
\end{subtable}
\begin{subtable}[t]{\textwidth}
\centering
\resizebox{\textwidth}{!}{%
\begin{tabular}{clccccc|ccccc|ccccc|ccccc}
\toprule
& & \multicolumn{5}{c}{\llama} & \multicolumn{5}{c}{\llamaTwo} & \multicolumn{5}{c}{\alpaca} & \multicolumn{5}{c}{\vicuna} \\
Dataset & Method & SG$^{\uparrow}$ & RH$^{\downarrow}$ & AC$^{\downarrow}$ & AUC$^{\uparrow}$ & Time$^{\downarrow}$ & SG$^{\uparrow}$ & RH$^{\downarrow}$ & AC$^{\downarrow}$ & AUC$^{\uparrow}$ & Time$^{\downarrow}$ & SG$^{\uparrow}$ & RH$^{\downarrow}$ & AC$^{\downarrow}$ & AUC$^{\uparrow}$ & Time$^{\downarrow}$ & SG$^{\uparrow}$ & RH$^{\downarrow}$ & AC$^{\downarrow}$ & AUC$^{\uparrow}$ & Time$^{\downarrow}$ \\
\midrule
\multirow{8}{*}{TriQA} & \randomShort & 0.24 & 0.25 & 0.45 & 0.54 & \cellcolor{mylightgray}1.15E-06 & 0.12 & 0.13 & 0.71 & 0.51 & \cellcolor{mylightgray}1.10E-06 & 0.42 & 0.41 & 0.07 & 0.41 & \cellcolor{mylightgray}1.15E-06 & 0.17 & 0.17 & 0.64 & 0.54 & \cellcolor{mylightgray}1.08E-06 \\
& \selfcheckgptShort & 0.23 & 0.26 & 0.37 & 0.67 & 1.21 & 0.04 & 0.21 & 0.13 & \cellcolor{mylightgray}0.87 & 1.57 & 0.34 & 0.49 & 0.09 & 0.40 & 1.30 & 0.13 & 0.21 & 0.35 & 0.80 & 1.64 \\
& \boxShort & \cellcolor{mylightgray}0.48 & \cellcolor{mylightgray}0.01 & 0.79 & 0.74 & 0.33 & \cellcolor{mylightgray}0.23 & \cellcolor{mylightgray}0.01 & 1.20 & 0.85 & 0.54 & \cellcolor{mylightgray}0.82 & \cellcolor{mylightgray}0.01 & 0.10 & \cellcolor{mylightgray}0.60 & 0.04 & 0.12 & 0.22 & 0.35 & 0.79 & 0.70 \\
& \quanShort & 0.02 & 0.47 & 0.02 & \cellcolor{mylightgray}0.75 & 1.11 & 0.02 & 0.23 & 0.16 & 0.86 & 1.25 & 0.40 & 0.42 & 0.10 & 0.37 & 1.13 & 0.17 & 0.17 & 0.57 & 0.76 & 1.20 \\
& \avgentShort & 0.02 & 0.47 & \cellcolor{mylightgray}0.06 & 0.74 & 1.28E-05 & 0.03 & 0.22 & \cellcolor{mylightgray}0.01 & \cellcolor{mylightgray}0.87 & 1.23E-05 & 0.16 & 0.66 & \cellcolor{mylightgray}0.00 & 0.55 & 1.28E-05 & 0.04 & 0.30 & \cellcolor{mylightgray}0.09 & \cellcolor{mylightgray}0.82 & 1.25E-05 \\
& \maxentShort & 0.34 & 0.15 & 0.73 & 0.53 & 8.13E-06 & 0.13 & 0.12 & 0.65 & 0.83 & 8.20E-06 & 0.43 & 0.40 & 0.05 & 0.44 & 8.48E-06 & 0.17 & 0.17 & 0.59 & 0.76 & 8.04E-06 \\
& \avglikShort & 0.18 & 0.31 & 0.40 & 0.64 & 8.92E-06 & 0.05 & 0.20 & 0.41 & 0.83 & 8.85E-06 & 0.21 & 0.61 & 0.05 & 0.48 & 9.18E-06 & 0.06 & 0.28 & 0.18 & 0.81 & 8.90E-06 \\
& \maxlikShort & 0.22 & 0.27 & 0.41 & 0.65 & 7.80E-06 & 0.08 & 0.17 & 0.47 & 0.83 & 7.91E-06 & 0.22 & 0.60 & 0.04 & 0.49 & 7.66E-06 & \cellcolor{mylightgray}0.20 & \cellcolor{mylightgray}0.14 & 0.63 & 0.76 & 7.79E-06 \\
\bottomrule
\end{tabular}
}
\end{subtable}
\begin{subtable}[t]{\textwidth}
\centering
\resizebox{\textwidth}{!}{%
\begin{tabular}{clccccc|ccccc|ccccc|ccccc}
\toprule
& & \multicolumn{5}{c}{\llama} & \multicolumn{5}{c}{\llamaTwo} & \multicolumn{5}{c}{\alpaca} & \multicolumn{5}{c}{\vicuna} \\
Dataset & Method & SG$^{\uparrow}$ & RH$^{\downarrow}$ & AC$^{\downarrow}$ & AUC$^{\uparrow}$ & Time$^{\downarrow}$ & SG$^{\uparrow}$ & RH$^{\downarrow}$ & AC$^{\downarrow}$ & AUC$^{\uparrow}$ & Time$^{\downarrow}$ & SG$^{\uparrow}$ & RH$^{\downarrow}$ & AC$^{\downarrow}$ & AUC$^{\uparrow}$ & Time$^{\downarrow}$ & SG$^{\uparrow}$ & RH$^{\downarrow}$ & AC$^{\downarrow}$ & AUC$^{\uparrow}$ & Time$^{\downarrow}$ \\
\midrule
\multirow{8}{*}{NQ} & \randomShort & 0.12 & 0.13 & 0.72 & 0.50 & \cellcolor{mylightgray}1.10E-06 & 0.11 & 0.09 & 0.80 & 0.56 & \cellcolor{mylightgray}1.07E-06 & 0.38 & 0.39 & 0.14 & 0.44 & \cellcolor{mylightgray}1.17E-06 & 0.19 & 0.21 & 0.58 & 0.58 & \cellcolor{mylightgray}1.07E-06 \\
& \selfcheckgptShort & 0.10 & 0.15 & 0.59 & 0.82 & 1.24 & 0.05 & 0.15 & 0.29 & 0.89 & 1.55 & 0.33 & 0.44 & 0.14 & 0.44 & 1.39 & 0.08 & 0.32 & 0.50 & 0.67 & 1.22 \\
& \boxShort & \cellcolor{mylightgray}0.24 & \cellcolor{mylightgray}0.01 & 1.39 & 0.70 & 1.34 & \cellcolor{mylightgray}0.19 & \cellcolor{mylightgray}0.00 & 1.53 & 0.84 & 0.61 & \cellcolor{mylightgray}0.70 & \cellcolor{mylightgray}0.07 & 0.31 & \cellcolor{mylightgray}0.62 & 0.11 & 0.07 & 0.33 & 0.27 & 0.74 & 0.79 \\
& \quanShort & 0.16 & 0.10 & 0.84 & 0.80 & 1.14 & 0.00 & 0.19 & \cellcolor{mylightgray}0.01 & \cellcolor{mylightgray}0.90 & 1.30 & 0.25 & 0.51 & 0.14 & 0.45 & 1.14 & 0.10 & 0.30 & 0.33 & 0.73 & 1.20 \\
& \avgentShort & 0.11 & 0.14 & \cellcolor{mylightgray}0.52 & \cellcolor{mylightgray}0.84 & 1.20E-05 & 0.03 & 0.17 & 0.14 & \cellcolor{mylightgray}0.90 & 1.22E-05 & 0.20 & 0.56 & \cellcolor{mylightgray}0.07 & 0.53 & 1.30E-05 & 0.23 & 0.18 & \cellcolor{mylightgray}0.64 & 0.68 & 1.20E-05 \\
& \maxentShort & 0.18 & 0.08 & 0.89 & 0.81 & 8.21E-06 & \cellcolor{mylightgray}0.19 & \cellcolor{mylightgray}0.00 & 1.54 & 0.83 & 8.27E-06 & 0.40 & 0.36 & 0.12 & 0.46 & 8.71E-06 & 0.18 & 0.22 & 0.39 & 0.74 & 8.07E-06 \\
& \avglikShort & 0.14 & 0.12 & 0.70 & 0.82 & 8.70E-06 & 0.09 & 0.11 & 0.58 & 0.88 & 8.88E-06 & 0.23 & 0.54 & 0.08 & 0.52 & 8.91E-06 & 0.10 & 0.31 & 0.26 & \cellcolor{mylightgray}0.75 & 8.71E-06 \\
& \maxlikShort & 0.16 & 0.10 & 0.82 & 0.81 & 7.65E-06 & 0.08 & 0.12 & 0.50 & 0.88 & 7.54E-06 & 0.27 & 0.50 & 0.08 & 0.51 & 7.97E-06 & \cellcolor{mylightgray}0.24 & \cellcolor{mylightgray}0.16 & 0.53 & 0.73 & 7.70E-06 \\
\bottomrule
\end{tabular}
}
\end{subtable}
\begin{subtable}[t]{\textwidth}
\centering
\resizebox{\textwidth}{!}{%
\begin{tabular}{clccccc|ccccc|ccccc|ccccc}
\toprule
& & \multicolumn{5}{c}{\llama} & \multicolumn{5}{c}{\llamaTwo} & \multicolumn{5}{c}{\alpaca} & \multicolumn{5}{c}{\vicuna} \\
Dataset & Method & SG$^{\uparrow}$ & RH$^{\downarrow}$ & AC$^{\downarrow}$ & AUC$^{\uparrow}$ & Time$^{\downarrow}$ & SG$^{\uparrow}$ & RH$^{\downarrow}$ & AC$^{\downarrow}$ & AUC$^{\uparrow}$ & Time$^{\downarrow}$ & SG$^{\uparrow}$ & RH$^{\downarrow}$ & AC$^{\downarrow}$ & AUC$^{\uparrow}$ & Time$^{\downarrow}$ & SG$^{\uparrow}$ & RH$^{\downarrow}$ & AC$^{\downarrow}$ & AUC$^{\uparrow}$ & Time$^{\downarrow}$ \\
\midrule
\multirow{8}{*}{RTP} & \randomShort & 0.20 & 0.23 & 0.51 & 0.58 & \cellcolor{mylightgray}1.29E-06 & 0.20 & 0.24 & 0.54 & 0.56 & \cellcolor{mylightgray}1.29E-06 & 0.21 & 0.20 & 0.57 & 0.58 & \cellcolor{mylightgray}1.44E-06 & 0.20 & 0.22 & 0.56 & 0.57 & \cellcolor{mylightgray}1.31E-06 \\
& \selfcheckgptShort & 0.21 & 0.22 & 0.55 & 0.67 & 1.35 & 0.16 & 0.27 & 0.46 & 0.68 & 2.58 & 0.17 & 0.25 & 0.58 & 0.66 & 1.36 & 0.20 & 0.22 & 0.48 & 0.70 & 1.40 \\
& \boxShort & 0.32 & 0.11 & 1.05 & 0.49 & 0.26     & 0.26 & 0.17 & 0.60 & 0.67 & 0.30     & 0.36 & 0.06 & 1.04 & 0.48 & 0.33     & 0.39 & 0.03 & 1.04 & 0.54 & 0.41 \\
& \quanShort & 0.04 & 0.39 & \cellcolor{mylightgray}0.15 & \cellcolor{mylightgray}0.75 & 1.08 & 0.04 & 0.39 & \cellcolor{mylightgray}0.08 & \cellcolor{mylightgray}0.77 & 1.20 & 0.28 & 0.13 & 0.74 & 0.65 & 1.14 & 0.23 & 0.19 & 0.53 & 0.70 & 1.18 \\
& \avgentShort & 0.18 & 0.25 & 0.52 & 0.67 & 1.32E-05 & 0.22 & 0.21 & 0.64 & 0.63 & 1.29E-05 & \cellcolor{mylightgray}0.41 & \cellcolor{mylightgray}0.01 & 1.06 & 0.62 & 1.31E-05 & 0.05 & 0.37 & \cellcolor{mylightgray}0.12 & 0.77 & 1.36E-05 \\
& \maxentShort & 0.30 & 0.13 & 0.71 & 0.66 & 8.48E-06 & 0.34 & 0.09 & 0.83 & 0.62 & 8.99E-06 & 0.23 & 0.18 & 0.59 & 0.69 & 9.19E-06 & \cellcolor{mylightgray}0.40 & \cellcolor{mylightgray}0.02 & 1.08 & \cellcolor{mylightgray}0.79 & 8.72E-06 \\
& \avglikShort & \cellcolor{mylightgray}0.35 & \cellcolor{mylightgray}0.08 & 0.82 & 0.64 & 9.38E-06 & \cellcolor{mylightgray}0.36 & \cellcolor{mylightgray}0.07 & 0.85 & 0.62 & 9.21E-06 & 0.23 & 0.19 & \cellcolor{mylightgray}0.52 & 0.71 & 9.47E-06 & 0.26 & 0.16 & 0.63 & 0.68 & 9.08E-06 \\
& \maxlikShort & 0.20 & 0.23 & 0.48 & 0.69 & 8.26E-06 & 0.24 & 0.19 & 0.56 & 0.67 & 7.83E-06 & 0.28 & 0.13 & 0.60 & \cellcolor{mylightgray}0.72 & 8.26E-06 & 0.29 & 0.13 & 0.70 & 0.67 & 7.80E-06 \\
\bottomrule
\end{tabular}
}
\end{subtable}
\begin{subtable}[t]{\textwidth}
\centering
\resizebox{\textwidth}{!}{%
\begin{tabular}{clccccc|ccccc|ccccc|ccccc}
\toprule
& & \multicolumn{5}{c}{\llama} & \multicolumn{5}{c}{\llamaTwo} & \multicolumn{5}{c}{\alpaca} & \multicolumn{5}{c}{\vicuna} \\
Dataset & Method & SG$^{\uparrow}$ & RH$^{\downarrow}$ & AC$^{\downarrow}$ & AUC$^{\uparrow}$ & Time$^{\downarrow}$ & SG$^{\uparrow}$ & RH$^{\downarrow}$ & AC$^{\downarrow}$ & AUC$^{\uparrow}$ & Time$^{\downarrow}$ & SG$^{\uparrow}$ & RH$^{\downarrow}$ & AC$^{\downarrow}$ & AUC$^{\uparrow}$ & Time$^{\downarrow}$ & SG$^{\uparrow}$ & RH$^{\downarrow}$ & AC$^{\downarrow}$ & AUC$^{\uparrow}$ & Time$^{\downarrow}$ \\
\midrule
\multirow{8}{*}{\wmt} & \randomShort & 0.32 & 0.28 & 0.32 & 0.58 & \cellcolor{mylightgray}1.22E-06 & 0.19 & 0.20 & 0.61 & 0.50 & \cellcolor{mylightgray}1.22E-06 & 0.20 & 0.20 & 0.53 & 0.51 & \cellcolor{mylightgray}1.22E-06 & 0.24 & 0.22 & 0.53 & 0.54 & \cellcolor{mylightgray}1.20E-06 \\
& \selfcheckgptShort & 0.19 & 0.40 & 0.33 & 0.55 & 0.95 & 0.15 & 0.24 & 0.52 & 0.71 & 1.08 & 0.16 & 0.24 & 0.53 & 0.69 & 2.89 & 0.18 & 0.28 & 0.42 & 0.67 & 1.22 \\
& \boxShort & \cellcolor{mylightgray}0.56 & \cellcolor{mylightgray}0.03 & 0.67 & 0.41 & 0.12 & \cellcolor{mylightgray}0.36 & \cellcolor{mylightgray}0.03 & 0.93 & 0.74 & 0.26 & \cellcolor{mylightgray}0.40 & \cellcolor{mylightgray}0.00 & 1.11 & 0.63 & 0.14 & \cellcolor{mylightgray}0.32 & \cellcolor{mylightgray}0.14 & 0.73 & 0.58 & 0.38 \\
& \quanShort & 0.01 & 0.58 & \cellcolor{mylightgray}0.00 & 0.71 & 1.22 & 0.01 & 0.37 & \cellcolor{mylightgray}0.06 & \cellcolor{mylightgray}0.80 & 1.20 & 0.28 & 0.12 & 0.82 & 0.64 & 1.14 & 0.24 & 0.22 & 0.57 & 0.62 & 1.18 \\
& \avgentShort & 0.34 & 0.25 & 0.11 & \cellcolor{mylightgray}0.72 & 1.25E-05 & 0.06 & 0.33 & 0.14 & 0.79 & 1.29E-05 & 0.12 & 0.28 & 0.18 & 0.78 & 1.22E-05 & 0.21 & 0.25 & 0.31 & 0.72 & 1.25E-05 \\
& \maxentShort & 0.48 & 0.12 & 0.37 & 0.63 & 8.71E-06 & 0.32 & 0.07 & 0.89 & 0.69 & 4.13E-05 & 0.39 & 0.01 & 1.06 & 0.71 & 8.32E-06 & 0.07 & 0.39 & 0.09 & 0.76 & 8.55E-06 \\
& \avglikShort & 0.42 & 0.18 & 0.44 & 0.54 & 8.71E-06 & 0.24 & 0.15 & 0.53 & 0.75 & 8.79E-06 & 0.10 & 0.30 & 0.16 & 0.78 & 9.04E-06 & 0.01 & 0.45 & \cellcolor{mylightgray}0.01 & \cellcolor{mylightgray}0.79 & 9.20E-06 \\
& \maxlikShort & 0.34 & 0.25 & 0.27 & 0.62 & 8.08E-06 & 0.20 & 0.19 & 0.43 & 0.76 & 7.71E-06 & 0.07 & 0.33 & \cellcolor{mylightgray}0.14 & \cellcolor{mylightgray}0.79 & 8.01E-06 & 0.15 & 0.31 & 0.13 & 0.77 & 7.69E-06 \\
\bottomrule
\end{tabular}
}
\end{subtable}
\end{table*}

\begin{table*}[!tb]
\centering
\caption{RQ2 - Experimental Results for the performance of online safety analysis methods in code generation task. (SG: Safety Gain; RH: Residual Hazard; AC: Availability Cost; Time in seconds.)}
\label{table:RQ2_code_results}

\resizebox{\textwidth}{!}{%
\begin{tabular}{clccccc|ccccc}
\toprule
& & \multicolumn{5}{c}{\mbpp} & \multicolumn{5}{c}{\humaneval}  \\
Model & Method & SG$^{\uparrow}$ & RH$^{\downarrow}$ & AC$^{\downarrow}$ & AUC$^{\uparrow}$ & Time$^{\downarrow}$ & SG$^{\uparrow}$ & RH$^{\downarrow}$ & AC$^{\downarrow}$ & AUC$^{\uparrow}$ & Time$^{\downarrow}$ \\
\midrule
\multirow{8}{*}{\codellama} & Random & 0.30 & 0.35 & 0.23 & 0.55 & \cellcolor{mylightgray}5.45E-06 & 0.21 & 0.33 & 0.50 & 0.51 & \cellcolor{mylightgray}1.71E-05  \\
& \selfcheckgpt & 0.16 & 0.48 & \cellcolor{mylightgray}0.09 & \cellcolor{mylightgray}0.63 & 1.08 & 0.33 & 0.21 & 0.30 & 0.67 & 2.07 \\
& \boxMon & 0.11 & 0.54 & 0.10 & 0.62 & 0.04 & 0.03 & 0.52 & 0.24 & 0.62 & 0.05 \\
& \quan & 0.23 & 0.42 & 0.16 & 0.59 & 1.04 & 0.27 & 0.27 & 0.31 & 0.64 & 1.14 \\
& \avgent & 0.30 & 0.35 & 0.29 & 0.50 & 1.75E-05 & 0.36 & 0.18 & 0.11 & 0.78 & 3.20E-05  \\
& \maxent & \cellcolor{mylightgray}0.64 & \cellcolor{mylightgray}0.01 & 0.53 & 0.53 & 1.57E-05 & \cellcolor{mylightgray}0.42 & \cellcolor{mylightgray}0.12 & 0.16 & \cellcolor{mylightgray}0.79 & 2.66E-05 \\
& \avglik & 0.05 & 0.60 & 0.11 & 0.61 & 1.40E-05 & 0.33 & 0.21 & 0.18 & 0.73 & 2.69E-05  \\
& \maxlik & 0.09 & 0.56 & 0.13 & 0.60 & 1.34E-05 & 0.24 & 0.30 & \cellcolor{mylightgray}0.07 & 0.75 & 2.36E-05 \\
\bottomrule
\end{tabular}
}
\end{table*}



We first assess the effectiveness of online safety analysis methods on NLP tasks for open-source LLMs.
The results, detailed in Table~\ref{table:RQ2_nlp_results}, lead to the following observations.

\begin{compactitem}[$\bullet$]
\item \textbf{\truthfulqa.} 
We observe that: 
(1) For \truthfulqa, \boxMon can achieve the highest SG and diminish the potential hazard for the model. 
However, AC is high due to the frequent reporting of danger.
It can be considered a conservative method, with the lowest AUC among all analysis methods for \llama (0.40) and \alpaca (0.30).
(2) From the perspective of AUC, the average entropy method gets the best performance.
The average AUC over four LLMs is 0.76, and on Vicuna, the AUC is the highest, with 0.79.
(3) We can see that the time cost of white-box methods and \selfcheckgpt is at least three orders of magnitudes higher than the grey-box methods.
This is expected since they either involve a relatively more complex reasoning (white-box methods) or higher computational requirement, e.g., \selfcheckgpt. 
\item \textbf{\triviaqa.}
In terms of SG, \boxMon is the best on three LLMs, i.e., \llama, \llamaTwo, and \alpaca, with a low RH as well, i.e., 0.01 for all models.
While \maxlik (0.20) is the best on \vicuna.
In addition, \avgent achieves the best overall models, with an average value of 0.06.
\item \textbf{\nqopen.} 
Similarly, \boxMon still gains the best performance in terms of SG on the three LLMs. 
This indicates that \boxMon accurately captures the pattern of the safe execution and might be the first choice of question answering tasks when safety is of great importance.
\avgent get the best AUC on \llama and \llamaTwo, and the best AC on \llama and \vicuna.
\item \textbf{\realtoxicityprompt.} 
We can see that the grey-box methods are better than other approaches in terms of SG (\avglik is the best on \llama and \llamaTwo, \avgent is the best on \alpaca, and \maxent is the best on \vicuna), means that they are more suitable when safety matters.
While \quan attains the best AC and AUC on \llama (0.15 and 0.75) and \llamaTwo (0.08 and 0.77).
\item \textbf{\wmt.} 
\boxMon scores the best SG on all LLMs, with an average value of 0.41, and in \alpaca, the corresponding RH even drops to 0.
The grey-box methods get the better AUC over other approaches, with an average value of 0.7250, while the AUC for black-box methods and white-box methods are 0.5937 and 0.6412, respectively. 
\end{compactitem}

For the code generation task, \codellama stands as the only open-source LLM accessible for our study. 
Therefore, we evaluate all the analysis methods using this model, with the results summarized in Table~\ref{table:RQ2_code_results}.

\begin{compactitem}[$\bullet$]
\item \textbf{\mbpp.}
We can see that \maxent achieves the best SG and RH in \mbpp, which is 0.64 and 0.01, respectively.
While \selfcheckgpt gets the best AC and AUC in \mbpp.
Grey-box methods are still advisable in code generation tasks, which strike a balance between performance and overhead.
\item \textbf{\humaneval.}
\maxent achieves the best SG and RH, which are 0.42 and 0.12.
The AUC of \maxent is high in \humaneval as well, which is 0.79, the highest over all safety analysis methods.
\end{compactitem}

To summarize, for question answering (\truthfulqa, \triviaqa, and \nqopen) and machine translation (\wmt) tasks, when safety is important, using \boxMon might be a good choice.
While for text continuation, the grey-box approaches might be more suitable, and \maxent is the best for code generation.
Moreover, in NLP tasks, \avgent achieves the lowest AC in 11/16 benchmark instances (model-dataset pairs).
In both NLP tasks and coding tasks, grey-box approaches are recommended when AUC and time cost are important, which can achieve a balance of performance and overhead.


\begin{tcolorbox}[size=title, colback=white, breakable]
    {\textbf{Answer to RQ2:}
    In our benchmark evaluation, in NLP tasks, \boxMon is recommended when safety is vital.
    While grey-box approaches can achieve a balance of AUC and time overhead, in both NLP and coding tasks.
    }
\end{tcolorbox}

\subsection{RQ3: \rqthree}
\label{subsec:rqthree_results}


Alongside the study on open-source models, we also investigate the effectiveness of the collected online safety analysis method in the context of closed-source LLMs (i.e., GPT series). 
These models usually have larger model sizes with superior understanding and generation capabilities compared to their open-source counterparts. 
zIt is worth noting that due to limited access to the internal information of GPT models, only black-box and grey-box analysis methods are applicable to GPTs (i.e., Random, Entropy-based, Likelihood-based and SelfCheckGPT).

\begin{figure*}[!htb]
    \centering
    \includegraphics[width=\linewidth]{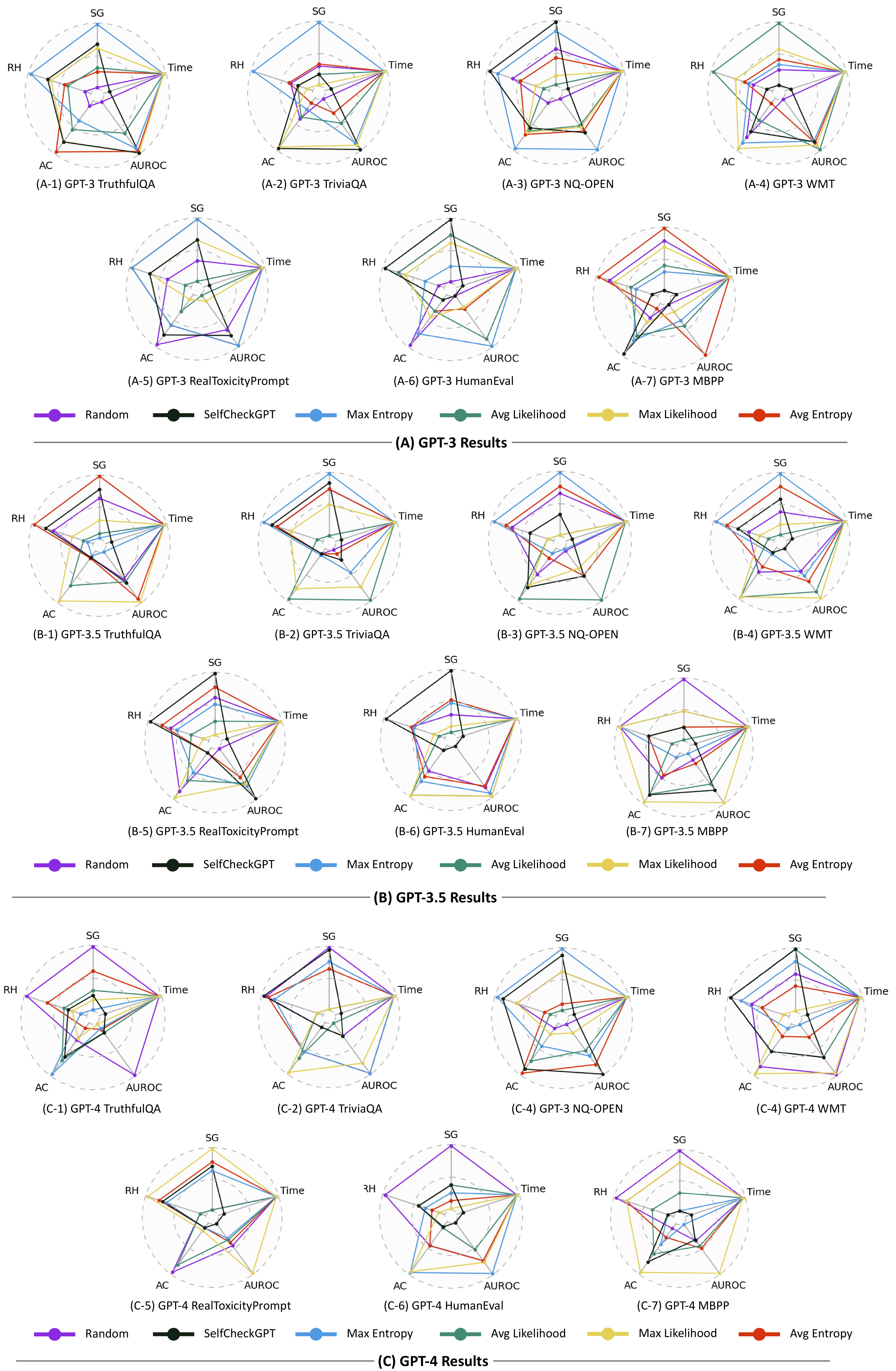}
    \caption{RQ3 - Radar plots of the performance of online safety analysis methods on closed-source LLMs. (SG: Safety Gain; RH: Residual Hazard; AC Availability Cost; Time in seconds.)
    }
    \label{fig:RQ3_radar_plot}
    \vspace{-10pt}
\end{figure*}

The radar plots in Figure~\ref{fig:RQ3_radar_plot} illustrate the performance of six applicable analysis methods w.r.t five designated metrics on three GPT models. 

\begin{compactitem}[$\bullet$]
\item \textbf{GPT-3}. The \maxent analysis method achieves superior results on question answering and text continuation tasks w.r.t SG, RH and AUC metrics. 
Surprisingly, the \avgent analysis method obtains the best scores in 4 out of 5 metrics in the MBPP dataset.
In contrast, the likelihood-based analysis methods present moderate results on question-answering and text continuation tasks but relatively better performance on coding and machine translation tasks.
Moreover, the black-box method, \selfcheckgpt, appears short in analyzing the machine translation task and suffers great overhead issues across all datasets.
\item \textbf{GPT-3.5}. In terms of GPT-3.5, the studied analysis methods show relatively similar results with GPT3. 
From Figure~\ref{fig:RQ3_radar_plot}(B), the likelihood-based analysis methods exhibit performant AUC, AC and Time scores. 
Namely, these two likelihood-based analysis methods are considered to have advantages in detecting faulty outputs during the generation with minor side effects on the performance of the LLM. 
From the perspective of SG and RH, the other grey-box approaches and entropy-based analysis methods outperform the other methods in question answering and machine translation tasks. 
In addition, \selfcheckgpt shows distinct advantages in the text continuation with the best scores on SG, RH and AUC but has inadequate Time and AC results on most tasks.
\item \textbf{GPT-4}. The likelihood-based methods show competitive performance on most tasks except question answering. 
In particular, as illustrated in Figure~\ref{fig:RQ3_radar_plot}(C), the \maxlik analysis method achieves the best or the nearly best AUC scores on coding, text continuation and translation tasks. 
Meanwhile, entropy-based analysis methods have the advantage of enhancing Safe Gain and reducing Residual Hazards across text-based tasks (except the \nqopen dataset).
In addition, the black-box methods, Random and \selfcheckgpt, exhibit moderate performance on all tasks.
\end{compactitem}

Overall, similar to the observations from the open-source models, the five analysis methods present inconsistent performance among different tasks, metrics and models.
In particular, the entropy-based methods have better performance regarding SG and RH on most tasks and GPT models; while the likelihood-based analysis methods present unique advantages on AUC and AC, in contrast.
The black-box approach, \selfcheckgpt, offers superior performance on specific tasks and models, but the concomitant overhead raises main concerns about the additional payload infused within the LLM operation. 
Drew on the experiment results, we consider the grey-box methods (i.e., entropy-based and likelihood-based) to have the most generalized and applicable capabilities for reflecting the performance of LLMs on certain metrics for various downstream tasks.


\begin{tcolorbox}[size=title, colback=white, breakable]
    {\textbf{Answer to RQ3:}
    For the studied closed-source GPT models, the entropy-based methods show advantages in ensuring safety gain and reducing residual hazards in most tasks.
    Meanwhile, the likelihood-based approaches exhibit better results in untruthful case detection and availability cost inhibition.
    }
\end{tcolorbox}

\section{Exploratory Study}
\label{sec:explore_study}

From the experiment results of RQ2 and RQ3, we notice that the examined online safety analysis methods have distinct advantages and drawbacks when tackling different downstream tasks and LLMs. 
Therefore, in this section, we describe an exploratory study, which contains simple but effective enhancements for online safety analysis methods - \emph{hybridization}.
The study design is detailed in Section~\ref{subsec:exploratory_design}, and the result is presented in Section~\ref{subsec:exploratory_result}.

\subsection{Study Design}
\label{subsec:exploratory_design}

\noindent \textbf{Motivation.}
Intuitively, it is reasonable that it is challenging for a single analysis method to handle diverse task scenarios and LLMs with different scales and characteristics. 
Therefore, our findings inspired us to consider that a \emph{hybrid analysis} approach may possibly leverage strengths from constituent analysis methods and obtain performance beyond that which could be achieved by any individual analysis method.
We initiate an early-stage exploration with three commonly used hybridization methods, namely, \emph{Random Selection},~\cite{ganaie2022ensemble}\emph{Majority Voting}~\cite{dietterich2002ensemble}, and \emph{Bagging}~\cite{sagi2018ensemble},
to probe their effectiveness in the online safety analysis for LLMs.

\noindent \textbf{Hybridization Methods.}
The details of hybridization methods are described as follows.

\begin{compactitem}[$\bullet$]

\item \textbf{Random Selection.}
As a basic hybridization method, the selected base analysis methods first perform their execution individually and provide their predictions.
Then, one judgment is randomly chosen as the prediction of the online safety analysis methods.

\item \textbf{Majority Voting.}
Majority voting hybridization combines the predictions of multiple base analysis methods by majority voting, where multiple individual models are combined to make a final prediction. 
Such a method is resilient to overfitting and able to handle imbalanced data.

\item \textbf{Bagging.} 
With advantages like the mitigation of overfitting and improved generalization, bagging involves constructing multiple instances of the same base online safety analysis methods on different subsets of the constructed data, sampled with replacement.
The final prediction is obtained by voting the predictions of all base analysis methods.
\end{compactitem}



\noindent \textbf{Experimental Design.}
We conduct the experiment on \truthfulqa with \vicuna and \mbpp with \codellama.
For random selection, we choose five online safety analysis methods as the base components: \random, \avgent, \avglik, \boxMon, and \selfcheckgpt.
For bagging hybridization, we evaluate \boxMon and \quan, due to the fact that the diversity of base methods comes from the different constructed data, and in this study, only white-box methods involve such constructed data.
We randomly extract 80\% of the safe data as the constructed data for the base analysis method, and five base analysis methods are constructed in our study.
For majority voting, we also choose the same five analysis methods as random selection.

\subsection{RQ4: \rqfour}
\label{subsec:exploratory_result}

Table~\ref{table:RQ5_experimentalResults} presents the evaluation results of three hybridization methods on question answering and code generation tasks, respectively. 

\begin{compactitem}[$\bullet$]
\item \textbf{TruthfulQA}. For the question-answering task, the proposed hybrid analysis methods do not show notable improvements compared with the individual counterparts.
Nevertheless, the Random Selection still manifests adequate performance compared to the single analysis methods. 
Namely, the Random Selection method achieves the best results on SG, RH and AUC, even if it is a basic hybridization solution.
In contrast, none of the individual analysis methods can obtain the best scores over multiple metrics simultaneously, which unveils that the hybrid analysis methods have the potential to balance the trade-off and satisfy multi-requirements at the same time. 

\item \textbf{MBPP}. In terms of the code generation task, the Bagging method with Quantitative analysis methods succeeds in producing more performant results in SG, RH, and AUC.
It is worth mentioning that even the base method of the aforementioned Bagging method, i.e., the Qualitative analysis method, does not achieve equivalent results on the MBPP dataset. 
Considering the rationale behind the Bagging approach, i.e., multiple base analysis methods trained on different data subsets, the Bagging method shows the prospect that hybridization methods can deliver advanced functionalities even with a group of same-type constituent analysis methods.
\end{compactitem}
In general, at least one of the hybridization methods can compete with or outperform the individual safety analysis methods across various metrics.
However, none of the studied hybridization methods can consistently offer performant results regardless of evaluation metrics and tasks. 
Therefore, the hybrid online safety analysis methods call for further study and investigation to probe advanced hybridization strategies with better stability and effectiveness.

\begin{tcolorbox}[size=title, colback=white, breakable]
    {\textbf{Answer to RQ4:}
    For both question-answering and code-generation tasks, hybrid safety analysis methods exhibit the potential to take advantage of constituent analysis methods and outperform individual ones over multiple evaluation perspectives. 
    }
\end{tcolorbox}

\begin{table*}[!tb]
\centering
\caption{RQ4 - Performance of different hybridization methods. 
(SG: Safety Gain; RH: Residual Hazard; AC Availability Cost; Time in seconds.) }
\label{table:RQ5_experimentalResults}
\resizebox{\textwidth}{!}{%
\vspace{-7pt}

\begin{tabular}{lccccc|ccccc}
\toprule
& \multicolumn{5}{c}{\vicuna-\truthfulqa} & \multicolumn{5}{c}{\codellama-\mbpp}  \\
Method & SG$^{\uparrow}$ & RH$^{\downarrow}$ & AC$^{\downarrow}$ & AUC$^{\uparrow}$ & Time$^{\downarrow}$ & SG$^{\uparrow}$ & RH$^{\downarrow}$ & AC$^{\downarrow}$ & AUC$^{\uparrow}$ & Time$^{\downarrow}$ \\
\midrule
\random & 0.14 & 0.20 & 0.68 & 0.72 & \cellcolor{mylightgray}3.20E-06 & 0.30 & 0.35 & 0.23 & 0.55 & \cellcolor{mylightgray}5.45E-06 \\
\selfcheckgpt & 0.09 & 0.24 & 0.46 & 0.76 & 1.43     & 0.16 & 0.48 & \cellcolor{mylightgray}0.09 & 0.63 & 1.08     \\
\avgent & 0.09 & 0.25 & \cellcolor{mylightgray}0.34 & 0.79 & 1.57E-05 & 0.30 & 0.35 & 0.29 & 0.50 & 1.75E-05 \\
\avglik  & 0.11 & 0.23 & 0.38 & 0.79 & 1.21E-05 & 0.05 & 0.60 & 0.11 & 0.61 & 1.40E-05 \\
\boxMon & \cellcolor{mylightgray}0.26 & 0.08 & 1.03 & 0.67 & 0.24     & 0.11 & 0.54 & 0.10 & 0.62 & 0.04     \\
\quan & 0.19 & 0.15 & 0.58 & 0.78 & 1.16     & 0.27 & \cellcolor{mylightgray}0.27 & 0.31 & 0.64 & 1.14     \\
\midrule
Random\_Hybridization & \cellcolor{mylightgray}0.26 & \cellcolor{mylightgray}0.07 & 0.56 & \cellcolor{mylightgray}0.81 & 1.43     & 0.12 & 0.53 & 0.24 & 0.52 & 1.08     \\
Voting\_Hybridization & 0.12 & 0.22 & 0.48 & 0.76 & 1.43     & 0.10 & 0.55 & 0.12 & \cellcolor{mylightgray}0.65 & 1.08     \\
Bagging\_Box\_Hybridization & 0.15 & 0.18 & 0.63 & 0.74 & 1.44     & 0.19 & 0.46 & 0.25 & 0.52 & 1.09     \\
Bagging\_Quantitative\_Hybridization & 0.17 & 0.16 & 0.61 & 0.76 & 1.44     & \cellcolor{mylightgray}0.35 & 0.30 & 0.21 & \cellcolor{mylightgray}0.65 & 1.15 \\
\bottomrule
\end{tabular}
}
\vspace{-10pt}
\end{table*}


\section{Discussion}
\label{sec:discussion}



\noindent \textbf{Guideline for developing LLM-specific online safety analysis method.}
As shown by the experimental results in Section~\ref{sec:empiricalstudy}, for analyzing open-source models, when safety is crucial, \boxMon might be more suitable than other methods.
Besides, considering the balance of safety, AC, AUC, and time, grey-box approaches, which take entropy and output likelihood into consideration, could be a good choice. 
In terms of closed-source LLMs, similar to the findings from the open-source models, the grey-box analysis methods show better generality and applicability in the context of different tasks and evaluation metrics.
The entropy-based methods have advantages over others in safeguarding the safety gain and inhibiting the residual hazard, while the likelihood methods gain momentum to detect unsafe cases and lower the availability cost.
Moreover, when developing and designing online safety analysis methods specifically for LLMs, unique attributes of LLMs, such as their auto-regressive nature, should be considered.
To enhance online safety analysis, techniques should be devised with the specific properties of LLMs in mind.


\noindent \textbf{Advanced hybrid online safety analysis method.} 
As mentioned in Section~\ref{sec:explore_study}, the proposed hybridization methods show distinct performance among different tasks and metrics; in other words, none of the studied hybrid analysis methods can retain consistent advantages. 
Note that the capability of the hybrid methods is greatly determined by the constituent analysis methods and the fusion strategy of their outputs.
However, the hybridization strategies studied do not take any information from the characteristics of individual analysis methods into account when selecting or fusing the outputs. 
The hybrid methods sometimes fail to identify the best candidate constituent analysis method under certain circumstances. 
Accordingly, more advanced hybridization solutions are needed to explicitly be aware of the capabilities of the constituent analysis methods under various cases to strategically generate the fused output to fulfill distinct LLM generation requirements simultaneously.

\noindent \textbf{Extend to other practical scenarios.}
LLM has also been deployed in various practical domains, such as autonomous driving~\cite{wu2024reality} and robotics  control~\cite{ren2023robots,song2023self,zhou2023isr}. 
For instance, in autonomous driving, LLMs can enhance the system's understanding of natural language commands, facilitating more efficient interactions between drivers and vehicles~\cite{wu2024reality}.
Moreover, recent researches~\cite{ren2023robots,song2023self,zhou2023isr} have shown a growing interest in integrating LLMs as high-level decision-makers in robotics systems, aiming to realize autonomous and intelligent control processes.
However, given the interaction between the physical system, e.g., cars and robots, with the real world, ensuring safety in these applications is of vital importance. 
This highlights the critical need for conducting comprehensive online safety analysis for LLM-enabled robotics applications.

\section{Threats to Validity}
\label{sec:threatToVal}

This section contains a discussion of the potential threats to the study's validity and the countermeasures that have been implemented.

\noindent\textbf{External Threat.}
This threat can be possibly from the considered online safety analysis methods, tasks and datasets, and LLMs.
For the online safety analysis methods, we collect eight representative analysis methods from the community of AI and software engineering. 
For the studied tasks and datasets, we consider representative tasks from NLP and coding, including question answering, text continuation, machine translation, and code generation.
The datasets have been widely studied for the quality assurance of LLMs.
For LLMs, the benchmark contains both open-source and closed-source models and is with state-of-the-art performance.

\noindent\textbf{Internal Threat.}
The implementation of the online safety analysis methods and evaluation metrics may result in internal threats.
To mitigate these threats, we implement the analysis methods and metrics based on the existing work~\cite{huang2023look,henzinger2019outside,guerin2022unifying,manakul2023selfcheckgpt,lukina2021into}, and with attentive verification conducted by two individuals with solid programming and LLMs knowledge.

\noindent\textbf{Construct Threat.}
The validity of the experimental results can be affected by the configuration of the online safety analysis method.
Given that there are few works addressing the gap of providing a benchmark for online safety analysis on LLMs, we follow the settings and hyper-parameters from existing work to perform the comparison.

\section{Related Work}
\label{sec:related_work}

\subsection{Online Safety Analysis for Deep Learning System}

Effectively performing online analysis for deep learning models is key in ensuring AI-driven systems' safe, secure, transparent, and reliable deployment~\cite{serban2020adoption, amershi2019software, liu2019secure}. This process aims to identify abnormal or erroneous states in AI models that could violate expected properties at runtime. Given that LLMs are highly complex and hard to interpret, assessing their correctness and reliability poses significant challenges. To address these issues, researchers from different communities have explored the issue from multiple perspectives. Related research efforts include but are not limited to selective classification~\cite{cortes2016learning, geifman2019selectivenet, geifman2017selective, fisch2022calibrated, xia2022augmenting}, Out-of-Distribution (OOD) detection~\cite{hendrycks2017a, lee2018simple, liang2018enhancing, hsu2020generalized, liu2020energy, berend2020cats, sun2021react, xia2022augmenting, ahn2023line}, misclassification or anomaly detection~\cite{sabokrou2018adversarially, hendrycks2018deep, kim2019guiding, zhang2020towards, granese2021doctor, chen2021detecting, nado2021uncertainty, zhu2023openmix, huang2023patchcensor, kim2023evaluating, huang2023look, gomes2024a, liu2024neuron}, and runtime monitoring~\cite{ferreira2021benchmarking, henzinger2019outside, aslansefat2020safeml, cheng2019runtime, stocco2020misbehaviour, xiao2021self, junges2021runtime, balakrishnan2021percemon, stocco2022thirdeye, guerin2022unifying, ayerdi2023metamorphic, konighofer2023online, zolfagharian2023smarla, xie2023mosaic, gronauer2024reinforcement, sun2024redriver}. They can be broadly classified into three categories~\cite{ferreira2021benchmarking}: input analysis~\cite{cortes2016learning, geifman2017selective, geifman2019selectivenet, sabokrou2018adversarially, hendrycks2018deep, stocco2020misbehaviour, chen2021detecting, fisch2022calibrated, zhu2023openmix}, DNNs' internal state analysis~\cite{lee2018simple, henzinger2019outside, sun2021react, aslansefat2020safeml, cheng2019runtime, xiao2021self, stocco2022thirdeye, ahn2023line, liu2024neuron}, and output analysis~\cite{liang2018enhancing, kim2019guiding, hsu2020generalized, liu2020energy, kim2023evaluating, ayerdi2023metamorphic, zhang2020towards, berend2020cats, granese2021doctor, nado2021uncertainty, xia2022augmenting, huang2023patchcensor, ferreira2021benchmarking, huang2023look, gomes2024a}. 

Input analysis adopts a data-driven approach for online safety analysis of DNNs, based on the assumption that inputs beyond a DNN's handling capability exhibit unique features and differ distributionally from normal inputs. This allows for the direct rejection of predictions on inputs outside the AI model's capabilities through real-time analysis. Related work includes designing classifiers that are capable of performing selective classification~\cite{cortes2016learning, geifman2019selectivenet, geifman2017selective, fisch2022calibrated} or training an additional DNN to identify abnormal inputs online~\cite{sabokrou2018adversarially, hendrycks2018deep, stocco2020misbehaviour, chen2021detecting, zhu2023openmix}, enhancing system robustness by preventing the processing of potentially problematic inputs. On the other hand, internal state analysis is more model-specific. Researchers find that it is possible to identify possible erroneous patterns by examining the neuron activations inside the neural networks. A representative solution is to compare the neuron activation distributions between training data and test instances to determine whether the test points are OOD~\cite{lee2018simple}. It is also possible to transform attention maps of DNNs into their confidence scores and identify inputs that can trigger errors~\cite{stocco2022thirdeye}. Finally, output analysis is closely related to uncertainty quantification and confidence estimation. These methods take the outputs of DNNs as indicators, attempting to estimate the possibility of errors. As an example, Hendrycks \etal established the first baseline using softmax prediction probability as the confidence score for OOD detection~\cite{hendrycks2017a}. 

While the aforementioned studies have shown success, they primarily focus on classification tasks that require time-invariant, standalone inferences. In contrast, LLMs operate in an auto-regressive, time-dependent manner, generating tokens sequentially. This process shares similarities with AI-enabled control systems in reinforcement learning, where decisions are made based on current world observations, system status, and historical actions. To ensure the safety of such dynamic systems at runtime, related research often involves predicting future event sequences and guiding the system to prevent it from entering unsafe states~\cite{junges2021runtime, xie2023mosaic, konighofer2023online, zolfagharian2023smarla, gomes2024a, gronauer2024reinforcement}. Under certain conditions, this process can be rigorously analyzed using mathematical models such as Markov Chains~\cite{junges2021runtime}, offering failure rate estimations for the current state. However, applying these methods to the online analysis of LLMs presents a challenge. Cyber-physical systems are characterized by states with well-defined meanings in the physical world. In contrast, the states within LLMs, defined by sequences of tokens and internal neural activations, lack direct correspondence to physical entities, making it challenging to directly transfer modelling approaches. Despite the difficulties, there are some early attempts to control the decoding process of LLMs to make the output better align with expectations. Cao \etal~\cite{cao2023systematic} proposed to perform dead-end analysis on LLMs' generation to avoid entering a dead-end state and thus making the output less toxic. Mudgal \etal~\cite{mudgal2023controlled} applied KL-regularized reinforcement learning on LLM decoding and demonstrated that the method could help increase dialog helpfulness and harmlessness. Our work is parallel to these efforts by focusing on analyzing the safety of the generation process rather than improving generation quality.









\subsection{Quality Assurance for LLM}
LLMs have significantly impacted natural language processing~\cite{chang2023survey, zhao2023survey, zhang2023sentiment, pu2023summarization} and software engineering~\cite{chen2021evaluating, lemieux2023codamosa, liu2023fill, prenner2022can, sobania2023analysis, nijkamp2023codegen, fan2023large, hou2024large,wang2024software}, and are believed to show early signs of Artificial General Intelligence (AGI)~\cite{bubeck2023sparks}. Despite their considerable achievements, research on effective quality assurance for these complex models is in its infancy. Existing studies have pointed out various serious safety issues for LLMs~\cite{sun2024trustllm}, such as hallucination~\cite{zhang2023siren, xu2024hallucination}, security~\cite{liu2023jailbreaking}, privacy~\cite{yao2024survey}, toxicity~\cite{gehman2020realtoxicityprompts}, fairness~\cite{li2023survey}, and robustness~\cite{wang2023large}. Addressing these issues is critical for LLMs to be safely deployed in critical environments. To bridge the gap, recent studies identify several promising approaches for detecting potential violations of specific properties in LLMs. These strategies fall into three main categories: detector-based methods~\cite{markov2023holistic, zhou-etal-2021-detecting, inan2023llama, bhatt2023purple, achintalwar2024detectors, zhang2024shieldlm}, uncertainty estimation for LLMs~\cite{manakul2023selfcheckgpt, lin2023generating, xiong2024can, kuhn2023semantic}, and self-refinement~\cite{huang2023large, shinn2024reflexion, chen2023teaching, pan2023automatically, gou2024critic, madaan2024self, zhou2023isr}. 

Detector-based methods train classifiers or fine-tune existing LLMs to identify harmful, erroneous, or unintended content in LLM inputs and outputs. These detectors operate under the assumption that the data distribution of misbehaviors is known and that related labelled data are available for training. This approach treats detection as a classical DNN training problem. With an appropriate architecture and sufficient data, these detectors can effectively identify unintended content. In an early effort towards this goal, OpenAI developed a systematic workflow that includes data collection, cleaning, and the integration of active and adversarial learning techniques~\cite{markov2023holistic}. This comprehensive approach led to the creation of a robust model for detecting undesired content, which is now available as a moderation API service. A similar workflow has also been adopted by IBM Research~\cite{achintalwar2024detectors}. In their workflow, the detectors are key to the system's continuous integration process, influencing the refinement of both pre-processing and tuning steps. Beyond these efforts, Meta has introduced Llama Guard~\cite{inan2023llama} and Purple Llama~\cite{bhatt2023purple}. Llama Guard is designed to safeguard human-AI interactions, ensuring these conversations remain safe and constructive. On the other hand, Purple Llama focuses on evaluating code security, helping to identify vulnerabilities and improve the safety of AI-generated code. 

The successes outlined above demonstrate that, with the prior knowledge of potential failure patterns and sufficient resources, it is feasible to construct effective detectors for quality assurance. However, this assumption may not always hold true. On the contrary, uncertainty estimation focuses on assessing the status of LLMs and conducting model-specific evaluations of their confidence levels. However, applying related methods to LLMs presents greater challenges than traditional DNN classifiers. This complexity arises from the sequential, time-dependent nature of the generation process in LLMs. In this research area, Manakul \etal introduced a black-box hallucination detection method that utilizes token-level prediction likelihood and entropy to identify inaccuracies in model outputs~\cite{manakul2023selfcheckgpt}. Kuhn \etal~\cite{kuhn2023semantic} suggested that instead of focusing on token-level confidence, assessing semantic uncertainty at the sentence level could provide a more accurate and insightful approach. Huang \etal~\cite{huang2023look} further explored this by conducting a large-scale empirical study on the effectiveness of token-level and sentence-level uncertainty estimation, finding that semantic uncertainty indeed offers superior performance. Xiong \etal~\cite{xiong2024can} critically evaluated the current state of uncertainty estimation, identifying several unresolved issues, including the challenge of overconfidence, which indicates that despite advancements, the methodology for uncertainty estimation in LLMs still requires significant refinement.

Given the versatility of LLMs in performing a variety of downstream tasks, it is viable to task LLMs by analyzing their own outputs. Research indicates that internal feedback mechanisms can enable LLMs to identify, detect, and even correct unintended behaviors autonomously. For instance, Chen \etal~\cite{chen2023teaching} introduced a self-debugging code generation framework that notably enhances baseline accuracy by up to $12\%$. Similarly, Shin \etal~\cite{shinn2024reflexion} demonstrated that LLMs could significantly improve performance through self-assessment of their past trajectories. Zhou \etal~\cite{zhou2023isr} incorporated additional validators into the iterative self-refinement process of LLMs, achieving an enhancement in their performance for long-horizon sequential task planning tasks.  Despite these advances, Huang \etal~\cite{huang2023large} argue that current self-reflection capabilities in LLMs are not yet perfect, suggesting it might be overly optimistic to depend heavily on these techniques at this stage.

Nevertheless, all the aforementioned methods focus on post-generation analysis of LLM outputs, whereas our work is dedicated to conducting online safety analysis for LLMs. 
This distinction underscores a fundamentally different approach: instead of evaluating outputs after generation, we aim to monitor and ensure safety in real-time as LLMs operate.

\subsection{Benchmark for LLM}

Developing benchmarks for LLMs is a crucial step in understanding the current limitations and providing insights for further improvements. 
Given the multifaceted capabilities of LLMs to tackle a wide range of tasks, their benchmarks are inherently more complex compared to traditional DL models.
These benchmarks encompass a variety of tasks, including natural language understanding~\cite{bang2023multitask, liang2023holistic, qin2023chatgpt}, reasoning~\cite{frieder2024mathematical, liu2023evaluating, fu2023chain}, and code generation~\cite{du2023classeval, chen2021evaluating, liu2024your}, evaluated from multiple perspectives, such as correctness~\cite{liang2023holistic}, factuality~\cite{lin2021truthfulqa, honovich2022true}, robustness~\cite{zhu2023promptbench, wang2023decodingtrust}, fairness~\cite{sun2024trustllm}, and privacy~\cite{hamid2023genaipabench}. 
These datasets may be human-labeled~\cite{santhanam2021rome, honovich2021q, chen2021evaluating}, illustrating the meticulous effort for better quality and relevance. 
Others are extracted from external resources, leveraging existing repositories of knowledge~\cite{liu2021token, thorne2018fever}. 
Some datasets are transformed from other datasets, undergoing modifications to better suit specific tasks or objectives~\cite{wang2021adversarial, wang-etal-2021-textflint}. 
Additionally, there are datasets that are labeled or generated by AI models themselves, reflecting a growing trend of leveraging AI to prepare data for evaluation~\cite{lin2021truthfulqa, gehman2020realtoxicityprompts}. 

Among the notable studies in this field, HELM~\cite{liang2023holistic} is an important work that provides an extensive evaluation of LLMs. HELM performed evaluation across seven metrics in $42$ different scenarios for $30$ language models, offering comprehensive insights into the capabilities and limitations of current LLMs. Another benchmark, DecodingTrust~\cite{wang2023decodingtrust}, focuses on assessing LLMs from eight perspectives of trustworthiness, highlighting the previously unknown vulnerabilities to trustworthiness threats. TrustLLM~\cite{sun2024trustllm} is a more recent benchmark that evaluates $16$ LLMs using over $30$ datasets across six critical dimensions: truthfulness, safety, fairness, robustness, privacy, and machine ethics. These benchmarks collectively contribute to a deeper understanding of LLM performance and ethical implications, guiding the development of more reliable and equitable AI systems. To the best of our knowledge, our study is the first to assess the effectiveness of various online safety analysis methods applied to both open-source and closed-source LLMs. Furthermore, our evaluation encompasses tasks within both the natural language processing and software engineering domains, offering a broad and diverse testing ground for future research.









\section{Conclusion}
\label{sec:conclusion}

In this paper, we introduce the first benchmark of online safety analysis methods for LLMs in different domains, which can be deemed as a base assessment framework for developing and enhancing new online safety analysis methods.
We start with a pilot study to show the presence of unsafe output indeed emerges at the early stage of generation and motivate the necessity of applying online safety analysis.
Based on the constructed benchmark, which contains a series of analysis methods, LLMs, datasets, and evaluation metrics, we systematically and extensively measure the effectiveness and performance of state-of-the-art online safety analysis methods. 
We also propose simple yet effective hybridization approaches, which can leverage the advantages of diverse individual analysis methods and achieve better performance.
We believe that the novel constructed benchmark and empirical evaluation can offer developers and academic researchers valuable insights on developing and understanding more LLM-specific online safety analysis methods and boost the development of quality assurance for LLM.

\bibliographystyle{unsrtnat}
\bibliography{ref}

\begin{thebibliography}{205}
\providecommand{\natexlab}[1]{#1}
\providecommand{\url}[1]{\texttt{#1}}
\expandafter\ifx\csname urlstyle\endcsname\relax
  \providecommand{\doi}[1]{doi: #1}\else
  \providecommand{\doi}{doi: \begingroup \urlstyle{rm}\Url}\fi

\bibitem[Achiam et~al.(2023)Achiam, Adler, Agarwal, Ahmad, Akkaya, Aleman,
  Almeida, Altenschmidt, Altman, Anadkat, et~al.]{achiam2023gpt}
Josh Achiam, Steven Adler, Sandhini Agarwal, Lama Ahmad, Ilge Akkaya,
  Florencia~Leoni Aleman, Diogo Almeida, Janko Altenschmidt, Sam Altman,
  Shyamal Anadkat, et~al.
\newblock Gpt-4 technical report.
\newblock \emph{arXiv preprint arXiv:2303.08774}, 2023.

\bibitem[Vaithilingam et~al.(2022)Vaithilingam, Zhang, and
  Glassman]{vaithilingam2022expectation}
Priyan Vaithilingam, Tianyi Zhang, and Elena~L Glassman.
\newblock Expectation vs. experience: Evaluating the usability of code
  generation tools powered by large language models.
\newblock In \emph{Chi conference on human factors in computing systems
  extended abstracts}, pages 1--7, 2022.

\bibitem[Ren et~al.(2023)Ren, Dixit, Bodrova, Singh, Tu, Brown, Xu, Takayama,
  Xia, Varley, et~al.]{ren2023robots}
Allen~Z Ren, Anushri Dixit, Alexandra Bodrova, Sumeet Singh, Stephen Tu, Noah
  Brown, Peng Xu, Leila Takayama, Fei Xia, Jake Varley, et~al.
\newblock Robots that ask for help: Uncertainty alignment for large language
  model planners.
\newblock \emph{arXiv preprint arXiv:2307.01928}, 2023.

\bibitem[Zhou et~al.(2023)Zhou, Song, Yao, Shu, and Ma]{zhou2023isr}
Zhehua Zhou, Jiayang Song, Kunpeng Yao, Zhan Shu, and Lei Ma.
\newblock Isr-llm: Iterative self-refined large language model for long-horizon
  sequential task planning.
\newblock \emph{arXiv preprint arXiv:2308.13724}, 2023.

\bibitem[Cascella et~al.(2023)Cascella, Montomoli, Bellini, and
  Bignami]{cascella2023evaluating}
Marco Cascella, Jonathan Montomoli, Valentina Bellini, and Elena Bignami.
\newblock Evaluating the feasibility of chatgpt in healthcare: an analysis of
  multiple clinical and research scenarios.
\newblock \emph{Journal of medical systems}, 47\penalty0 (1):\penalty0 33,
  2023.

\bibitem[Wu et~al.(2023)Wu, Irsoy, Lu, Dabravolski, Dredze, Gehrmann, Kambadur,
  Rosenberg, and Mann]{wu2023bloomberggpt}
Shijie Wu, Ozan Irsoy, Steven Lu, Vadim Dabravolski, Mark Dredze, Sebastian
  Gehrmann, Prabhanjan Kambadur, David Rosenberg, and Gideon Mann.
\newblock Bloomberggpt: A large language model for finance.
\newblock \emph{arXiv preprint arXiv:2303.17564}, 2023.

\bibitem[Wenzek et~al.(2019)Wenzek, Lachaux, Conneau, Chaudhary, Guzm{\'a}n,
  Joulin, and Grave]{wenzek2019ccnet}
Guillaume Wenzek, Marie-Anne Lachaux, Alexis Conneau, Vishrav Chaudhary,
  Francisco Guzm{\'a}n, Armand Joulin, and Edouard Grave.
\newblock Ccnet: Extracting high quality monolingual datasets from web crawl
  data.
\newblock \emph{arXiv preprint arXiv:1911.00359}, 2019.

\bibitem[Raffel et~al.(2020)Raffel, Shazeer, Roberts, Lee, Narang, Matena,
  Zhou, Li, and Liu]{raffel2020exploring}
Colin Raffel, Noam Shazeer, Adam Roberts, Katherine Lee, Sharan Narang, Michael
  Matena, Yanqi Zhou, Wei Li, and Peter~J Liu.
\newblock Exploring the limits of transfer learning with a unified text-to-text
  transformer.
\newblock \emph{Journal of machine learning research}, 21\penalty0
  (140):\penalty0 1--67, 2020.

\bibitem[Gao et~al.(2020)Gao, Biderman, Black, Golding, Hoppe, Foster, Phang,
  He, Thite, Nabeshima, et~al.]{gao2020pile}
Leo Gao, Stella Biderman, Sid Black, Laurence Golding, Travis Hoppe, Charles
  Foster, Jason Phang, Horace He, Anish Thite, Noa Nabeshima, et~al.
\newblock The pile: An 800gb dataset of diverse text for language modeling.
\newblock \emph{arXiv preprint arXiv:2101.00027}, 2020.

\bibitem[Goertzel(2014)]{goertzel2014artificial}
Ben Goertzel.
\newblock Artificial general intelligence: concept, state of the art, and
  future prospects.
\newblock \emph{Journal of Artificial General Intelligence}, 5\penalty0
  (1):\penalty0 1--48, 2014.

\bibitem[Ouyang et~al.(2022)Ouyang, Wu, Jiang, Almeida, Wainwright, Mishkin,
  Zhang, Agarwal, Slama, Ray, et~al.]{ouyang2022training}
Long Ouyang, Jeffrey Wu, Xu~Jiang, Diogo Almeida, Carroll Wainwright, Pamela
  Mishkin, Chong Zhang, Sandhini Agarwal, Katarina Slama, Alex Ray, et~al.
\newblock Training language models to follow instructions with human feedback.
\newblock \emph{Advances in neural information processing systems},
  35:\penalty0 27730--27744, 2022.

\bibitem[Huang et~al.(2023{\natexlab{a}})Huang, Ruan, Huang, Jin, Dong, Wu,
  Bensalem, Mu, Qi, Zhao, et~al.]{huang2023survey}
Xiaowei Huang, Wenjie Ruan, Wei Huang, Gaojie Jin, Yi~Dong, Changshun Wu,
  Saddek Bensalem, Ronghui Mu, Yi~Qi, Xingyu Zhao, et~al.
\newblock A survey of safety and trustworthiness of large language models
  through the lens of verification and validation.
\newblock \emph{arXiv preprint arXiv:2305.11391}, 2023{\natexlab{a}}.

\bibitem[Inan et~al.(2023)Inan, Upasani, Chi, Rungta, Iyer, Mao, Tontchev, Hu,
  Fuller, Testuggine, et~al.]{inan2023llama}
Hakan Inan, Kartikeya Upasani, Jianfeng Chi, Rashi Rungta, Krithika Iyer,
  Yuning Mao, Michael Tontchev, Qing Hu, Brian Fuller, Davide Testuggine,
  et~al.
\newblock Llama guard: Llm-based input-output safeguard for human-ai
  conversations.
\newblock \emph{arXiv preprint arXiv:2312.06674}, 2023.

\bibitem[Touvron et~al.(2023{\natexlab{a}})Touvron, Lavril, Izacard, Martinet,
  Lachaux, Lacroix, Rozi{\`e}re, Goyal, Hambro, Azhar,
  et~al.]{touvron2023llama}
Hugo Touvron, Thibaut Lavril, Gautier Izacard, Xavier Martinet, Marie-Anne
  Lachaux, Timoth{\'e}e Lacroix, Baptiste Rozi{\`e}re, Naman Goyal, Eric
  Hambro, Faisal Azhar, et~al.
\newblock Llama: Open and efficient foundation language models.
\newblock \emph{arXiv preprint arXiv:2302.13971}, 2023{\natexlab{a}}.

\bibitem[Ji et~al.(2023)Ji, Lee, Frieske, Yu, Su, Xu, Ishii, Bang, Madotto, and
  Fung]{ji2023survey}
Ziwei Ji, Nayeon Lee, Rita Frieske, Tiezheng Yu, Dan Su, Yan Xu, Etsuko Ishii,
  Ye~Jin Bang, Andrea Madotto, and Pascale Fung.
\newblock Survey of hallucination in natural language generation.
\newblock \emph{ACM Computing Surveys}, 55\penalty0 (12):\penalty0 1--38, 2023.

\bibitem[Gehman et~al.(2020)Gehman, Gururangan, Sap, Choi, and
  Smith]{gehman2020realtoxicityprompts}
Samuel Gehman, Suchin Gururangan, Maarten Sap, Yejin Choi, and Noah~A Smith.
\newblock Realtoxicityprompts: Evaluating neural toxic degeneration in language
  models.
\newblock \emph{arXiv preprint arXiv:2009.11462}, 2020.

\bibitem[Li et~al.(2023{\natexlab{a}})Li, Du, Song, Wang, and
  Wang]{li2023survey}
Yingji Li, Mengnan Du, Rui Song, Xin Wang, and Ying Wang.
\newblock A survey on fairness in large language models.
\newblock \emph{arXiv preprint arXiv:2308.10149}, 2023{\natexlab{a}}.

\bibitem[Wang et~al.(2023{\natexlab{a}})Wang, Ma, Yu, Gui, Zhang, Huang, Ma,
  Chang, Zhang, Shen, et~al.]{wang2023large}
Haoyu Wang, Guozheng Ma, Cong Yu, Ning Gui, Linrui Zhang, Zhiqi Huang, Suwei
  Ma, Yongzhe Chang, Sen Zhang, Li~Shen, et~al.
\newblock Are large language models really robust to word-level perturbations?
\newblock \emph{arXiv preprint arXiv:2309.11166}, 2023{\natexlab{a}}.

\bibitem[Liu et~al.(2021)Liu, Zhang, Brockett, Mao, Sui, Chen, and
  Dolan]{liu2021token}
Tianyu Liu, Yizhe Zhang, Chris Brockett, Yi~Mao, Zhifang Sui, Weizhu Chen, and
  Bill Dolan.
\newblock A token-level reference-free hallucination detection benchmark for
  free-form text generation.
\newblock \emph{arXiv preprint arXiv:2104.08704}, 2021.

\bibitem[Rateike et~al.(2023)Rateike, Cintas, Wamburu, Akumu, and
  Speakman]{rateike2023weakly}
Miriam Rateike, Celia Cintas, John Wamburu, Tanya Akumu, and Skyler Speakman.
\newblock Weakly supervised detection of hallucinations in llm activations.
\newblock \emph{arXiv preprint arXiv:2312.02798}, 2023.

\bibitem[News(2024)]{lawyerhallucination}
Global News.
\newblock B.c. lawyer who used fake, ai-generated cases faces law society
  probe, possible costs, 2024.
\newblock URL
  \url{https://globalnews.ca/news/10263897/fake-ai-cases-b-c-supreme-court/}.

\bibitem[Ousidhoum et~al.(2021)Ousidhoum, Zhao, Fang, Song, and
  Yeung]{ousidhoum2021probing}
Nedjma Ousidhoum, Xinran Zhao, Tianqing Fang, Yangqiu Song, and Dit-Yan Yeung.
\newblock Probing toxic content in large pre-trained language models.
\newblock In \emph{Proceedings of the 59th Annual Meeting of the Association
  for Computational Linguistics and the 11th International Joint Conference on
  Natural Language Processing (Volume 1: Long Papers)}, pages 4262--4274, 2021.

\bibitem[Pei et~al.(2017)Pei, Cao, Yang, and Jana]{pei2017deepxplore}
Kexin Pei, Yinzhi Cao, Junfeng Yang, and Suman Jana.
\newblock Deepxplore: Automated whitebox testing of deep learning systems.
\newblock In \emph{proceedings of the 26th Symposium on Operating Systems
  Principles}, pages 1--18, 2017.

\bibitem[Ma et~al.(2018)Ma, Juefei-Xu, Zhang, Sun, Xue, Li, Chen, Su, Li, Liu,
  et~al.]{ma2018deepgauge}
Lei Ma, Felix Juefei-Xu, Fuyuan Zhang, Jiyuan Sun, Minhui Xue, Bo~Li, Chunyang
  Chen, Ting Su, Li~Li, Yang Liu, et~al.
\newblock Deepgauge: Multi-granularity testing criteria for deep learning
  systems.
\newblock In \emph{Proceedings of the 33rd ACM/IEEE international conference on
  automated software engineering}, pages 120--131, 2018.

\bibitem[Kim et~al.(2019)Kim, Feldt, and Yoo]{kim2019guiding}
Jinhan Kim, Robert Feldt, and Shin Yoo.
\newblock Guiding deep learning system testing using surprise adequacy.
\newblock In \emph{2019 IEEE/ACM 41st International Conference on Software
  Engineering (ICSE)}, pages 1039--1049. IEEE, 2019.

\bibitem[Li et~al.(2023{\natexlab{b}})Li, Qi, Liu, Di, Liu, Pei, Yi, and
  Zhou]{li2023trustworthy}
Bo~Li, Peng Qi, Bo~Liu, Shuai Di, Jingen Liu, Jiquan Pei, Jinfeng Yi, and Bowen
  Zhou.
\newblock Trustworthy ai: From principles to practices.
\newblock \emph{ACM Computing Surveys}, 55\penalty0 (9):\penalty0 1--46,
  2023{\natexlab{b}}.

\bibitem[Riccio et~al.(2020)Riccio, Jahangirova, Stocco, Humbatova, Weiss, and
  Tonella]{riccio2020testing}
Vincenzo Riccio, Gunel Jahangirova, Andrea Stocco, Nargiz Humbatova, Michael
  Weiss, and Paolo Tonella.
\newblock Testing machine learning based systems: a systematic mapping.
\newblock \emph{Empirical Software Engineering}, 25:\penalty0 5193--5254, 2020.

\bibitem[Katz et~al.(2017)Katz, Barrett, Dill, Julian, and
  Kochenderfer]{katz2017reluplex}
Guy Katz, Clark Barrett, David~L Dill, Kyle Julian, and Mykel~J Kochenderfer.
\newblock Reluplex: An efficient smt solver for verifying deep neural networks.
\newblock In \emph{Computer Aided Verification: 29th International Conference,
  CAV 2017, Heidelberg, Germany, July 24-28, 2017, Proceedings, Part I 30},
  pages 97--117. Springer, 2017.

\bibitem[Gehr et~al.(2018)Gehr, Mirman, Drachsler-Cohen, Tsankov, Chaudhuri,
  and Vechev]{gehr2018ai2}
Timon Gehr, Matthew Mirman, Dana Drachsler-Cohen, Petar Tsankov, Swarat
  Chaudhuri, and Martin Vechev.
\newblock Ai2: Safety and robustness certification of neural networks with
  abstract interpretation.
\newblock In \emph{2018 IEEE symposium on security and privacy (SP)}, pages
  3--18. IEEE, 2018.

\bibitem[Lwakatare et~al.(2020)Lwakatare, Raj, Crnkovic, Bosch, and
  Olsson]{lwakatare2020large}
Lucy~Ellen Lwakatare, Aiswarya Raj, Ivica Crnkovic, Jan Bosch, and
  Helena~Holmstr{\"o}m Olsson.
\newblock Large-scale machine learning systems in real-world industrial
  settings: A review of challenges and solutions.
\newblock \emph{Information and software technology}, 127:\penalty0 106368,
  2020.

\bibitem[Gao et~al.(2022)Gao, Feng, Yin, Liu, Chen, and Xu]{gao2022adaptive}
Xinyu Gao, Yang Feng, Yining Yin, Zixi Liu, Zhenyu Chen, and Baowen Xu.
\newblock Adaptive test selection for deep neural networks.
\newblock In \emph{Proceedings of the 44th International Conference on Software
  Engineering}, pages 73--85, 2022.

\bibitem[Yang et~al.(2022{\natexlab{a}})Yang, Shi, Asyrofi, and
  Lo]{yang2022revisiting}
Zhou Yang, Jieke Shi, Muhammad~Hilmi Asyrofi, and David Lo.
\newblock Revisiting neuron coverage metrics and quality of deep neural
  networks.
\newblock In \emph{2022 IEEE International Conference on Software Analysis,
  Evolution and Reengineering (SANER)}, pages 408--419. IEEE,
  2022{\natexlab{a}}.

\bibitem[Riccio and Tonella(2023)]{riccio2023and}
Vincenzo Riccio and Paolo Tonella.
\newblock When and why test generators for deep learning produce invalid
  inputs: an empirical study.
\newblock In \emph{2023 IEEE/ACM 45th International Conference on Software
  Engineering (ICSE)}, pages 1161--1173. IEEE, 2023.

\bibitem[Wang et~al.(2022)Wang, Qiu, Rong, Ye, Li, Li, and Zhang]{wang2022bet}
Jialai Wang, Han Qiu, Yi~Rong, Hengkai Ye, Qi~Li, Zongpeng Li, and Chao Zhang.
\newblock Bet: black-box efficient testing for convolutional neural networks.
\newblock In \emph{Proceedings of the 31st ACM SIGSOFT International Symposium
  on Software Testing and Analysis}, pages 164--175, 2022.

\bibitem[Wang et~al.(2021{\natexlab{a}})Wang, Chen, Sun, Ma, Wang, Sun, and
  Cheng]{wang2021robot}
Jingyi Wang, Jialuo Chen, Youcheng Sun, Xingjun Ma, Dongxia Wang, Jun Sun, and
  Peng Cheng.
\newblock Robot: Robustness-oriented testing for deep learning systems.
\newblock In \emph{2021 IEEE/ACM 43rd International Conference on Software
  Engineering (ICSE)}, pages 300--311. IEEE, 2021{\natexlab{a}}.

\bibitem[Wang et~al.(2021{\natexlab{b}})Wang, You, Chen, Zhang, Dong, and
  Zhang]{wang2021prioritizing}
Zan Wang, Hanmo You, Junjie Chen, Yingyi Zhang, Xuyuan Dong, and Wenbin Zhang.
\newblock Prioritizing test inputs for deep neural networks via mutation
  analysis.
\newblock In \emph{2021 IEEE/ACM 43rd International Conference on Software
  Engineering (ICSE)}, pages 397--409. IEEE, 2021{\natexlab{b}}.

\bibitem[Feng et~al.(2020)Feng, Shi, Gao, Wan, Fang, and
  Chen]{feng2020deepgini}
Yang Feng, Qingkai Shi, Xinyu Gao, Jun Wan, Chunrong Fang, and Zhenyu Chen.
\newblock Deepgini: prioritizing massive tests to enhance the robustness of
  deep neural networks.
\newblock In \emph{Proceedings of the 29th ACM SIGSOFT International Symposium
  on Software Testing and Analysis}, pages 177--188, 2020.

\bibitem[Tian et~al.(2018)Tian, Pei, Jana, and Ray]{tian2018deeptest}
Yuchi Tian, Kexin Pei, Suman Jana, and Baishakhi Ray.
\newblock Deeptest: Automated testing of deep-neural-network-driven autonomous
  cars.
\newblock In \emph{Proceedings of the 40th international conference on software
  engineering}, pages 303--314, 2018.

\bibitem[Sotoudeh and Thakur(2021)]{sotoudeh2021provable}
Matthew Sotoudeh and Aditya~V Thakur.
\newblock Provable repair of deep neural networks.
\newblock In \emph{Proceedings of the 42nd ACM SIGPLAN International Conference
  on Programming Language Design and Implementation}, pages 588--603, 2021.

\bibitem[Stocco et~al.(2020)Stocco, Weiss, Calzana, and
  Tonella]{stocco2020misbehaviour}
Andrea Stocco, Michael Weiss, Marco Calzana, and Paolo Tonella.
\newblock Misbehaviour prediction for autonomous driving systems.
\newblock In \emph{Proceedings of the ACM/IEEE 42nd international conference on
  software engineering}, pages 359--371, 2020.

\bibitem[Sun et~al.(2022)Sun, Sun, Pham, and Shi]{sun2022causality}
Bing Sun, Jun Sun, Long~H Pham, and Jie Shi.
\newblock Causality-based neural network repair.
\newblock In \emph{Proceedings of the 44th International Conference on Software
  Engineering}, pages 338--349, 2022.

\bibitem[Kim et~al.(2023{\natexlab{a}})Kim, Humbatova, Jahangirova, Tonella,
  and Yoo]{kim2023repairing}
Jinhan Kim, Nargiz Humbatova, Gunel Jahangirova, Paolo Tonella, and Shin Yoo.
\newblock Repairing dnn architecture: Are we there yet?
\newblock In \emph{2023 IEEE Conference on Software Testing, Verification and
  Validation (ICST)}, pages 234--245. IEEE, 2023{\natexlab{a}}.

\bibitem[Yang et~al.(2022{\natexlab{b}})Yang, Yamaguchi, Tran, Hoxha, Johnson,
  and Prokhorov]{yang2022neural}
Xiaodong Yang, Tom Yamaguchi, Hoang-Dung Tran, Bardh Hoxha, Taylor~T Johnson,
  and Danil Prokhorov.
\newblock Neural network repair with reachability analysis.
\newblock In \emph{International Conference on Formal Modeling and Analysis of
  Timed Systems}, pages 221--236. Springer, 2022{\natexlab{b}}.

\bibitem[Huang et~al.(2020)Huang, Kroening, Ruan, Sharp, Sun, Thamo, Wu, and
  Yi]{huang2020survey}
Xiaowei Huang, Daniel Kroening, Wenjie Ruan, James Sharp, Youcheng Sun, Emese
  Thamo, Min Wu, and Xinping Yi.
\newblock A survey of safety and trustworthiness of deep neural networks:
  Verification, testing, adversarial attack and defence, and interpretability.
\newblock \emph{Computer Science Review}, 37:\penalty0 100270, 2020.

\bibitem[Sun et~al.(2021{\natexlab{a}})Sun, Sun, Dai, and
  Zhang]{sun2021probabilistic}
Bing Sun, Jun Sun, Ting Dai, and Lijun Zhang.
\newblock Probabilistic verification of neural networks against group fairness.
\newblock In \emph{Formal Methods: 24th International Symposium, FM 2021,
  Virtual Event, November 20--26, 2021, Proceedings 24}, pages 83--102.
  Springer, 2021{\natexlab{a}}.

\bibitem[Henzinger et~al.(2019)Henzinger, Lukina, and
  Schilling]{henzinger2019outside}
Thomas~A Henzinger, Anna Lukina, and Christian Schilling.
\newblock Outside the box: Abstraction-based monitoring of neural networks.
\newblock \emph{arXiv preprint arXiv:1911.09032}, 2019.

\bibitem[Lukina et~al.(2021)Lukina, Schilling, and Henzinger]{lukina2021into}
Anna Lukina, Christian Schilling, and Thomas~A Henzinger.
\newblock Into the unknown: Active monitoring of neural networks.
\newblock In \emph{International Conference on Runtime Verification}, pages
  42--61. Springer, 2021.

\bibitem[Cheng et~al.(2019)Cheng, N{\"u}hrenberg, and
  Yasuoka]{cheng2019runtime}
Chih-Hong Cheng, Georg N{\"u}hrenberg, and Hirotoshi Yasuoka.
\newblock Runtime monitoring neuron activation patterns.
\newblock In \emph{2019 Design, Automation \& Test in Europe Conference \&
  Exhibition (DATE)}, pages 300--303. IEEE, 2019.

\bibitem[Guerin et~al.(2022{\natexlab{a}})Guerin, Ferreira, Delmas, and
  Guiochet]{guerin2022unifying}
Joris Guerin, Raul~Sena Ferreira, Kevin Delmas, and J{\'e}r{\'e}mie Guiochet.
\newblock Unifying evaluation of machine learning safety monitors.
\newblock In \emph{2022 IEEE 33rd International Symposium on Software
  Reliability Engineering (ISSRE)}, pages 414--422. IEEE, 2022{\natexlab{a}}.

\bibitem[Guerin et~al.(2022{\natexlab{b}})Guerin, Delmas, and
  Guiochet]{guerin2022evaluation}
Joris Guerin, Kevin Delmas, and J{\'e}r{\'e}mie Guiochet.
\newblock Evaluation of runtime monitoring for uav emergency landing.
\newblock In \emph{2022 International Conference on Robotics and Automation
  (ICRA)}, pages 9703--9709. IEEE, 2022{\natexlab{b}}.

\bibitem[Zolfagharian et~al.(2023)Zolfagharian, Abdellatif, Briand,
  et~al.]{zolfagharian2023smarla}
Amirhossein Zolfagharian, Manel Abdellatif, Lionel~C Briand, et~al.
\newblock Smarla: A safety monitoring approach for deep reinforcement learning
  agents.
\newblock \emph{arXiv preprint arXiv:2308.02594}, 2023.

\bibitem[Aslansefat et~al.(2020)Aslansefat, Sorokos, Whiting,
  Tavakoli~Kolagari, and Papadopoulos]{aslansefat2020safeml}
Koorosh Aslansefat, Ioannis Sorokos, Declan Whiting, Ramin Tavakoli~Kolagari,
  and Yiannis Papadopoulos.
\newblock Safeml: safety monitoring of machine learning classifiers through
  statistical difference measures.
\newblock In \emph{International Symposium on Model-Based Safety and
  Assessment}, pages 197--211. Springer, 2020.

\bibitem[Stocco et~al.(2022)Stocco, Nunes, d'Amorim, and
  Tonella]{stocco2022thirdeye}
Andrea Stocco, Paulo~J Nunes, Marcelo d'Amorim, and Paolo Tonella.
\newblock Thirdeye: Attention maps for safe autonomous driving systems.
\newblock In \emph{Proceedings of the 37th IEEE/ACM International Conference on
  Automated Software Engineering}, pages 1--12, 2022.

\bibitem[Zhao et~al.(2023)Zhao, Zhou, Li, Tang, Wang, Hou, Min, Zhang, Zhang,
  Dong, et~al.]{zhao2023survey}
Wayne~Xin Zhao, Kun Zhou, Junyi Li, Tianyi Tang, Xiaolei Wang, Yupeng Hou,
  Yingqian Min, Beichen Zhang, Junjie Zhang, Zican Dong, et~al.
\newblock A survey of large language models.
\newblock \emph{arXiv preprint arXiv:2303.18223}, 2023.

\bibitem[Markov et~al.(2023)Markov, Zhang, Agarwal, Nekoul, Lee, Adler, Jiang,
  and Weng]{markov2023holistic}
Todor Markov, Chong Zhang, Sandhini Agarwal, Florentine~Eloundou Nekoul,
  Theodore Lee, Steven Adler, Angela Jiang, and Lilian Weng.
\newblock A holistic approach to undesired content detection in the real world.
\newblock In \emph{Proceedings of the AAAI Conference on Artificial
  Intelligence}, volume~37, pages 15009--15018, 2023.

\bibitem[Zhou et~al.(2021)Zhou, Neubig, Gu, Diab, Guzm{\'a}n, Zettlemoyer, and
  Ghazvininejad]{zhou-etal-2021-detecting}
Chunting Zhou, Graham Neubig, Jiatao Gu, Mona Diab, Francisco Guzm{\'a}n, Luke
  Zettlemoyer, and Marjan Ghazvininejad.
\newblock Detecting hallucinated content in conditional neural sequence
  generation.
\newblock In \emph{Findings of the Association for Computational Linguistics:
  ACL-IJCNLP 2021}, pages 1393--1404, Online, August 2021. Association for
  Computational Linguistics.
\newblock \doi{10.18653/v1/2021.findings-acl.120}.
\newblock URL \url{https://aclanthology.org/2021.findings-acl.120}.

\bibitem[Bhatt et~al.(2023)Bhatt, Chennabasappa, Nikolaidis, Wan, Evtimov,
  Gabi, Song, Ahmad, Aschermann, Fontana, et~al.]{bhatt2023purple}
Manish Bhatt, Sahana Chennabasappa, Cyrus Nikolaidis, Shengye Wan, Ivan
  Evtimov, Dominik Gabi, Daniel Song, Faizan Ahmad, Cornelius Aschermann,
  Lorenzo Fontana, et~al.
\newblock Purple llama cyberseceval: A secure coding benchmark for language
  models.
\newblock \emph{arXiv preprint arXiv:2312.04724}, 2023.

\bibitem[Achintalwar et~al.(2024)Achintalwar, Garcia, Anaby-Tavor, Baldini,
  Berger, Bhattacharjee, Bouneffouf, Chaudhury, Chen, Chiazor,
  et~al.]{achintalwar2024detectors}
Swapnaja Achintalwar, Adriana~Alvarado Garcia, Ateret Anaby-Tavor, Ioana
  Baldini, Sara~E Berger, Bishwaranjan Bhattacharjee, Djallel Bouneffouf,
  Subhajit Chaudhury, Pin-Yu Chen, Lamogha Chiazor, et~al.
\newblock Detectors for safe and reliable llms: Implementations, uses, and
  limitations.
\newblock \emph{arXiv preprint arXiv:2403.06009}, 2024.

\bibitem[Zhang et~al.(2024)Zhang, Lu, Ma, Zhang, Li, Ke, Sun, Sha, Sui, Wang,
  et~al.]{zhang2024shieldlm}
Zhexin Zhang, Yida Lu, Jingyuan Ma, Di~Zhang, Rui Li, Pei Ke, Hao Sun, Lei Sha,
  Zhifang Sui, Hongning Wang, et~al.
\newblock Shieldlm: Empowering llms as aligned, customizable and explainable
  safety detectors.
\newblock \emph{arXiv preprint arXiv:2402.16444}, 2024.

\bibitem[Manakul et~al.(2023)Manakul, Liusie, and
  Gales]{manakul2023selfcheckgpt}
Potsawee Manakul, Adian Liusie, and Mark~JF Gales.
\newblock Selfcheckgpt: Zero-resource black-box hallucination detection for
  generative large language models.
\newblock \emph{arXiv preprint arXiv:2303.08896}, 2023.

\bibitem[Lin et~al.(2023)Lin, Trivedi, and Sun]{lin2023generating}
Zhen Lin, Shubhendu Trivedi, and Jimeng Sun.
\newblock Generating with confidence: Uncertainty quantification for black-box
  large language models.
\newblock \emph{arXiv preprint arXiv:2305.19187}, 2023.

\bibitem[Xiong et~al.(2024)Xiong, Hu, Lu, LI, Fu, He, and Hooi]{xiong2024can}
Miao Xiong, Zhiyuan Hu, Xinyang Lu, YIFEI LI, Jie Fu, Junxian He, and Bryan
  Hooi.
\newblock Can {LLM}s express their uncertainty? an empirical evaluation of
  confidence elicitation in {LLM}s.
\newblock In \emph{The Twelfth International Conference on Learning
  Representations}, 2024.
\newblock URL \url{https://openreview.net/forum?id=gjeQKFxFpZ}.

\bibitem[Kuhn et~al.(2023)Kuhn, Gal, and Farquhar]{kuhn2023semantic}
Lorenz Kuhn, Yarin Gal, and Sebastian Farquhar.
\newblock Semantic uncertainty: Linguistic invariances for uncertainty
  estimation in natural language generation.
\newblock In \emph{The Eleventh International Conference on Learning
  Representations}, 2023.
\newblock URL \url{https://openreview.net/forum?id=VD-AYtP0dve}.

\bibitem[Devlin et~al.(2019)Devlin, Chang, Lee, and
  Toutanova]{devlin-etal-2019-bert}
Jacob Devlin, Ming-Wei Chang, Kenton Lee, and Kristina Toutanova.
\newblock {BERT}: Pre-training of deep bidirectional transformers for language
  understanding.
\newblock In Jill Burstein, Christy Doran, and Thamar Solorio, editors,
  \emph{Proceedings of the 2019 Conference of the North {A}merican Chapter of
  the Association for Computational Linguistics: Human Language Technologies,
  Volume 1 (Long and Short Papers)}, pages 4171--4186, Minneapolis, Minnesota,
  June 2019. Association for Computational Linguistics.
\newblock \doi{10.18653/v1/N19-1423}.
\newblock URL \url{https://aclanthology.org/N19-1423}.

\bibitem[Brown et~al.(2020)Brown, Mann, Ryder, Subbiah, Kaplan, Dhariwal,
  Neelakantan, Shyam, Sastry, Askell, et~al.]{brown2020language}
Tom Brown, Benjamin Mann, Nick Ryder, Melanie Subbiah, Jared~D Kaplan, Prafulla
  Dhariwal, Arvind Neelakantan, Pranav Shyam, Girish Sastry, Amanda Askell,
  et~al.
\newblock Language models are few-shot learners.
\newblock \emph{Advances in neural information processing systems},
  33:\penalty0 1877--1901, 2020.

\bibitem[Touvron et~al.(2023{\natexlab{b}})Touvron, Martin, Stone, Albert,
  Almahairi, Babaei, Bashlykov, Batra, Bhargava, Bhosale,
  et~al.]{touvron2023llama2}
Hugo Touvron, Louis Martin, Kevin Stone, Peter Albert, Amjad Almahairi, Yasmine
  Babaei, Nikolay Bashlykov, Soumya Batra, Prajjwal Bhargava, Shruti Bhosale,
  et~al.
\newblock Llama 2: Open foundation and fine-tuned chat models.
\newblock \emph{arXiv preprint arXiv:2307.09288}, 2023{\natexlab{b}}.

\bibitem[Vaswani et~al.(2017)Vaswani, Shazeer, Parmar, Uszkoreit, Jones, Gomez,
  Kaiser, and Polosukhin]{vaswani2017attention}
Ashish Vaswani, Noam Shazeer, Niki Parmar, Jakob Uszkoreit, Llion Jones,
  Aidan~N Gomez, {\L}ukasz Kaiser, and Illia Polosukhin.
\newblock Attention is all you need.
\newblock \emph{Advances in neural information processing systems}, 30, 2017.

\bibitem[Kaplan et~al.(2020)Kaplan, McCandlish, Henighan, Brown, Chess, Child,
  Gray, Radford, Wu, and Amodei]{kaplan2020scaling}
Jared Kaplan, Sam McCandlish, Tom Henighan, Tom~B Brown, Benjamin Chess, Rewon
  Child, Scott Gray, Alec Radford, Jeffrey Wu, and Dario Amodei.
\newblock Scaling laws for neural language models.
\newblock \emph{arXiv preprint arXiv:2001.08361}, 2020.

\bibitem[Dosovitskiy et~al.(2021)Dosovitskiy, Beyer, Kolesnikov, Weissenborn,
  Zhai, Unterthiner, Dehghani, Minderer, Heigold, Gelly, Uszkoreit, and
  Houlsby]{dosovitskiy2021an}
Alexey Dosovitskiy, Lucas Beyer, Alexander Kolesnikov, Dirk Weissenborn,
  Xiaohua Zhai, Thomas Unterthiner, Mostafa Dehghani, Matthias Minderer, Georg
  Heigold, Sylvain Gelly, Jakob Uszkoreit, and Neil Houlsby.
\newblock An image is worth 16x16 words: Transformers for image recognition at
  scale.
\newblock In \emph{International Conference on Learning Representations}, 2021.
\newblock URL \url{https://openreview.net/forum?id=YicbFdNTTy}.

\bibitem[Ba et~al.(2016)Ba, Kiros, and Hinton]{ba2016layer}
Jimmy~Lei Ba, Jamie~Ryan Kiros, and Geoffrey~E Hinton.
\newblock Layer normalization.
\newblock \emph{arXiv preprint arXiv:1607.06450}, 2016.

\bibitem[He et~al.(2016)He, Zhang, Ren, and Sun]{he2016deep}
Kaiming He, Xiangyu Zhang, Shaoqing Ren, and Jian Sun.
\newblock Deep residual learning for image recognition.
\newblock In \emph{Proceedings of the IEEE conference on computer vision and
  pattern recognition}, pages 770--778, 2016.

\bibitem[Germann(2003)]{germann2003greedy}
Ulrich Germann.
\newblock Greedy decoding for statistical machine translation in almost linear
  time.
\newblock In \emph{Proceedings of the 2003 Human Language Technology Conference
  of the North American Chapter of the Association for Computational
  Linguistics}, pages 72--79, 2003.

\bibitem[Sun et~al.(2024{\natexlab{a}})Sun, Huang, Wang, Wu, Zhang, Gao, Huang,
  Lyu, Zhang, Li, et~al.]{sun2024trustllm}
Lichao Sun, Yue Huang, Haoran Wang, Siyuan Wu, Qihui Zhang, Chujie Gao, Yixin
  Huang, Wenhan Lyu, Yixuan Zhang, Xiner Li, et~al.
\newblock Trustllm: Trustworthiness in large language models.
\newblock \emph{arXiv preprint arXiv:2401.05561}, 2024{\natexlab{a}}.

\bibitem[Wang et~al.(2023{\natexlab{b}})Wang, Chen, Pei, Xie, Kang, Zhang, Xu,
  Xiong, Dutta, Schaeffer, Truong, Arora, Mazeika, Hendrycks, Lin, Cheng,
  Koyejo, Song, and Li]{wang2023decodingtrust}
Boxin Wang, Weixin Chen, Hengzhi Pei, Chulin Xie, Mintong Kang, Chenhui Zhang,
  Chejian Xu, Zidi Xiong, Ritik Dutta, Rylan Schaeffer, Sang~T. Truong, Simran
  Arora, Mantas Mazeika, Dan Hendrycks, Zinan Lin, Yu~Cheng, Sanmi Koyejo, Dawn
  Song, and Bo~Li.
\newblock Decodingtrust: A comprehensive assessment of trustworthiness in {GPT}
  models.
\newblock In \emph{Thirty-seventh Conference on Neural Information Processing
  Systems Datasets and Benchmarks Track}, 2023{\natexlab{b}}.
\newblock URL \url{https://openreview.net/forum?id=kaHpo8OZw2}.

\bibitem[Lin et~al.(2021)Lin, Hilton, and Evans]{lin2021truthfulqa}
Stephanie Lin, Jacob Hilton, and Owain Evans.
\newblock Truthfulqa: Measuring how models mimic human falsehoods.
\newblock \emph{arXiv preprint arXiv:2109.07958}, 2021.

\bibitem[Yao et~al.(2024)Yao, Duan, Xu, Cai, Sun, and Zhang]{yao2024survey}
Yifan Yao, Jinhao Duan, Kaidi Xu, Yuanfang Cai, Zhibo Sun, and Yue Zhang.
\newblock A survey on large language model (llm) security and privacy: The
  good, the bad, and the ugly.
\newblock \emph{High-Confidence Computing}, page 100211, 2024.

\bibitem[Papineni et~al.(2002{\natexlab{a}})Papineni, Roukos, Ward, and
  Zhu]{papineni-etal-2002-bleu}
Kishore Papineni, Salim Roukos, Todd Ward, and Wei-Jing Zhu.
\newblock {B}leu: a method for automatic evaluation of machine translation.
\newblock In Pierre Isabelle, Eugene Charniak, and Dekang Lin, editors,
  \emph{Proceedings of the 40th Annual Meeting of the Association for
  Computational Linguistics}, pages 311--318, Philadelphia, Pennsylvania, USA,
  July 2002{\natexlab{a}}. Association for Computational Linguistics.
\newblock \doi{10.3115/1073083.1073135}.
\newblock URL \url{https://aclanthology.org/P02-1040}.

\bibitem[Hendrycks et~al.(2020)Hendrycks, Burns, Basart, Critch, Li, Song, and
  Steinhardt]{hendrycks2020aligning}
Dan Hendrycks, Collin Burns, Steven Basart, Andrew Critch, Jerry Li, Dawn Song,
  and Jacob Steinhardt.
\newblock Aligning ai with shared human values.
\newblock \emph{arXiv preprint arXiv:2008.02275}, 2020.

\bibitem[Bolukbasi et~al.(2016)Bolukbasi, Chang, Zou, Saligrama, and
  Kalai]{bolukbasi2016man}
Tolga Bolukbasi, Kai-Wei Chang, James~Y Zou, Venkatesh Saligrama, and Adam~T
  Kalai.
\newblock Man is to computer programmer as woman is to homemaker? debiasing
  word embeddings.
\newblock \emph{Advances in neural information processing systems}, 29, 2016.

\bibitem[Azaria and Mitchell(2023)]{azaria2023internal}
Amos Azaria and Tom Mitchell.
\newblock The internal state of an llm knows when it's lying, 2023.

\bibitem[Liu et~al.(2023{\natexlab{a}})Liu, Casper, Hadfield-Menell, and
  Andreas]{liu2023cognitive}
Kevin Liu, Stephen Casper, Dylan Hadfield-Menell, and Jacob Andreas.
\newblock Cognitive dissonance: Why do language model outputs disagree with
  internal representations of truthfulness?
\newblock \emph{arXiv preprint arXiv:2312.03729}, 2023{\natexlab{a}}.

\bibitem[Welbl et~al.(2021)Welbl, Glaese, Uesato, Dathathri, Mellor, Hendricks,
  Anderson, Kohli, Coppin, and Huang]{welbl2021challenges}
Johannes Welbl, Amelia Glaese, Jonathan Uesato, Sumanth Dathathri, John Mellor,
  Lisa~Anne Hendricks, Kirsty Anderson, Pushmeet Kohli, Ben Coppin, and Po-Sen
  Huang.
\newblock Challenges in detoxifying language models.
\newblock \emph{arXiv preprint arXiv:2109.07445}, 2021.

\bibitem[Madiega(2021)]{madiega2021artificial}
Tambiama Madiega.
\newblock Artificial intelligence act.
\newblock \emph{European Parliament: European Parliamentary Research Service},
  2021.

\bibitem[Zhang et~al.(2020{\natexlab{a}})Zhang, Harman, Ma, and
  Liu]{zhang2020machine}
Jie~M Zhang, Mark Harman, Lei Ma, and Yang Liu.
\newblock Machine learning testing: Survey, landscapes and horizons.
\newblock \emph{IEEE Transactions on Software Engineering}, 48\penalty0
  (1):\penalty0 1--36, 2020{\natexlab{a}}.

\bibitem[Braiek and Khomh(2020)]{braiek2020testing}
Houssem~Ben Braiek and Foutse Khomh.
\newblock On testing machine learning programs.
\newblock \emph{Journal of Systems and Software}, 164:\penalty0 110542, 2020.

\bibitem[Sun et~al.(2018)Sun, Huang, Kroening, Sharp, Hill, and
  Ashmore]{sun2018testing}
Youcheng Sun, Xiaowei Huang, Daniel Kroening, James Sharp, Matthew Hill, and
  Rob Ashmore.
\newblock Testing deep neural networks.
\newblock \emph{arXiv preprint arXiv:1803.04792}, 2018.

\bibitem[Wang et~al.(2019)Wang, Dong, Sun, Wang, and
  Zhang]{wang2019adversarial}
Jingyi Wang, Guoliang Dong, Jun Sun, Xinyu Wang, and Peixin Zhang.
\newblock Adversarial sample detection for deep neural network through model
  mutation testing.
\newblock In \emph{2019 IEEE/ACM 41st International Conference on Software
  Engineering (ICSE)}, pages 1245--1256. IEEE, 2019.

\bibitem[Wang et~al.(2021{\natexlab{c}})Wang, Zhang, Xu, Lin, Jana, Hsieh, and
  Kolter]{wang2021beta}
Shiqi Wang, Huan Zhang, Kaidi Xu, Xue Lin, Suman Jana, Cho-Jui Hsieh, and
  J~Zico Kolter.
\newblock Beta-crown: Efficient bound propagation with per-neuron split
  constraints for neural network robustness verification.
\newblock \emph{Advances in Neural Information Processing Systems},
  34:\penalty0 29909--29921, 2021{\natexlab{c}}.

\bibitem[Tran et~al.(2020)Tran, Yang, Manzanas~Lopez, Musau, Nguyen, Xiang,
  Bak, and Johnson]{tran2020nnv}
Hoang-Dung Tran, Xiaodong Yang, Diego Manzanas~Lopez, Patrick Musau, Luan~Viet
  Nguyen, Weiming Xiang, Stanley Bak, and Taylor~T Johnson.
\newblock Nnv: the neural network verification tool for deep neural networks
  and learning-enabled cyber-physical systems.
\newblock In \emph{International Conference on Computer Aided Verification},
  pages 3--17. Springer, 2020.

\bibitem[Shriver et~al.(2021)Shriver, Elbaum, and Dwyer]{shriver2021dnnv}
David Shriver, Sebastian Elbaum, and Matthew~B Dwyer.
\newblock Dnnv: A framework for deep neural network verification.
\newblock In \emph{International Conference on Computer Aided Verification},
  pages 137--150. Springer, 2021.

\bibitem[Huang et~al.(2017)Huang, Kwiatkowska, Wang, and Wu]{huang2017safety}
Xiaowei Huang, Marta Kwiatkowska, Sen Wang, and Min Wu.
\newblock Safety verification of deep neural networks.
\newblock In \emph{Computer Aided Verification: 29th International Conference,
  CAV 2017, Heidelberg, Germany, July 24-28, 2017, Proceedings, Part I 30},
  pages 3--29. Springer, 2017.

\bibitem[Wang et~al.(2018)Wang, Pei, Whitehouse, Yang, and
  Jana]{wang2018efficient}
Shiqi Wang, Kexin Pei, Justin Whitehouse, Junfeng Yang, and Suman Jana.
\newblock Efficient formal safety analysis of neural networks.
\newblock \emph{Advances in neural information processing systems}, 31, 2018.

\bibitem[Bommasani et~al.(2021)Bommasani, Hudson, Adeli, Altman, Arora, von
  Arx, Bernstein, Bohg, Bosselut, Brunskill,
  et~al.]{bommasani2021opportunities}
Rishi Bommasani, Drew~A Hudson, Ehsan Adeli, Russ Altman, Simran Arora, Sydney
  von Arx, Michael~S Bernstein, Jeannette Bohg, Antoine Bosselut, Emma
  Brunskill, et~al.
\newblock On the opportunities and risks of foundation models.
\newblock \emph{arXiv preprint arXiv:2108.07258}, 2021.

\bibitem[Schwarting et~al.(2018)Schwarting, Alonso-Mora, and
  Rus]{schwarting2018planning}
Wilko Schwarting, Javier Alonso-Mora, and Daniela Rus.
\newblock Planning and decision-making for autonomous vehicles.
\newblock \emph{Annual Review of Control, Robotics, and Autonomous Systems},
  1:\penalty0 187--210, 2018.

\bibitem[Wang et~al.(2023{\natexlab{c}})Wang, Zhao, Ouyang, Wang, and
  Shen]{wang2023chatcad}
Sheng Wang, Zihao Zhao, Xi~Ouyang, Qian Wang, and Dinggang Shen.
\newblock Chatcad: Interactive computer-aided diagnosis on medical image using
  large language models.
\newblock \emph{arXiv preprint arXiv:2302.07257}, 2023{\natexlab{c}}.

\bibitem[Bartocci et~al.(2018)Bartocci, Deshmukh, Donz{\'e}, Fainekos, Maler,
  Ni{\v{c}}kovi{\'c}, and Sankaranarayanan]{bartocci2018specification}
Ezio Bartocci, Jyotirmoy Deshmukh, Alexandre Donz{\'e}, Georgios Fainekos, Oded
  Maler, Dejan Ni{\v{c}}kovi{\'c}, and Sriram Sankaranarayanan.
\newblock Specification-based monitoring of cyber-physical systems: a survey on
  theory, tools and applications.
\newblock \emph{Lectures on Runtime Verification: Introductory and Advanced
  Topics}, pages 135--175, 2018.

\bibitem[S{\'a}nchez et~al.(2019)S{\'a}nchez, Schneider, Ahrendt, Bartocci,
  Bianculli, Colombo, Falcone, Francalanza, Krsti{\'c}, Louren{\c{c}}o,
  et~al.]{sanchez2019survey}
C{\'e}sar S{\'a}nchez, Gerardo Schneider, Wolfgang Ahrendt, Ezio Bartocci,
  Domenico Bianculli, Christian Colombo, Yli{\`e}s Falcone, Adrian Francalanza,
  Sran Krsti{\'c}, Joao~M Louren{\c{c}}o, et~al.
\newblock A survey of challenges for runtime verification from advanced
  application domains (beyond software).
\newblock \emph{Formal Methods in System Design}, 54:\penalty0 279--335, 2019.

\bibitem[Ratasich et~al.(2019)Ratasich, Khalid, Geissler, Grosu, Shafique, and
  Bartocci]{ratasich2019roadmap}
Denise Ratasich, Faiq Khalid, Florian Geissler, Radu Grosu, Muhammad Shafique,
  and Ezio Bartocci.
\newblock A roadmap toward the resilient internet of things for cyber-physical
  systems.
\newblock \emph{IEEE Access}, 7:\penalty0 13260--13283, 2019.

\bibitem[Falcone et~al.(2021)Falcone, Krsti{\'c}, Reger, and
  Traytel]{falcone2021taxonomy}
Yli{\`e}s Falcone, Sran Krsti{\'c}, Giles Reger, and Dmitriy Traytel.
\newblock A taxonomy for classifying runtime verification tools.
\newblock \emph{International Journal on Software Tools for Technology
  Transfer}, 23\penalty0 (2):\penalty0 255--284, 2021.

\bibitem[Gal and Ghahramani(2016)]{gal2016dropout}
Yarin Gal and Zoubin Ghahramani.
\newblock Dropout as a bayesian approximation: Representing model uncertainty
  in deep learning.
\newblock In \emph{international conference on machine learning}, pages
  1050--1059. PMLR, 2016.

\bibitem[Salay et~al.(2019)Salay, Angus, and Czarnecki]{salay2019safety}
Rick Salay, Matt Angus, and Krzysztof Czarnecki.
\newblock A safety analysis method for perceptual components in automated
  driving.
\newblock In \emph{2019 IEEE 30th International Symposium on Software
  Reliability Engineering (ISSRE)}, pages 24--34. IEEE, 2019.

\bibitem[Huang et~al.(2023{\natexlab{b}})Huang, Song, Wang, Chen, and
  Ma]{huang2023look}
Yuheng Huang, Jiayang Song, Zhijie Wang, Huaming Chen, and Lei Ma.
\newblock Look before you leap: An exploratory study of uncertainty measurement
  for large language models.
\newblock \emph{arXiv preprint arXiv:2307.10236}, 2023{\natexlab{b}}.

\bibitem[Austin et~al.(2021)Austin, Odena, Nye, Bosma, Michalewski, Dohan,
  Jiang, Cai, Terry, Le, et~al.]{austin2021program}
Jacob Austin, Augustus Odena, Maxwell Nye, Maarten Bosma, Henryk Michalewski,
  David Dohan, Ellen Jiang, Carrie Cai, Michael Terry, Quoc Le, et~al.
\newblock Program synthesis with large language models.
\newblock \emph{arXiv preprint arXiv:2108.07732}, 2021.

\bibitem[Google(2024)]{perspectiveapi}
Google.
\newblock Perspective api, 2024.
\newblock URL
  \url{https://support.perspectiveapi.com/s/about-the-api?language=en_US}.

\bibitem[Krishna and Murty(1999)]{krishna1999genetic}
K~Krishna and M~Narasimha Murty.
\newblock Genetic k-means algorithm.
\newblock \emph{IEEE Transactions on Systems, Man, and Cybernetics, Part B
  (Cybernetics)}, 29\penalty0 (3):\penalty0 433--439, 1999.

\bibitem[Taori et~al.(2023{\natexlab{a}})Taori, Gulrajani, Zhang, Dubois, Li,
  Guestrin, Liang, and Hashimoto]{alpaca}
Rohan Taori, Ishaan Gulrajani, Tianyi Zhang, Yann Dubois, Xuechen Li, Carlos
  Guestrin, Percy Liang, and Tatsunori~B. Hashimoto.
\newblock Stanford alpaca: An instruction-following llama model.
\newblock \url{https://github.com/tatsu-lab/stanford_alpaca},
  2023{\natexlab{a}}.

\bibitem[Chiang et~al.(2023)Chiang, Li, Lin, Sheng, Wu, Zhang, Zheng, Zhuang,
  Zhuang, Gonzalez, Stoica, and Xing]{vicuna2023}
Wei-Lin Chiang, Zhuohan Li, Zi~Lin, Ying Sheng, Zhanghao Wu, Hao Zhang, Lianmin
  Zheng, Siyuan Zhuang, Yonghao Zhuang, Joseph~E. Gonzalez, Ion Stoica, and
  Eric~P. Xing.
\newblock Vicuna: An open-source chatbot impressing gpt-4 with 90\%* chatgpt
  quality, March 2023.
\newblock URL \url{https://lmsys.org/blog/2023-03-30-vicuna/}.

\bibitem[ShareGPT(2024)]{sharegpt}
ShareGPT.
\newblock Sharegpt, 2024.
\newblock URL \url{https://sharegpt.com/}.

\bibitem[Roziere et~al.(2023)Roziere, Gehring, Gloeckle, Sootla, Gat, Tan, Adi,
  Liu, Remez, Rapin, et~al.]{roziere2023code}
Baptiste Roziere, Jonas Gehring, Fabian Gloeckle, Sten Sootla, Itai Gat,
  Xiaoqing~Ellen Tan, Yossi Adi, Jingyu Liu, Tal Remez, J{\'e}r{\'e}my Rapin,
  et~al.
\newblock Code llama: Open foundation models for code.
\newblock \emph{arXiv preprint arXiv:2308.12950}, 2023.

\bibitem[OpenAI(2023)]{gpt35}
OpenAI.
\newblock Gpt 3.5, 2023.
\newblock URL \url{https://platform.openai.com/docs/models/gpt-3-5}.

\bibitem[Taori et~al.(2023{\natexlab{b}})Taori, Gulrajani, Zhang, Dubois, Li,
  Guestrin, Liang, and Hashimoto]{taori2023alpaca}
Rohan Taori, Ishaan Gulrajani, Tianyi Zhang, Yann Dubois, Xuechen Li, Carlos
  Guestrin, Percy Liang, and Tatsunori~B Hashimoto.
\newblock Alpaca: A strong, replicable instruction-following model.
\newblock \emph{Stanford Center for Research on Foundation Models.
  https://crfm. stanford. edu/2023/03/13/alpaca. html}, 3\penalty0
  (6):\penalty0 7, 2023{\natexlab{b}}.

\bibitem[Joshi et~al.(2017)Joshi, Choi, Weld, and
  Zettlemoyer]{joshi2017triviaqa}
Mandar Joshi, Eunsol Choi, Daniel~S Weld, and Luke Zettlemoyer.
\newblock Triviaqa: A large scale distantly supervised challenge dataset for
  reading comprehension.
\newblock \emph{arXiv preprint arXiv:1705.03551}, 2017.

\bibitem[Kwiatkowski et~al.(2019)Kwiatkowski, Palomaki, Redfield, Collins,
  Parikh, Alberti, Epstein, Polosukhin, Devlin, Lee,
  et~al.]{kwiatkowski2019natural}
Tom Kwiatkowski, Jennimaria Palomaki, Olivia Redfield, Michael Collins, Ankur
  Parikh, Chris Alberti, Danielle Epstein, Illia Polosukhin, Jacob Devlin,
  Kenton Lee, et~al.
\newblock Natural questions: a benchmark for question answering research.
\newblock \emph{Transactions of the Association for Computational Linguistics},
  7:\penalty0 453--466, 2019.

\bibitem[Bojar et~al.(2014)Bojar, Buck, Federmann, Haddow, Koehn, Leveling,
  Monz, Pecina, Post, Saint-Amand, et~al.]{bojar2014findings}
Ond{\v{r}}ej Bojar, Christian Buck, Christian Federmann, Barry Haddow, Philipp
  Koehn, Johannes Leveling, Christof Monz, Pavel Pecina, Matt Post, Herve
  Saint-Amand, et~al.
\newblock Findings of the 2014 workshop on statistical machine translation.
\newblock In \emph{Proceedings of the ninth workshop on statistical machine
  translation}, pages 12--58, 2014.

\bibitem[Papineni et~al.(2002{\natexlab{b}})Papineni, Roukos, Ward, and
  Zhu]{papineni2002bleu}
Kishore Papineni, Salim Roukos, Todd Ward, and Wei-Jing Zhu.
\newblock Bleu: a method for automatic evaluation of machine translation.
\newblock In \emph{Proceedings of the 40th annual meeting of the Association
  for Computational Linguistics}, pages 311--318, 2002{\natexlab{b}}.

\bibitem[Chen et~al.(2021{\natexlab{a}})Chen, Tworek, Jun, Yuan, Pinto, Kaplan,
  Edwards, Burda, Joseph, Brockman, et~al.]{chen2021evaluating}
Mark Chen, Jerry Tworek, Heewoo Jun, Qiming Yuan, Henrique Ponde de~Oliveira
  Pinto, Jared Kaplan, Harri Edwards, Yuri Burda, Nicholas Joseph, Greg
  Brockman, et~al.
\newblock Evaluating large language models trained on code.
\newblock \emph{arXiv preprint arXiv:2107.03374}, 2021{\natexlab{a}}.

\bibitem[Gu{\'e}rin et~al.(2023)Gu{\'e}rin, Delmas, Ferreira, and
  Guiochet]{guerin2023out}
Joris Gu{\'e}rin, Kevin Delmas, Raul Ferreira, and J{\'e}r{\'e}mie Guiochet.
\newblock Out-of-distribution detection is not all you need.
\newblock In \emph{Proceedings of the AAAI conference on artificial
  intelligence}, volume~37, pages 14829--14837, 2023.

\bibitem[Fawcett(2006)]{fawcett2006introduction}
Tom Fawcett.
\newblock An introduction to roc analysis.
\newblock \emph{Pattern recognition letters}, 27\penalty0 (8):\penalty0
  861--874, 2006.

\bibitem[Xie et~al.(2023)Xie, Song, Zhou, Zhang, and Ma]{xie2023mosaic}
Xuan Xie, Jiayang Song, Zhehua Zhou, Fuyuan Zhang, and Lei Ma.
\newblock Mosaic: Model-based safety analysis framework for ai-enabled
  cyber-physical systems.
\newblock \emph{arXiv preprint arXiv:2305.03882}, 2023.

\bibitem[Zhang et~al.(2022)Zhang, Arcaini, and Xie]{zhang2022online}
Zhenya Zhang, Paolo Arcaini, and Xuan Xie.
\newblock Online reset for signal temporal logic monitoring.
\newblock \emph{IEEE Transactions on Computer-Aided Design of Integrated
  Circuits and Systems}, 41\penalty0 (11):\penalty0 4421--4432, 2022.

\bibitem[Deshmukh et~al.(2017)Deshmukh, Donz{\'e}, Ghosh, Jin, Juniwal, and
  Seshia]{deshmukh2017robust}
Jyotirmoy~V Deshmukh, Alexandre Donz{\'e}, Shromona Ghosh, Xiaoqing Jin, Garvit
  Juniwal, and Sanjit~A Seshia.
\newblock Robust online monitoring of signal temporal logic.
\newblock \emph{Formal Methods in System Design}, 51:\penalty0 5--30, 2017.

\bibitem[Ganaie et~al.(2022)Ganaie, Hu, Malik, Tanveer, and
  Suganthan]{ganaie2022ensemble}
Mudasir~A Ganaie, Minghui Hu, Ashwani~Kumar Malik, Muhammad Tanveer, and
  Ponnuthurai~N Suganthan.
\newblock Ensemble deep learning: A review.
\newblock \emph{Engineering Applications of Artificial Intelligence},
  115:\penalty0 105151, 2022.

\bibitem[Dietterich et~al.(2002)]{dietterich2002ensemble}
Thomas~G Dietterich et~al.
\newblock Ensemble learning.
\newblock \emph{The handbook of brain theory and neural networks}, 2\penalty0
  (1):\penalty0 110--125, 2002.

\bibitem[Sagi and Rokach(2018)]{sagi2018ensemble}
Omer Sagi and Lior Rokach.
\newblock Ensemble learning: A survey.
\newblock \emph{Wiley interdisciplinary reviews: data mining and knowledge
  discovery}, 8\penalty0 (4):\penalty0 e1249, 2018.

\bibitem[Wu et~al.(2024)Wu, Lu, Arrieta, Yue, and Ali]{wu2024reality}
Jiahui Wu, Chengjie Lu, Aitor Arrieta, Tao Yue, and Shaukat Ali.
\newblock Reality bites: Assessing the realism of driving scenarios with large
  language models.
\newblock \emph{arXiv preprint arXiv:2403.09906}, 2024.

\bibitem[Song et~al.(2023)Song, Zhou, Liu, Fang, Shu, and Ma]{song2023self}
Jiayang Song, Zhehua Zhou, Jiawei Liu, Chunrong Fang, Zhan Shu, and Lei Ma.
\newblock Self-refined large language model as automated reward function
  designer for deep reinforcement learning in robotics.
\newblock \emph{arXiv preprint arXiv:2309.06687}, 2023.

\bibitem[Serban et~al.(2020)Serban, van~der Blom, Hoos, and
  Visser]{serban2020adoption}
Alex Serban, Koen van~der Blom, Holger Hoos, and Joost Visser.
\newblock Adoption and effects of software engineering best practices in
  machine learning.
\newblock In \emph{Proceedings of the 14th ACM/IEEE International Symposium on
  Empirical Software Engineering and Measurement (ESEM)}, pages 1--12, 2020.

\bibitem[Amershi et~al.(2019)Amershi, Begel, Bird, DeLine, Gall, Kamar,
  Nagappan, Nushi, and Zimmermann]{amershi2019software}
Saleema Amershi, Andrew Begel, Christian Bird, Robert DeLine, Harald Gall, Ece
  Kamar, Nachiappan Nagappan, Besmira Nushi, and Thomas Zimmermann.
\newblock Software engineering for machine learning: A case study.
\newblock In \emph{2019 IEEE/ACM 41st International Conference on Software
  Engineering: Software Engineering in Practice (ICSE-SEIP)}, pages 291--300.
  IEEE, 2019.

\bibitem[Liu et~al.(2019)Liu, Ma, and Zhao]{liu2019secure}
Yang Liu, Lei Ma, and Jianjun Zhao.
\newblock Secure deep learning engineering: A road towards quality assurance of
  intelligent systems.
\newblock In \emph{Formal Methods and Software Engineering: 21st International
  Conference on Formal Engineering Methods, ICFEM 2019, Shenzhen, China,
  November 5--9, 2019, Proceedings 21}, pages 3--15. Springer, 2019.

\bibitem[Cortes et~al.(2016)Cortes, DeSalvo, and Mohri]{cortes2016learning}
Corinna Cortes, Giulia DeSalvo, and Mehryar Mohri.
\newblock Learning with rejection.
\newblock In \emph{Algorithmic Learning Theory: 27th International Conference,
  ALT 2016, Bari, Italy, October 19-21, 2016, Proceedings 27}, pages 67--82.
  Springer, 2016.

\bibitem[Geifman and El-Yaniv(2019)]{geifman2019selectivenet}
Yonatan Geifman and Ran El-Yaniv.
\newblock Selectivenet: A deep neural network with an integrated reject option.
\newblock In \emph{International conference on machine learning}, pages
  2151--2159. PMLR, 2019.

\bibitem[Geifman and El-Yaniv(2017)]{geifman2017selective}
Yonatan Geifman and Ran El-Yaniv.
\newblock Selective classification for deep neural networks.
\newblock \emph{Advances in neural information processing systems}, 30, 2017.

\bibitem[Fisch et~al.(2022)Fisch, Jaakkola, and Barzilay]{fisch2022calibrated}
Adam Fisch, Tommi~S. Jaakkola, and Regina Barzilay.
\newblock Calibrated selective classification.
\newblock \emph{Transactions on Machine Learning Research}, 2022.
\newblock ISSN 2835-8856.
\newblock URL \url{https://openreview.net/forum?id=zFhNBs8GaV}.

\bibitem[Xia and Bouganis(2022)]{xia2022augmenting}
Guoxuan Xia and Christos-Savvas Bouganis.
\newblock Augmenting softmax information for selective classification with
  out-of-distribution data.
\newblock In \emph{Proceedings of the Asian Conference on Computer Vision},
  pages 1995--2012, 2022.

\bibitem[Hendrycks and Gimpel(2017)]{hendrycks2017a}
Dan Hendrycks and Kevin Gimpel.
\newblock A baseline for detecting misclassified and out-of-distribution
  examples in neural networks.
\newblock In \emph{International Conference on Learning Representations}, 2017.
\newblock URL \url{https://openreview.net/forum?id=Hkg4TI9xl}.

\bibitem[Lee et~al.(2018)Lee, Lee, Lee, and Shin]{lee2018simple}
Kimin Lee, Kibok Lee, Honglak Lee, and Jinwoo Shin.
\newblock A simple unified framework for detecting out-of-distribution samples
  and adversarial attacks.
\newblock \emph{Advances in neural information processing systems}, 31, 2018.

\bibitem[Liang et~al.(2018)Liang, Li, and Srikant]{liang2018enhancing}
Shiyu Liang, Yixuan Li, and R.~Srikant.
\newblock Enhancing the reliability of out-of-distribution image detection in
  neural networks.
\newblock In \emph{International Conference on Learning Representations}, 2018.
\newblock URL \url{https://openreview.net/forum?id=H1VGkIxRZ}.

\bibitem[Hsu et~al.(2020)Hsu, Shen, Jin, and Kira]{hsu2020generalized}
Yen-Chang Hsu, Yilin Shen, Hongxia Jin, and Zsolt Kira.
\newblock Generalized odin: Detecting out-of-distribution image without
  learning from out-of-distribution data.
\newblock In \emph{Proceedings of the IEEE/CVF conference on computer vision
  and pattern recognition}, pages 10951--10960, 2020.

\bibitem[Liu et~al.(2020)Liu, Wang, Owens, and Li]{liu2020energy}
Weitang Liu, Xiaoyun Wang, John Owens, and Yixuan Li.
\newblock Energy-based out-of-distribution detection.
\newblock \emph{Advances in neural information processing systems},
  33:\penalty0 21464--21475, 2020.

\bibitem[Berend et~al.(2020)Berend, Xie, Ma, Zhou, Liu, Xu, and
  Zhao]{berend2020cats}
David Berend, Xiaofei Xie, Lei Ma, Lingjun Zhou, Yang Liu, Chi Xu, and Jianjun
  Zhao.
\newblock Cats are not fish: Deep learning testing calls for
  out-of-distribution awareness.
\newblock In \emph{Proceedings of the 35th IEEE/ACM international conference on
  automated software engineering}, pages 1041--1052, 2020.

\bibitem[Sun et~al.(2021{\natexlab{b}})Sun, Guo, and Li]{sun2021react}
Yiyou Sun, Chuan Guo, and Yixuan Li.
\newblock React: Out-of-distribution detection with rectified activations.
\newblock \emph{Advances in Neural Information Processing Systems},
  34:\penalty0 144--157, 2021{\natexlab{b}}.

\bibitem[Ahn et~al.(2023)Ahn, Park, and Kim]{ahn2023line}
Yong~Hyun Ahn, Gyeong-Moon Park, and Seong~Tae Kim.
\newblock Line: Out-of-distribution detection by leveraging important neurons.
\newblock In \emph{2023 IEEE/CVF Conference on Computer Vision and Pattern
  Recognition (CVPR)}, pages 19852--19862. IEEE, 2023.

\bibitem[Sabokrou et~al.(2018)Sabokrou, Khalooei, Fathy, and
  Adeli]{sabokrou2018adversarially}
Mohammad Sabokrou, Mohammad Khalooei, Mahmood Fathy, and Ehsan Adeli.
\newblock Adversarially learned one-class classifier for novelty detection.
\newblock In \emph{Proceedings of the IEEE conference on computer vision and
  pattern recognition}, pages 3379--3388, 2018.

\bibitem[Hendrycks et~al.(2019)Hendrycks, Mazeika, and
  Dietterich]{hendrycks2018deep}
Dan Hendrycks, Mantas Mazeika, and Thomas Dietterich.
\newblock Deep anomaly detection with outlier exposure.
\newblock In \emph{International Conference on Learning Representations}, 2019.
\newblock URL \url{https://openreview.net/forum?id=HyxCxhRcY7}.

\bibitem[Zhang et~al.(2020{\natexlab{b}})Zhang, Xie, Ma, Du, Hu, Liu, Zhao, and
  Sun]{zhang2020towards}
Xiyue Zhang, Xiaofei Xie, Lei Ma, Xiaoning Du, Qiang Hu, Yang Liu, Jianjun
  Zhao, and Meng Sun.
\newblock Towards characterizing adversarial defects of deep learning software
  from the lens of uncertainty.
\newblock In \emph{Proceedings of the ACM/IEEE 42nd International Conference on
  Software Engineering}, pages 739--751, 2020{\natexlab{b}}.

\bibitem[Granese et~al.(2021)Granese, Romanelli, Gorla, Palamidessi, and
  Piantanida]{granese2021doctor}
Federica Granese, Marco Romanelli, Daniele Gorla, Catuscia Palamidessi, and
  Pablo Piantanida.
\newblock Doctor: A simple method for detecting misclassification errors.
\newblock \emph{Advances in Neural Information Processing Systems},
  34:\penalty0 5669--5681, 2021.

\bibitem[Chen et~al.(2021{\natexlab{b}})Chen, Liu, Avci, Wu, Liang, and
  Jha]{chen2021detecting}
Jiefeng Chen, Frederick Liu, Besim Avci, Xi~Wu, Yingyu Liang, and Somesh Jha.
\newblock Detecting errors and estimating accuracy on unlabeled data with
  self-training ensembles.
\newblock \emph{Advances in Neural Information Processing Systems},
  34:\penalty0 14980--14992, 2021{\natexlab{b}}.

\bibitem[Nado et~al.(2021)Nado, Band, Collier, Djolonga, Dusenberry, Farquhar,
  Feng, Filos, Havasi, Jenatton, et~al.]{nado2021uncertainty}
Zachary Nado, Neil Band, Mark Collier, Josip Djolonga, Michael~W Dusenberry,
  Sebastian Farquhar, Qixuan Feng, Angelos Filos, Marton Havasi, Rodolphe
  Jenatton, et~al.
\newblock Uncertainty baselines: Benchmarks for uncertainty \& robustness in
  deep learning.
\newblock \emph{arXiv preprint arXiv:2106.04015}, 2021.

\bibitem[Zhu et~al.(2023{\natexlab{a}})Zhu, Cheng, Zhang, and
  Liu]{zhu2023openmix}
Fei Zhu, Zhen Cheng, Xu-Yao Zhang, and Cheng-Lin Liu.
\newblock Openmix: Exploring outlier samples for misclassification detection.
\newblock In \emph{Proceedings of the IEEE/CVF Conference on Computer Vision
  and Pattern Recognition}, pages 12074--12083, 2023{\natexlab{a}}.

\bibitem[Huang et~al.(2023{\natexlab{c}})Huang, Ma, and
  Li]{huang2023patchcensor}
Yuheng Huang, Lei Ma, and Yuanchun Li.
\newblock Patchcensor: Patch robustness certification for transformers via
  exhaustive testing.
\newblock \emph{ACM Transactions on Software Engineering and Methodology},
  32\penalty0 (6):\penalty0 1--34, 2023{\natexlab{c}}.

\bibitem[Kim et~al.(2023{\natexlab{b}})Kim, Feldt, and Yoo]{kim2023evaluating}
Jinhan Kim, Robert Feldt, and Shin Yoo.
\newblock Evaluating surprise adequacy for deep learning system testing.
\newblock \emph{ACM Transactions on Software Engineering and Methodology},
  32\penalty0 (2):\penalty0 1--29, 2023{\natexlab{b}}.

\bibitem[Gomes et~al.(2024)Gomes, Romanelli, Pichler, and
  Piantanida]{gomes2024a}
Eduardo Dadalto~C{\^a}mara Gomes, Marco Romanelli, Georg Pichler, and Pablo
  Piantanida.
\newblock A data-driven measure of relative uncertainty for misclassification
  detection.
\newblock In \emph{The Twelfth International Conference on Learning
  Representations}, 2024.
\newblock URL \url{https://openreview.net/forum?id=ruGY8v10mK}.

\bibitem[Liu et~al.(2024{\natexlab{a}})Liu, TIAN, Li, Ma, and
  Wang]{liu2024neuron}
Yibing Liu, Chris~XING TIAN, Haoliang Li, Lei Ma, and Shiqi Wang.
\newblock Neuron activation coverage: Rethinking out-of-distribution detection
  and generalization.
\newblock In \emph{The Twelfth International Conference on Learning
  Representations}, 2024{\natexlab{a}}.
\newblock URL \url{https://openreview.net/forum?id=SNGXbZtK6Q}.

\bibitem[Ferreira et~al.(2021)Ferreira, Arlat, Guiochet, and
  Waeselynck]{ferreira2021benchmarking}
Raul~Sena Ferreira, Jean Arlat, J{\'e}r{\'e}mie Guiochet, and H{\'e}l{\`e}ne
  Waeselynck.
\newblock Benchmarking safety monitors for image classifiers with machine
  learning.
\newblock In \emph{2021 IEEE 26th Pacific Rim International Symposium on
  Dependable Computing (PRDC)}, pages 7--16. IEEE, 2021.

\bibitem[Xiao et~al.(2021)Xiao, Beschastnikh, Rosenblum, Sun, Elbaum, Lin, and
  Dong]{xiao2021self}
Yan Xiao, Ivan Beschastnikh, David~S Rosenblum, Changsheng Sun, Sebastian
  Elbaum, Yun Lin, and Jin~Song Dong.
\newblock Self-checking deep neural networks in deployment. in 2021 ieee/acm
  43rd international conference on software engineering (icse).
\newblock \emph{IEEE, 372{\'s}384}, 2021.

\bibitem[Junges et~al.(2021)Junges, Torfah, and Seshia]{junges2021runtime}
Sebastian Junges, Hazem Torfah, and Sanjit~A Seshia.
\newblock Runtime monitors for markov decision processes.
\newblock In \emph{International Conference on Computer Aided Verification},
  pages 553--576. Springer, 2021.

\bibitem[Balakrishnan et~al.(2021)Balakrishnan, Deshmukh, Hoxha, Yamaguchi, and
  Fainekos]{balakrishnan2021percemon}
Anand Balakrishnan, Jyotirmoy Deshmukh, Bardh Hoxha, Tomoya Yamaguchi, and
  Georgios Fainekos.
\newblock Percemon: online monitoring for perception systems.
\newblock In \emph{Runtime Verification: 21st International Conference, RV
  2021, Virtual Event, October 11--14, 2021, Proceedings 21}, pages 297--308.
  Springer, 2021.

\bibitem[Ayerdi et~al.(2023)Ayerdi, Iriarte, Valle, Roman, Illarramendi, and
  Arrieta]{ayerdi2023metamorphic}
Jon Ayerdi, Asier Iriarte, Pablo Valle, Ibai Roman, Miren Illarramendi, and
  Aitor Arrieta.
\newblock Metamorphic runtime monitoring of autonomous driving systems.
\newblock \emph{arXiv preprint arXiv:2310.07414}, 2023.

\bibitem[K{\"o}nighofer et~al.(2023)K{\"o}nighofer, Rudolf, Palmisano, Tappler,
  and Bloem]{konighofer2023online}
Bettina K{\"o}nighofer, Julian Rudolf, Alexander Palmisano, Martin Tappler, and
  Roderick Bloem.
\newblock Online shielding for reinforcement learning.
\newblock \emph{Innovations in Systems and Software Engineering}, 19\penalty0
  (4):\penalty0 379--394, 2023.

\bibitem[Gronauer et~al.(2024)Gronauer, Haider, da~Roza, and
  Diepold]{gronauer2024reinforcement}
Sven Gronauer, Tom Haider, Felippe~Schmoeller da~Roza, and Klaus Diepold.
\newblock Reinforcement learning with ensemble model predictive safety
  certification.
\newblock \emph{arXiv preprint arXiv:2402.04182}, 2024.

\bibitem[Sun et~al.(2024{\natexlab{b}})Sun, Poskitt, Zhang, and
  Sun]{sun2024redriver}
Yang Sun, Christopher~M Poskitt, Xiaodong Zhang, and Jun Sun.
\newblock Redriver: Runtime enforcement for autonomous vehicles.
\newblock \emph{arXiv preprint arXiv:2401.02253}, 2024{\natexlab{b}}.

\bibitem[Cao et~al.(2023)Cao, Fatemi, Cheung, and Shabanian]{cao2023systematic}
Meng Cao, Mehdi Fatemi, Jackie~CK Cheung, and Samira Shabanian.
\newblock Systematic rectification of language models via dead-end analysis.
\newblock In \emph{The Eleventh International Conference on Learning
  Representations}, 2023.
\newblock URL \url{https://openreview.net/forum?id=k8_yVW3Wqln}.

\bibitem[Mudgal et~al.(2023)Mudgal, Lee, Ganapathy, Li, Wang, Huang, Chen,
  Cheng, Collins, Strohman, et~al.]{mudgal2023controlled}
Sidharth Mudgal, Jong Lee, Harish Ganapathy, YaGuang Li, Tao Wang, Yanping
  Huang, Zhifeng Chen, Heng-Tze Cheng, Michael Collins, Trevor Strohman, et~al.
\newblock Controlled decoding from language models.
\newblock \emph{arXiv preprint arXiv:2310.17022}, 2023.

\bibitem[Chang et~al.(2023)Chang, Wang, Wang, Wu, Yang, Zhu, Chen, Yi, Wang,
  Wang, et~al.]{chang2023survey}
Yupeng Chang, Xu~Wang, Jindong Wang, Yuan Wu, Linyi Yang, Kaijie Zhu, Hao Chen,
  Xiaoyuan Yi, Cunxiang Wang, Yidong Wang, et~al.
\newblock A survey on evaluation of large language models.
\newblock \emph{ACM Transactions on Intelligent Systems and Technology}, 2023.

\bibitem[Zhang et~al.(2023{\natexlab{a}})Zhang, Deng, Liu, Pan, and
  Bing]{zhang2023sentiment}
Wenxuan Zhang, Yue Deng, Bing Liu, Sinno~Jialin Pan, and Lidong Bing.
\newblock Sentiment analysis in the era of large language models: A reality
  check.
\newblock \emph{arXiv preprint arXiv:2305.15005}, 2023{\natexlab{a}}.

\bibitem[Pu et~al.(2023)Pu, Gao, and Wan]{pu2023summarization}
Xiao Pu, Mingqi Gao, and Xiaojun Wan.
\newblock Summarization is (almost) dead.
\newblock \emph{arXiv preprint arXiv:2309.09558}, 2023.

\bibitem[Lemieux et~al.(2023)Lemieux, Inala, Lahiri, and
  Sen]{lemieux2023codamosa}
Caroline Lemieux, Jeevana~Priya Inala, Shuvendu~K Lahiri, and Siddhartha Sen.
\newblock Codamosa: Escaping coverage plateaus in test generation with
  pre-trained large language models.
\newblock In \emph{International conference on software engineering (ICSE)},
  2023.

\bibitem[Liu et~al.(2023{\natexlab{b}})Liu, Chen, Wang, Che, Huang, Hu, and
  Wang]{liu2023fill}
Zhe Liu, Chunyang Chen, Junjie Wang, Xing Che, Yuekai Huang, Jun Hu, and Qing
  Wang.
\newblock Fill in the blank: Context-aware automated text input generation for
  mobile gui testing.
\newblock In \emph{2023 IEEE/ACM 45th International Conference on Software
  Engineering (ICSE)}, pages 1355--1367. IEEE, 2023{\natexlab{b}}.

\bibitem[Prenner et~al.(2022)Prenner, Babii, and Robbes]{prenner2022can}
Julian~Aron Prenner, Hlib Babii, and Romain Robbes.
\newblock Can openai's codex fix bugs? an evaluation on quixbugs.
\newblock In \emph{Proceedings of the Third International Workshop on Automated
  Program Repair}, pages 69--75, 2022.

\bibitem[Sobania et~al.(2023)Sobania, Briesch, Hanna, and
  Petke]{sobania2023analysis}
Dominik Sobania, Martin Briesch, Carol Hanna, and Justyna Petke.
\newblock An analysis of the automatic bug fixing performance of chatgpt.
\newblock \emph{arXiv preprint arXiv:2301.08653}, 2023.

\bibitem[Nijkamp et~al.(2023)Nijkamp, Pang, Hayashi, Tu, Wang, Zhou, Savarese,
  and Xiong]{nijkamp2023codegen}
Erik Nijkamp, Bo~Pang, Hiroaki Hayashi, Lifu Tu, Huan Wang, Yingbo Zhou, Silvio
  Savarese, and Caiming Xiong.
\newblock Codegen: An open large language model for code with multi-turn
  program synthesis, 2023.

\bibitem[Fan et~al.(2023)Fan, Gokkaya, Harman, Lyubarskiy, Sengupta, Yoo, and
  Zhang]{fan2023large}
Angela Fan, Beliz Gokkaya, Mark Harman, Mitya Lyubarskiy, Shubho Sengupta, Shin
  Yoo, and Jie~M Zhang.
\newblock Large language models for software engineering: Survey and open
  problems.
\newblock \emph{arXiv preprint arXiv:2310.03533}, 2023.

\bibitem[Hou et~al.(2024)Hou, Zhao, Liu, Yang, Wang, Li, Luo, Lo, Grundy, and
  Wang]{hou2024large}
Xinyi Hou, Yanjie Zhao, Yue Liu, Zhou Yang, Kailong Wang, Li~Li, Xiapu Luo,
  David Lo, John Grundy, and Haoyu Wang.
\newblock Large language models for software engineering: A systematic
  literature review, 2024.

\bibitem[Wang et~al.(2024)Wang, Huang, Chen, Liu, Wang, and
  Wang]{wang2024software}
Junjie Wang, Yuchao Huang, Chunyang Chen, Zhe Liu, Song Wang, and Qing Wang.
\newblock Software testing with large language models: Survey, landscape, and
  vision.
\newblock \emph{IEEE Transactions on Software Engineering}, 2024.

\bibitem[Bubeck et~al.(2023)Bubeck, Chandrasekaran, Eldan, Gehrke, Horvitz,
  Kamar, Lee, Lee, Li, Lundberg, et~al.]{bubeck2023sparks}
S{\'e}bastien Bubeck, Varun Chandrasekaran, Ronen Eldan, Johannes Gehrke, Eric
  Horvitz, Ece Kamar, Peter Lee, Yin~Tat Lee, Yuanzhi Li, Scott Lundberg,
  et~al.
\newblock Sparks of artificial general intelligence: Early experiments with
  gpt-4.
\newblock \emph{arXiv preprint arXiv:2303.12712}, 2023.

\bibitem[Zhang et~al.(2023{\natexlab{b}})Zhang, Li, Cui, Cai, Liu, Fu, Huang,
  Zhao, Zhang, Chen, et~al.]{zhang2023siren}
Yue Zhang, Yafu Li, Leyang Cui, Deng Cai, Lemao Liu, Tingchen Fu, Xinting
  Huang, Enbo Zhao, Yu~Zhang, Yulong Chen, et~al.
\newblock Siren's song in the ai ocean: a survey on hallucination in large
  language models.
\newblock \emph{arXiv preprint arXiv:2309.01219}, 2023{\natexlab{b}}.

\bibitem[Xu et~al.(2024)Xu, Jain, and Kankanhalli]{xu2024hallucination}
Ziwei Xu, Sanjay Jain, and Mohan Kankanhalli.
\newblock Hallucination is inevitable: An innate limitation of large language
  models.
\newblock \emph{arXiv preprint arXiv:2401.11817}, 2024.

\bibitem[Liu et~al.(2023{\natexlab{c}})Liu, Deng, Xu, Li, Zheng, Zhang, Zhao,
  Zhang, and Liu]{liu2023jailbreaking}
Yi~Liu, Gelei Deng, Zhengzi Xu, Yuekang Li, Yaowen Zheng, Ying Zhang, Lida
  Zhao, Tianwei Zhang, and Yang Liu.
\newblock Jailbreaking chatgpt via prompt engineering: An empirical study.
\newblock \emph{arXiv preprint arXiv:2305.13860}, 2023{\natexlab{c}}.

\bibitem[Huang et~al.(2023{\natexlab{d}})Huang, Chen, Mishra, Zheng, Yu, Song,
  and Zhou]{huang2023large}
Jie Huang, Xinyun Chen, Swaroop Mishra, Huaixiu~Steven Zheng, Adams~Wei Yu,
  Xinying Song, and Denny Zhou.
\newblock Large language models cannot self-correct reasoning yet.
\newblock \emph{arXiv preprint arXiv:2310.01798}, 2023{\natexlab{d}}.

\bibitem[Shinn et~al.(2024)Shinn, Cassano, Gopinath, Narasimhan, and
  Yao]{shinn2024reflexion}
Noah Shinn, Federico Cassano, Ashwin Gopinath, Karthik Narasimhan, and Shunyu
  Yao.
\newblock Reflexion: Language agents with verbal reinforcement learning.
\newblock \emph{Advances in Neural Information Processing Systems}, 36, 2024.

\bibitem[Chen et~al.(2023)Chen, Lin, Sch{\"a}rli, and Zhou]{chen2023teaching}
Xinyun Chen, Maxwell Lin, Nathanael Sch{\"a}rli, and Denny Zhou.
\newblock Teaching large language models to self-debug.
\newblock \emph{arXiv preprint arXiv:2304.05128}, 2023.

\bibitem[Pan et~al.(2023)Pan, Saxon, Xu, Nathani, Wang, and
  Wang]{pan2023automatically}
Liangming Pan, Michael Saxon, Wenda Xu, Deepak Nathani, Xinyi Wang, and
  William~Yang Wang.
\newblock Automatically correcting large language models: Surveying the
  landscape of diverse self-correction strategies.
\newblock \emph{arXiv preprint arXiv:2308.03188}, 2023.

\bibitem[Gou et~al.(2024)Gou, Shao, Gong, yelong shen, Yang, Duan, and
  Chen]{gou2024critic}
Zhibin Gou, Zhihong Shao, Yeyun Gong, yelong shen, Yujiu Yang, Nan Duan, and
  Weizhu Chen.
\newblock {CRITIC}: Large language models can self-correct with
  tool-interactive critiquing.
\newblock In \emph{The Twelfth International Conference on Learning
  Representations}, 2024.
\newblock URL \url{https://openreview.net/forum?id=Sx038qxjek}.

\bibitem[Madaan et~al.(2024)Madaan, Tandon, Gupta, Hallinan, Gao, Wiegreffe,
  Alon, Dziri, Prabhumoye, Yang, et~al.]{madaan2024self}
Aman Madaan, Niket Tandon, Prakhar Gupta, Skyler Hallinan, Luyu Gao, Sarah
  Wiegreffe, Uri Alon, Nouha Dziri, Shrimai Prabhumoye, Yiming Yang, et~al.
\newblock Self-refine: Iterative refinement with self-feedback.
\newblock \emph{Advances in Neural Information Processing Systems}, 36, 2024.

\bibitem[Bang et~al.(2023)Bang, Cahyawijaya, Lee, Dai, Su, Wilie, Lovenia, Ji,
  Yu, Chung, Do, Xu, and Fung]{bang2023multitask}
Yejin Bang, Samuel Cahyawijaya, Nayeon Lee, Wenliang Dai, Dan Su, Bryan Wilie,
  Holy Lovenia, Ziwei Ji, Tiezheng Yu, Willy Chung, Quyet~V. Do, Yan Xu, and
  Pascale Fung.
\newblock A multitask, multilingual, multimodal evaluation of chatgpt on
  reasoning, hallucination, and interactivity, 2023.

\bibitem[Liang et~al.(2023)Liang, Bommasani, Lee, Tsipras, Soylu, Yasunaga,
  Zhang, Narayanan, Wu, Kumar, Newman, Yuan, Yan, Zhang, Cosgrove, Manning, Re,
  Acosta-Navas, Hudson, Zelikman, Durmus, Ladhak, Rong, Ren, Yao, WANG,
  Santhanam, Orr, Zheng, Yuksekgonul, Suzgun, Kim, Guha, Chatterji, Khattab,
  Henderson, Huang, Chi, Xie, Santurkar, Ganguli, Hashimoto, Icard, Zhang,
  Chaudhary, Wang, Li, Mai, Zhang, and Koreeda]{liang2023holistic}
Percy Liang, Rishi Bommasani, Tony Lee, Dimitris Tsipras, Dilara Soylu,
  Michihiro Yasunaga, Yian Zhang, Deepak Narayanan, Yuhuai Wu, Ananya Kumar,
  Benjamin Newman, Binhang Yuan, Bobby Yan, Ce~Zhang, Christian~Alexander
  Cosgrove, Christopher~D Manning, Christopher Re, Diana Acosta-Navas,
  Drew~Arad Hudson, Eric Zelikman, Esin Durmus, Faisal Ladhak, Frieda Rong,
  Hongyu Ren, Huaxiu Yao, Jue WANG, Keshav Santhanam, Laurel Orr, Lucia Zheng,
  Mert Yuksekgonul, Mirac Suzgun, Nathan Kim, Neel Guha, Niladri~S. Chatterji,
  Omar Khattab, Peter Henderson, Qian Huang, Ryan~Andrew Chi, Sang~Michael Xie,
  Shibani Santurkar, Surya Ganguli, Tatsunori Hashimoto, Thomas Icard, Tianyi
  Zhang, Vishrav Chaudhary, William Wang, Xuechen Li, Yifan Mai, Yuhui Zhang,
  and Yuta Koreeda.
\newblock Holistic evaluation of language models.
\newblock \emph{Transactions on Machine Learning Research}, 2023.
\newblock ISSN 2835-8856.
\newblock URL \url{https://openreview.net/forum?id=iO4LZibEqW}.
\newblock Featured Certification, Expert Certification.

\bibitem[Qin et~al.(2023)Qin, Zhang, Zhang, Chen, Yasunaga, and
  Yang]{qin2023chatgpt}
Chengwei Qin, Aston Zhang, Zhuosheng Zhang, Jiaao Chen, Michihiro Yasunaga, and
  Diyi Yang.
\newblock Is chatgpt a general-purpose natural language processing task solver?
\newblock \emph{arXiv preprint arXiv:2302.06476}, 2023.

\bibitem[Frieder et~al.(2024)Frieder, Pinchetti, Griffiths, Salvatori,
  Lukasiewicz, Petersen, and Berner]{frieder2024mathematical}
Simon Frieder, Luca Pinchetti, Ryan-Rhys Griffiths, Tommaso Salvatori, Thomas
  Lukasiewicz, Philipp Petersen, and Julius Berner.
\newblock Mathematical capabilities of chatgpt.
\newblock \emph{Advances in Neural Information Processing Systems}, 36, 2024.

\bibitem[Liu et~al.(2023{\natexlab{d}})Liu, Ning, Teng, Liu, Zhou, and
  Zhang]{liu2023evaluating}
Hanmeng Liu, Ruoxi Ning, Zhiyang Teng, Jian Liu, Qiji Zhou, and Yue Zhang.
\newblock Evaluating the logical reasoning ability of chatgpt and gpt-4.
\newblock \emph{arXiv preprint arXiv:2304.03439}, 2023{\natexlab{d}}.

\bibitem[Fu et~al.(2023)Fu, Ou, Chen, Wan, Peng, and Khot]{fu2023chain}
Yao Fu, Litu Ou, Mingyu Chen, Yuhao Wan, Hao Peng, and Tushar Khot.
\newblock Chain-of-thought hub: A continuous effort to measure large language
  models' reasoning performance.
\newblock \emph{arXiv preprint arXiv:2305.17306}, 2023.

\bibitem[Du et~al.(2023)Du, Liu, Wang, Wang, Liu, Chen, Feng, Sha, Peng, and
  Lou]{du2023classeval}
Xueying Du, Mingwei Liu, Kaixin Wang, Hanlin Wang, Junwei Liu, Yixuan Chen,
  Jiayi Feng, Chaofeng Sha, Xin Peng, and Yiling Lou.
\newblock Classeval: A manually-crafted benchmark for evaluating llms on
  class-level code generation.
\newblock \emph{arXiv preprint arXiv:2308.01861}, 2023.

\bibitem[Liu et~al.(2024{\natexlab{b}})Liu, Xia, Wang, and Zhang]{liu2024your}
Jiawei Liu, Chunqiu~Steven Xia, Yuyao Wang, and Lingming Zhang.
\newblock Is your code generated by chatgpt really correct? rigorous evaluation
  of large language models for code generation.
\newblock \emph{Advances in Neural Information Processing Systems}, 36,
  2024{\natexlab{b}}.

\bibitem[Honovich et~al.(2022)Honovich, Aharoni, Herzig, Taitelbaum, Kukliansy,
  Cohen, Scialom, Szpektor, Hassidim, and Matias]{honovich2022true}
Or~Honovich, Roee Aharoni, Jonathan Herzig, Hagai Taitelbaum, Doron Kukliansy,
  Vered Cohen, Thomas Scialom, Idan Szpektor, Avinatan Hassidim, and Yossi
  Matias.
\newblock True: Re-evaluating factual consistency evaluation.
\newblock \emph{arXiv preprint arXiv:2204.04991}, 2022.

\bibitem[Zhu et~al.(2023{\natexlab{b}})Zhu, Wang, Zhou, Wang, Chen, Wang, Yang,
  Ye, Gong, Zhang, et~al.]{zhu2023promptbench}
Kaijie Zhu, Jindong Wang, Jiaheng Zhou, Zichen Wang, Hao Chen, Yidong Wang,
  Linyi Yang, Wei Ye, Neil~Zhenqiang Gong, Yue Zhang, et~al.
\newblock Promptbench: Towards evaluating the robustness of large language
  models on adversarial prompts.
\newblock \emph{arXiv preprint arXiv:2306.04528}, 2023{\natexlab{b}}.

\bibitem[Hamid et~al.(2023)Hamid, Samidi, Finin, Pappachan, and
  Yus]{hamid2023genaipabench}
Aamir Hamid, Hemanth~Reddy Samidi, Tim Finin, Primal Pappachan, and Roberto
  Yus.
\newblock Genaipabench: A benchmark for generative ai-based privacy assistants.
\newblock \emph{arXiv preprint arXiv:2309.05138}, 2023.

\bibitem[Santhanam et~al.(2021)Santhanam, Hedayatnia, Gella, Padmakumar, Kim,
  Liu, and Hakkani-T{\"u}r]{santhanam2021rome}
Sashank Santhanam, Behnam Hedayatnia, Spandana Gella, Aishwarya Padmakumar,
  Seokhwan Kim, Yang Liu, and Dilek Hakkani-T{\"u}r.
\newblock Rome was built in 1776: A case study on factual correctness in
  knowledge-grounded response generation.
\newblock 2021.

\bibitem[Honovich et~al.(2021)Honovich, Choshen, Aharoni, Neeman, Szpektor, and
  Abend]{honovich2021q}
Or~Honovich, Leshem Choshen, Roee Aharoni, Ella Neeman, Idan Szpektor, and Omri
  Abend.
\newblock Q2: Evaluating factual consistency in knowledge-grounded dialogues
  via question generation and question answering.
\newblock \emph{arXiv preprint arXiv:2104.08202}, 2021.

\bibitem[Thorne et~al.(2018)Thorne, Vlachos, Christodoulopoulos, and
  Mittal]{thorne2018fever}
James Thorne, Andreas Vlachos, Christos Christodoulopoulos, and Arpit Mittal.
\newblock Fever: a large-scale dataset for fact extraction and verification.
\newblock \emph{arXiv preprint arXiv:1803.05355}, 2018.

\bibitem[Wang et~al.(2021{\natexlab{d}})Wang, Xu, Wang, Gan, Cheng, Gao,
  Awadallah, and Li]{wang2021adversarial}
Boxin Wang, Chejian Xu, Shuohang Wang, Zhe Gan, Yu~Cheng, Jianfeng Gao,
  Ahmed~Hassan Awadallah, and Bo~Li.
\newblock Adversarial glue: A multi-task benchmark for robustness evaluation of
  language models.
\newblock \emph{arXiv preprint arXiv:2111.02840}, 2021{\natexlab{d}}.

\bibitem[Wang et~al.(2021{\natexlab{e}})Wang, Liu, Gui, Zhang, Zou, Zhou, Ye,
  Zhang, Zheng, Pang, Wu, Li, Zhang, Ma, Fei, Cai, Zhao, Hu, Yan, Tan, Hu,
  Bian, Liu, Qin, Zhu, Xing, Fu, Zhang, Peng, Zheng, Zhou, Wei, Qiu, and
  Huang]{wang-etal-2021-textflint}
Xiao Wang, Qin Liu, Tao Gui, Qi~Zhang, Yicheng Zou, Xin Zhou, Jiacheng Ye,
  Yongxin Zhang, Rui Zheng, Zexiong Pang, Qinzhuo Wu, Zhengyan Li, Chong Zhang,
  Ruotian Ma, Zichu Fei, Ruijian Cai, Jun Zhao, Xingwu Hu, Zhiheng Yan, Yiding
  Tan, Yuan Hu, Qiyuan Bian, Zhihua Liu, Shan Qin, Bolin Zhu, Xiaoyu Xing,
  Jinlan Fu, Yue Zhang, Minlong Peng, Xiaoqing Zheng, Yaqian Zhou, Zhongyu Wei,
  Xipeng Qiu, and Xuanjing Huang.
\newblock {T}ext{F}lint: Unified multilingual robustness evaluation toolkit for
  natural language processing.
\newblock In Heng Ji, Jong~C. Park, and Rui Xia, editors, \emph{Proceedings of
  the 59th Annual Meeting of the Association for Computational Linguistics and
  the 11th International Joint Conference on Natural Language Processing:
  System Demonstrations}, pages 347--355, Online, August 2021{\natexlab{e}}.
  Association for Computational Linguistics.
\newblock \doi{10.18653/v1/2021.acl-demo.41}.
\newblock URL \url{https://aclanthology.org/2021.acl-demo.41}.

\bibitem[Zhuang et~al.(2021)Zhuang, Wayne, Ya, and Jun]{zhuang2021robustly}
Liu Zhuang, Lin Wayne, Shi Ya, and Zhao Jun.
\newblock A robustly optimized bert pre-training approach with post-training.
\newblock In \emph{Proceedings of the 20th chinese national conference on
  computational linguistics}, pages 1218--1227, 2021.

\bibitem[Ahmed et~al.(2020)Ahmed, Seraj, and Islam]{ahmed2020k}
Mohiuddin Ahmed, Raihan Seraj, and Syed Mohammed~Shamsul Islam.
\newblock The k-means algorithm: A comprehensive survey and performance
  evaluation.
\newblock \emph{Electronics}, 9\penalty0 (8):\penalty0 1295, 2020.

\bibitem[Arora et~al.(2021)Arora, Huang, and He]{arora2021types}
Udit Arora, William Huang, and He~He.
\newblock Types of out-of-distribution texts and how to detect them.
\newblock \emph{arXiv preprint arXiv:2109.06827}, 2021.

\bibitem[Raunak et~al.(2021)Raunak, Menezes, and
  Junczys-Dowmunt]{raunak2021curious}
Vikas Raunak, Arul Menezes, and Marcin Junczys-Dowmunt.
\newblock The curious case of hallucinations in neural machine translation,
  2021.

\bibitem[Post(2018)]{post2018call}
Matt Post.
\newblock A call for clarity in reporting bleu scores.
\newblock \emph{arXiv preprint arXiv:1804.08771}, 2018.

\end{thebibliography}


\clearpage

\appendix

\section*{Collected Safety Analysis Methods}
\label{appendix:monitor}

In this section, we present more details about the online safety analysis methods studied.
The safety analysis methods can be divided into three categories: black-box (\random and \selfcheckgpt), white-box (\boxMon and \quan), and grey-box (\avgent, \maxent, \avglik, and \maxlik).

\begin{compactitem}[$\bullet$]
\item \textbf{Random}: Random method is a naive black prediction method.

\item \textbf{\selfcheckgpt}~\cite{manakul2023selfcheckgpt}: It is a representative example of black-box methods, which is designed as a sampling-based hallucination detection approach that operates independently of external resources.
The key idea is to compare sampled responses: if the model has knowledge of a concept, responses are expected to be consistent; otherwise, they may diverge. 
Several comparison methods are used to decide whether there exists a divergence, e.g., BERTScore~\cite{zhuang2021robustly}.
Here, we focus on the best-performing method \selfcheckgpt with prompt.
In particular, another LLM is queried to assess whether the given sentence is supported by the context using the prompt: \textit{Context: ...
Sentence: ... Is the sentence supported by the context above?}
In this work, to adapt to diverse safety requirements, we modify the prompt to the queried LLM according to different safety requirements. 
An example of its application in toxicity detection is that, given a partial output, we ask the queried LLM using this prompt: \textit{Are the following sentences toxic? ...} 
If the queried LLM returns \textit{Yes}, we consider the safety requirement is violated, otherwise, it is safetisfied. 
We choose \alpaca as the query LLM due to the budget limit.

\item \textbf{\boxMon}~\cite{henzinger2019outside}: 
Box-based method~\cite{henzinger2019outside} is a typical example of this type of technique.
It is proposed to analyze neural networks to detect novel inputs.
The method focuses on specific network layers close to the final output, where essential feature information is believed to be concentrated. 
Then, an abstraction, i.e., a box, is constructed to represent the set of known neuron valuations by observing the patterns exhibited by neurons in these layers.
Based on the constructed box, the analyzer is trained to recognize typical input patterns for each class during runtime. 
If the observed pattern deviates significantly from the expected behavior, the method raises a warning about a possible novelty in the input.
To perform an online safety analysis for LLM, we build the box on the collected safe input.
At runtime, if a new coming prompt triggers generation that is out-of-box, we consider it as an unsafe one and raise a warning.
we count the number of tokens that are out of the box and compare them to a predefined threshold to judge whether it is abnormal.

\item \textbf{\quan}~\cite{lukina2021into}: This approach is proposed for novelty detection.
Compared to traditional qualitative methods (like box-based), it assigns a numerical score to each observed input-output pair.
Similar to the \boxMon, it first collects a set of internal states of the selected layer by running a small set of prompts.
Then, it performs clustering, e.g., KMeans~\cite{ahmed2020k} on the states set and gets the centers of different clustering.
At runtime, given a new input, for the selected token, it measures the distances between the current states and different centers.
The minimum distance is compared with a pre-defined threshold: if the distance is larger, it is considered to be a potential violation of trustworthiness.


\item \textbf{\avgent}~\cite{arora2021types,huang2023look,manakul2023selfcheckgpt}: Huang et al.~\cite{huang2023look} investigate the correlation between uncertainty and LLMs’ performance on NLP tasks and code tasks. 
Manakul et al.~\cite{manakul2023selfcheckgpt} demonstrate that entropy and token likelihood can be used to detect hallucinations.
Hence, we adapt entropy and token likelihood into the online safety analysis for LLM.
We compute the sentence-level average entropy and compare it with a pre-defined threshold, to decide whether it could be a violation.
\begin{equation}
    \label{eq:avg_entropy}
    \small
    Avg(\mathcal{H})_i = \frac{1}{J}\sum_{j} \mathcal{H}_{ij},
\end{equation}
where $\mathcal{H}_{ij}$ is the entropy of this token's probability distribution.
The thresholds are different for each pair of tasks and models.
To ease the presentation, we put the threshold on the project website.

\item \textbf{\maxent}~\cite{arora2021types,huang2023look,manakul2023selfcheckgpt}:
The maximum entropy is computed as
\begin{equation}
    \label{eq:max_entropy}
    \small
    Max(\mathcal{H})_i = \underset{j}{max} [\mathcal{H}_{ij}].
\end{equation}

\item \textbf{\avglik}~\cite{huang2023look,manakul2023selfcheckgpt}:
The average likelihood is computed as
\begin{equation}
    \label{eq:avg_prob}
    \small
    Avg(-\log p)_i = -\frac{1}{J}\sum_{j} \log p_{ij},
\end{equation}
where $p_{ij}$ is the probability of a token at position $j$ of a sentence $i$.

\item \textbf{\maxlik}~\cite{huang2023look,manakul2023selfcheckgpt}:
The maximum likelihood is computed as
\begin{equation}
    \label{eq:max_prob}
    \small
    Max(-\log p)_i = \underset{j}{max} ( -\log p_{ij}).
\end{equation}

\end{compactitem}

\section*{Collected Metrics}
\label{appendix:metrics}
The full definition of the collected metrics is as follows.
The goal of the evaluation metrics is to evaluate the online safety analysis methods from different perspectives.

For Safety Gain (SG), Residual Hazard (RH), and Availability Cost (AC), we follow the definitions of Guerin et al.~\cite{guerin2022unifying}.
The original definition is for single inference of classic DNNs and it can be adapted to the context of LLM.

SG is used to measure the safety addition from the method, which is defined as 
\begin{equation}\label{eq:SG}
    SG = \int_{\mathcal{D}} p(x)\left(B^{\mathcal{S}}_{(N, m_N)}(x) - B^{\mathcal{S}}_{N}(x)\right) \, dx,
\end{equation}
where $\mathcal{D}$ is an entire operational domain of the ML model, $B^{\mathcal{S}}_{\_}$ is the safety return of the model running with/without the online safety analysis method, $N$ is the ML model, and $m_N$ is the online safety analysis method, $(N, m_N)$ is the model running under the supervision of the online safety analysis method.
The safety return $B^{\mathcal{S}}_{\_}$ is a measurement of the safety of the model.
In particular, in our study, it is defined as:
\begin{equation}\label{eq:returnHazard_classif_maintext}
B^{\mathcal{S}}_{(N,m_N)}(x) = 
\begin{cases} 
    0 & \text{else,}\\
    1 & \text{if } \textit{unsafe} \text{ and }  m_N(x) = 1.
\end{cases}
\end{equation}
In other words, the online safety analysis method gets safety scores only if it successfully reports an unsafe case.
$m_N(x) = 1$ means the online safety analysis method reports the model is in an unsafe scenario.
$B^{\mathcal{S}}_{N}(x) = 0$ in the study.
A greater SG means the online safety analysis method provides more safety to the execution of the model.
This metric represents the safety benefits of using the online safety analysis method. It focuses on how the online safety analysis method helps in preventing hazardous situations by detecting prediction errors and raising alerts when necessary. A higher Safety Gain indicates that the online safety analysis method is effectively improving the safety of the system.

RH is a measurement of the remaining unsafety, which is defined as:
\begin{equation}\label{eq:RH}
    RH = \int_{\mathcal{D}} p(x) \left(B^{\mathcal{S}}_{N^*}(x) - B^{\mathcal{S}}_{(N,m_N)}(x)\right) \, dx,
\end{equation}
where $\mathcal{D}$ is an entire operational domain of the ML model, $B^{\mathcal{S}}_{\_}$ is the safety return of the model running with/without the online safety analysis method, $N^*$ is the ideal ML model that can avoid all unsafe cases, and $m_N$ is the online safety analysis method.
$B^{\mathcal{S}}_{N^*}(x)$ is defined as:
\begin{equation}\label{eq:returnHazard_classif}
B^{\mathcal{S}}_{f^*}(x) = 1 \text{ if } \textit{unsafe}.
\end{equation}
A greater RH indicates that the online safety analysis method misses more dangerous events during the execution of the online safety analysis method.
The Residual Hazard metric measures the remaining safety gaps despite using the online safety analysis method. It compares the safety of the monitored model against the safety of an optimal model. A lower Residual Hazard value indicates that the online safety analysis method is successful in reducing the amount of hazard still present in the system.

AC is the decrease of the system performance, which is defined as:
\begin{equation}\label{eq:AC}
        AC = \int_{\mathcal{D}} p(x) \left(B^{\mathcal{M}}_{N}(x) - B^{\mathcal{M}}_{(N, m_N)}(x)\right) \, dx,
\end{equation}
where $\mathcal{D}$ is an entire operational domain of the ML model, $B^{\mathcal{M}}_{\_}$ is the mission return of the model running with/without the online safety analysis method, $N$ is the ML model, and $m_N$ is the online safety analysis method, $(N, m_N)$ is the model running under the supervision of the online safety analysis method.
A greater AC implies that the operation of the online safety analysis method is more costly. 
$B^{\mathcal{M}}_{\_}$ is the mission return by applying the online safety analysis method on the system, which is defined as:
\begin{equation}\label{eq:missionreturn_classif}
B^{\mathcal{M}}_{(N,m_N)}(x) = 
\begin{cases} 
    0.2 & \text{else,}\\
    -2 & \text{if } \textit{safe} \text{ and }  m_N(x) = 1.
\end{cases}
\end{equation}
That is to say, if the online safety analysis method performs a wrong reporting, the mission return will decrease by 2, while when the reporting is correct, the mission return will increase by 0.2.
$B^{\mathcal{M}}_{N}(x) = 0$ in our case.
Availability Cost quantifies the negative impact of the online safety analysis method on the system's performance. It evaluates how the online safety analysis method affects the system's ability to perform its mission by comparing the availability of the monitored model with the availability of the initial system. A lower Availability Cost suggests that the online safety analysis method is minimizing the performance impact on the system.

We also include classic metrics, i.e.,
Area Under the receiver operating characteristic Curve (AUC) and time cost.
AUC~\cite{fawcett2006introduction} is a traditional classification task metric that summarizes the binary classifier's performance.
The ROC curve is a graphical representation of the true positive rate against the false positive rate for various threshold values. AUC quantifies the overall performance of the model by calculating the area under this curve, with a value ranging from 0 to 1. 
A higher AUC value indicates better model performance.
Moreover, time cost~\cite{xie2023mosaic,zhang2022online,deshmukh2017robust} is another important metric to measure the overhead of the online safety analysis method.
The overhead of the online safety analysis method should be as small as possible, so that it does not bring additional debt to the model.

\section*{Collected Models}
\label{appendix:model}

For open-source model,
$max\_new\_token$ is set to 200 (for \mbpp and \humaneval) and 100, otherwise.
$temperature$ is set to be 1.

For the closed-source models, we use the following specific model type: \textsf{davinci-002} (GPT-3), \textsf
{gpt-3.5-turbo-0125} (GPT-3.5), and \textsf{gpt-4-0125-preview} (GPT-4).
We set the following parameters when calling the model, $max\_new\_token$ is set to 500 (for \mbpp and \humaneval) and 100, otherwise.
$temperature$ is set to be 1.

\section*{Collected Datasets}
\label{appendix:dataset}

\begin{compactitem}[$\bullet$]
\item \textbf{Open-source model evaluation}: For RealToxicityPrompt, we use a version provided by Wang et al.~\cite{wang2023decodingtrust}, which contain 1,200 toxic task prompts and 1,200 non-toxic task prompts.
\item \textbf{Closed-source model evaluation}: For closed-source model evaluation, due to the budget limit, for each dataset, we randomly choose 100 instances as the evaluation object.
\end{compactitem}

\begin{table}[!tb]
\caption{The collected NLP and programming tasks.}
\centering
\footnotesize
\renewcommand{\arraystretch}{1.1}
\label{table:appendix:studied_tasks}
  \resizebox{\columnwidth}{!}{
    \begin{tabular}{llcc}
    \toprule
    \textbf{Dataset} & \textbf{Task Domain}  & \multicolumn{1}{p{1.4cm}}{\centering \textbf{Size} \\ \textbf{(\#~instances)}}  & \multicolumn{1}{p{2cm}}{\centering \textbf{Trustworthiness} \\ \textbf{Perspectives}}  \ \\ \midrule
    {\bf TruthfulQA} & Question Answering  & 816 & Truthfulness \\
    {\bf TriviaQA} & Question Answering  & 3,605 & Truthfulness \\
    {\bf Natural Question} & Question Answering & 3,605 & Truthfulness \\
    {\bf RealToxicityPrompt} & Text Continuation & 2,400 & Toxicity\\
    {\bf WMT 2014}    & Machine Translation & 3,003 & BLEU\\
    {\bf MBPP} & Code Generation & 500 & Pass@1\\
    {\bf HumanEval} & Code Generation & 164 & Pass@1\\
    \bottomrule
    \end{tabular}
  }
\end{table}

\section*{Safety}
\label{appendix:safety}
The assessment of the trustworthiness is conducted as follows.
\begin{compactitem}[$\bullet$]
\item \textbf{Truthfulness}: Truthfulness of LLM typically refers to the accuracy and fidelity of generated information~\cite{raunak2021curious}. 
The ground-truth of the instance is judged by a fine-tuned GPT-3 model (GPT-judge)~\cite{lin2021truthfulqa}, which is the gold standard in the field, to
categorize each instance as truthful or untruthful.
\item \textbf{Toxicity}: 
Toxicity~\cite{gehman2020realtoxicityprompts} refers to the presence of harmful, offensive, or inappropriate language in textual data or generated outputs.
We use Perspective API~\cite{perspectiveapi}, an automated online model, accessed by Google APIs, for detecting hate speech and toxic words.
It returns the probability that the instance is toxic and we use 0.5 as the classification threshold.  
\item \textbf{BLEU}: The BLEU (Bilingual Evaluation Understudy)~\cite{post2018call} is a metric used to evaluate the quality of machine-translated text by comparing it to one or more human reference translations.
\item \textbf{Pass@1}: Pass@1~\cite{chen2021evaluating} is a common metric to evaluate the quality of the code generation.
It evaluates the quality of the top suggestion or the first completion proposed by the model.

\end{compactitem}

\section*{More Examples of Pilot Study}

\noindent\textbf{Example 2}:
In \mbpp with code generation task, an instance is with the prompt:
\begin{small}
\begin{verbatim}
------------------------------------------------
Write a python function to check whether the 
given array is monotonic or not. 
assert is_Monotonic([6, 5, 4, 4]) == True.
------------------------------------------------
\end{verbatim}
\end{small}
The prompt is given as the input to \codellama and the first $25\%$ of the output is shown in Listing~\ref{lst:codellamaoutputexample}.
When the length of the input array is 2 
, the function simply returns \texttt{True}, which is wrong.
\begin{figure}[h]
    \begin{lstlisting}[caption={Example of code generation.}, captionpos=b, language=Python, label=lst:codellamaoutputexample, basicstyle=\scriptsize\ttfamily, linewidth=0.45\textwidth]
def is_monotonic(arr):
  if len(arr) == 1:
    return True
  if len(arr) == 2:
    return True
  if len(arr) == 3:
    ...
    \end{lstlisting}
\vspace{-10pt}
\end{figure}

\end{document}